\documentclass[10pt]{article}

\usepackage[none,bottom,dark]{draftcopy}
\usepackage{amssymb,amsmath,amsthm,mathrsfs}
\usepackage[dvips]{graphics,epsfig}
\usepackage{verbatim}
\usepackage{colordvi,color}
\usepackage{psfrag}
\usepackage{colordvi,color,pspicture}
\usepackage{url}
\usepackage[authoryear,sort&compress]{natbib}

\usepackage{natbib}

\newcommand{\Y}{\mathcal{Y}}
\newcommand{\X}{\mathcal{X}}
\newcommand{\NY}{N_{\mathcal{Y}}}
\newcommand{\G}{\Gamma}

\newcommand{\N}{\mathcal N}
\newcommand{\V}{\mathcal{V}}
\newcommand{\A}{\mathcal{A}}
\newcommand{\R}{\mathbb{R}}
\newcommand{\I}{\mathbf{I}}
\newcommand{\y}{\mathsf{y}}
\newcommand{\U}{\mathcal{U}}

\newcommand{\E}[1]{\mathbf{E}\,#1}
\newcommand{\Var}[1]{\mathbf{Var}\,#1}
\newcommand{\Cov}[1]{\mathbf{Cov}\,#1}

\DeclareMathOperator{\argsup}{argsup}
\DeclareMathOperator{\PAE}{PAE}
\DeclareMathOperator{\HLAE}{HLAE}

\begin{document}
\title{Relative Density of the Random $r$-Factor Proximity Catch Digraph for Testing
Spatial Patterns of Segregation and Association}
\author{\it Elvan Ceyhan, Carey E. Priebe, \& John C. Wierman\\
\it Johns Hopkins University, Baltimore}
\date{\today}

\maketitle
\begin{abstract}
Statistical pattern classification methods based on data-random
 graphs were introduced recently.
In this approach, a
random directed graph is constructed from the data using the
relative positions of the data points from various classes.
Different random graphs result from different definitions of the
proximity region associated with each data point and different
graph statistics can be employed for data reduction.
The approach used in this article is based on a parameterized family
of proximity maps determining an associated family of data-random digraphs.
The relative arc density of the digraph is used
as the summary statistic, providing an alternative to the
domination number employed previously.
An important advantage of
the relative arc density is that, properly re-scaled, it is a
$U$-statistic, facilitating analytic study of its asymptotic
distribution using standard $U$-statistic central limit theory.
The approach is illustrated with an application to the testing of
spatial patterns of segregation and association.
Knowledge of the asymptotic distribution allows evaluation of the Pitman and
Hodges-Lehmann asymptotic efficacies, and selection of the proximity map
parameter to optimize efficiency.
Furthermore the approach presented here also
has the advantage of validity for data in any dimension.
\end{abstract}

\section{Introduction}
Classification and clustering have received considerable attention in the statistical literature.
In recent years, a new classification approach has been developed which is based on the relative
 positions of the data points from various classes.
Priebe et al. introduced the class cover catch digraphs (CCCD) in $\R$ and gave the exact and
the asymptotic distribution of the domination number of the CCCD \cite{priebe:2001}.
DeVinney et al. \cite{devinney:2002a}, Marchette and Priebe \cite{marchette:2003},
Priebe et al. \cite{priebe:2003a}, \cite{priebe:2003b} applied the concept in higher
dimensions and demonstrated relatively good performance of CCCD in classification.
The methods employed involve data reduction (condensing) by using approximate
minimum dominating sets as prototype sets (since finding the exact minimum
dominating set is an NP-hard problem ---in particular for CCCD).
Furthermore the exact and the asymptotic distribution of the domination number
of the CCCD are not analytically tractable in multiple dimensions.

Ceyhan and Priebe introduced the central similarity proximity map and $r$-factor
proximity maps and the associated random digraphs in \cite{ceyhan:CS-JSM-2003}
and \cite{ceyhan:TR-dom-num-NPE-spatial}, respectively.
In both cases, the space is partitioned by the Delaunay tessellation which
is the Delaunay triangulation in $\R^2$.
In each triangle, a family of
data-random proximity catch digraphs is constructed based on the proximity
of the points to each other.
The advantages of the $r$-factor proximity catch
digraphs are that an exact minimum dominating set can be found in polynomial
time and the asymptotic distribution of the domination number is analytically tractable.
The latter is then used to test segregation and association of
points of different classes in \cite{ceyhan:TR-dom-num-NPE-spatial}.
Segregation and assocation are two patterns that describe the
spatial relation between two or more classes.
See Section \ref{sec:null-and-alt} for more detail.

In this article, we employ a different statistic, namely the relative (arc) density,
that is the proportion of all possible arcs (directed edges) which are present
in the data random digraph.
This test statistic has the advantage that, properly rescaled, it is a $U$-statistic.
Two plain classes of alternative hypotheses, for segregation and association,
are defined in Section \ref{sec:null-and-alt}.
The asymptotic distributions under both the null and the alternative hypotheses
are determined in Section \ref{sec:asy-norm} by using standard $U$-statistic central limit theory.
Pitman and Hodges-Lehman asymptotic efficacies are analyzed in Sections \ref{sec:Pitman}
and \ref{sec:Hodges-Lehmann}, respectively.
This test is related to the available tests of segregation and association
in the ecology literature, such as Pielou's test and Ripley's test.
See discussion in Section \ref{sec:discussion} for more detail.
Our approach is valid for data in any dimension, but for simplicity
of expression and visualization, will be described for two-dimensional data.

\section{Preliminaries}
\subsection{Proximity Maps}
Let $(\Omega,\mathcal{M})$ be a measurable space and
consider a function $N:\Omega \times 2^{\Omega} \rightarrow 2^{\Omega}$,
where $2^{\Omega}$ represents the power set of $\Omega$.
Then given $\Y \subseteq \Omega$,
the {\em proximity map}
$\NY(\cdot) = N(\cdot,\Y): \Omega \rightarrow \wp(\Omega)$
associates with each point $x \in \Omega$
a {\em proximity region} $\NY(x) \subset \Omega$.
Typically, $N$ is chosen
to satisfy $x \in \NY(x)$ for all $x \in \Omega$.
 The use of the adjective \emph{proximity} comes form thinking of
the region $\NY(x)$ as representing a neighborhood of points ``close" to $x$.
(\cite{toussaint:1980,jaromczyk:1992}.)

\subsection{$r$-Factor Proximity Maps}
\begin{figure} [ht]
    \centering
   \scalebox{.4}{\input{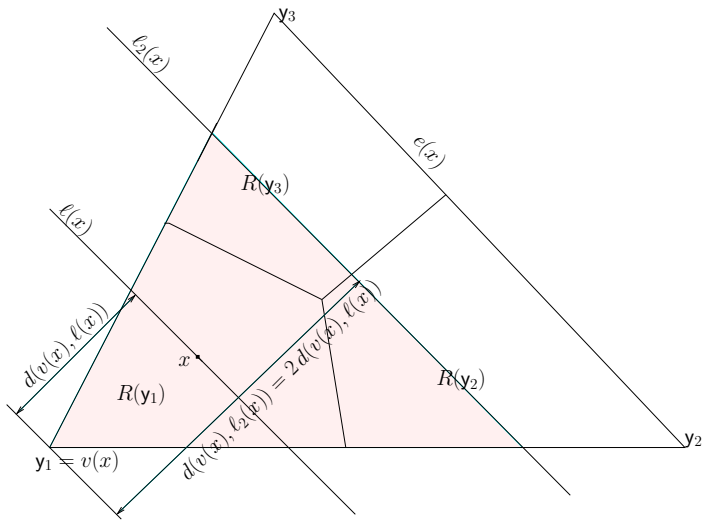}}
    \caption{Construction of $r$-factor proximity region, $\NY^2(x)$ (shaded region).}
\label{fig:ProxMapDef}
    \end{figure}

We now briefly define $r$-factor proximity maps.
(See Ceyhan and Priebe \cite{ceyhan:TR-dom-num-NPE-spatial} for more details).
Let $\Omega = \R^2$
and let $\Y = \{\y_1,\y_2,\y_3\} \subset \R^2$
be three non-collinear points.
Denote by $T(\Y)$ the triangle ---including the interior---
formed by the three points
(i.e. $T(\Y)$ is the convex hull of $\Y$).
For $r \in [1,\infty]$,
define $\NY^r$ to be the {\em r-factor} proximity map as follows;
see also Figure \ref{fig:ProxMapDef}.
Using line segments from the center of mass (centroid) of $T(\Y)$ to the midpoints of its edges,
we partition $T(\Y)$ into ``vertex regions" $R(\y_1)$, $R(\y_2)$, and $R(\y_3)$.
For $x \in T(\Y) \setminus \Y$, let $v(x) \in \Y$ be the
vertex in whose region $x$ falls, so $x \in R(v(x))$.
If $x$ falls on the boundary of two vertex regions,
 we assign $v(x)$ arbitrarily to one of the adjacent regions.
Let $e(x)$ be the edge of $T(\Y)$ opposite $v(x)$.
Let $\ell(x)$ be the line parallel to $e(x)$ through $x$.
Let $d(v(x),\ell(x))$ be the Euclidean (perpendicular) distance from $v(x)$ to $\ell(x)$.
For $r \in [1,\infty)$, let $\ell_r(x)$ be the line parallel to $e(x)$
such that $d(v(x),\ell_r(x)) = rd(v(x),\ell(x))$ and $d(\ell(x),\ell_r(x)) < d(v(x),\ell_r(x))$.
Let $T_r(x)$ be
the triangle similar to
and with the same orientation as $T(\Y)$
having $v(x)$ as a vertex
and $\ell_r(x)$ as the opposite edge.
Then the {\em r-factor} proximity region
$\NY^r(x)$ is defined to be $T_r(x) \cap T(\Y)$.
Notice that $r \ge 1$ implies $x \in \NY^r(x)$.
Note also that
$\lim_{r \rightarrow \infty} \NY^r(x) = T(\Y)$
for all $x \in T(\Y) \setminus \Y$,
so we define $\NY^{\infty}(x) = T(\Y)$ for all such $x$.
For $x \in \Y$, we define $\NY^r(x) = \{x\}$ for all $r \in [1,\infty]$.

\subsection{Data-Random Proximity Catch Digraphs}
\label{ref:data-random-PCD}
If $\X_n:=\{X_1,X_2,\cdots,X_n\}$ is a set of $\Omega$-valued random variables,
then the $\NY(X_i),\; i=1,\cdots,n$, are random sets.
If the $X_i$ are independent and identically distributed,
then so are the random sets $\NY(X_i)$.

In the case of an $r$-factor proximity map, notice that if $X_i \stackrel{iid}{\sim} F$
and $F$ has a non-degenerate two-dimensional
probability density function $f$ with support$(f) \subseteq T(\Y)$,
 then the special case in the construction
of $\NY^r$ ---
$X$ falls on the boundary of two vertex regions ---
occurs with probability zero.

The proximities of the data points to each other are used to construct a digraph.
A digraph is a directed graph; i.e. a graph with directed edges from one vertex
to another based on a binary relation.
Define the data-random proximity catch digraph $D$
with vertex set $\V=\{X_1,\cdots,X_n\}$
and arc set $\A$ by
$(X_i,X_j) \in \A \iff X_j \in \NY(X_i)$.  Since this relationship is not symmetric,
a digraph is needed rather than a graph.
The random digraph $D$ depends on
the (joint) distribution of the $X_i$ and on
the map $\NY$.

\subsection{Relative Density}
The \emph{relative arc density} of a digraph $D=(\V,\A)$
of order $|\V| = n$,
denoted $\rho(D)$,
is defined as
$$
\rho(D) = \frac{|\A|}{n(n-1)}
$$
where $|\cdot|$ denotes the set cardinality functional \cite{janson:2000}.

Thus $\rho(D)$ represents the ratio of the number of arcs
in the digraph $D$ to the number of arcs in the complete symmetric
digraph of order $n$, which is $n(n-1)$.
For brevity of notation we use \emph{relative density} rather than relative arc density henceforth.

If $X_1,\cdots,X_n \stackrel{iid}{\sim} F$
the relative density
of the associated data-random proximity catch digraph $D$,
denoted $\rho(\X_n;h,\NY)$, is a $U$-statistic,
\begin{eqnarray}
\rho(\X_n;h,\NY) =
  \frac{1}{n(n-1)}
    \sum\hspace*{-0.1 in}\sum_{i < j \hspace*{0.25 in}}   \hspace*{-0.1 in}
      \hspace*{-0.1 in}\;\;h(X_i,X_j;\NY)
\end{eqnarray}
where
\begin{eqnarray}
h(X_i,X_j;\NY)&=& \I\{(X_i,X_j) \in \A\}+ \I\{(X_j,X_i) \in \A\} \nonumber \\
       &=& \I\{X_j \in \NY(X_i)\}+ \I\{X_i \in \NY(X_j)\},
\end{eqnarray}
where $\I(\cdot)$ is the indicator function.
We denote $h(X_i,X_j;\NY)$ as $h_{ij}$ for brevity of notation.
Although the digraph is asymmetric, $h_{ij}$ is defined as
the number of arcs in $D$ between vertices $X_i$ and $X_j$,
in order to produce a symmetric kernel with finite variance \cite{lehmann:1988}.

The random variable $\rho_n := \rho(\X_n;h,\NY)$ depends on $n$ and $\NY$ explicitly
and on $F$ implicitly.
The expectation $\E[\rho_n]$, however, is independent of $n$
and depends on only $F$ and $\NY$:
\begin{eqnarray}
0 \leq \E[\rho_n] = \frac{1}{2}\E[h_{12}] \leq 1 \text{ for all $n\ge 2$}.
\end{eqnarray}
The variance $\Var[\rho_n]$ simplifies to
\begin{eqnarray}
\label{eq:var-rho}
0 \leq
  \Var[\rho_n] =
     \frac{1}{2n(n-1)} \Var[h_{12}] +
     \frac{n-2}{n(n-1)} \Cov[h_{12},h_{13}]
  \leq 1/4.
\end{eqnarray}
A central limit theorem for $U$-statistics
\cite{lehmann:1988}
yields
\begin{eqnarray}
\sqrt{n}\bigl(\rho_n-\E[\rho_n]\bigr) \stackrel{\mathcal{L}}{\longrightarrow} \N\bigl(0,\Cov[h_{12},h_{13}]\bigr)
\end{eqnarray}
provided $\Cov[h_{12},h_{13}] > 0$.
The asymptotic variance of $\rho_n$, $\Cov[h_{12},h_{13}]$,
depends on only $F$ and $\NY$.
Thus, we need determine only
$\E[h_{12}]$
and
$\Cov[h_{12},h_{13}]$
in order to obtain the normal approximation
\begin{eqnarray}
\rho_n \stackrel{\text{approx}}{\sim}
\N\bigl(\E[\rho_n],\Var[\rho_n]\bigr) =
\N\left(\frac{\E[h_{12}]}{2},\frac{\Cov[h_{12},h_{13}]}{n}\right) \text{ for large $n$}.
\end{eqnarray}

\subsection{Null and Alternative Hypotheses}
\label{sec:null-and-alt}
In a two class setting,
the phenomenon known as {\em segregation} occurs when members of
one class have a tendency to repel members of the other class.
For instance, it may be the case that one type of plant
does not grow well in the vicinity of another type of plant,
and vice versa.
This implies, in our notation,
that $X_i$ are unlikely to be located near any elements of $\Y$.
Alternatively, association occurs when members of one class
have a tendency to attract members of the other class,
as in symbiotic species, so that the $X_i$ will tend to
cluster around the elements of $\Y$, for example.
See, for instance, \cite{dixon:1994}, \cite{coomes:1999}.
The null hypothesis for spatial patterns have been a contraversial topic in ecology
from the early days.
Gotelli and Graves \cite{gotelli:1996} have collected a voluminous literature to present a
comprehensive analysis of the use and misuse of null models in ecology community.
They also define and attempt to clarify the null model concept as
``a pattern-generating model that is based on randomization of ecological data or
random sampling from a known or imagined distribution. . . .
The randomization is designed to produce a pattern that would be expected
in the absence of a particular ecological mechanism."
In other words, the hypothesized null models can be viewed as``thought experiments,"
which is conventially used in the physical sciences, and
these models provide a statistical baseline for the analysis of the patterns.
For statistical testing for segregation and association, the null hypothesis we consider is a type of
{\em complete spatial randomness};
that is,
$$H_0: X_i \stackrel{iid}{\sim} \U(T(\Y))$$
where $\U(T(\Y))$ is the uniform distribution on $T(\Y)$.
If it is desired to have the sample size be a random variable,
we may consider a spatial Poisson point process on $T(\Y)$
as our null hypothesis.

We define two classes of alternatives,
$H^S_{\epsilon}$ and $H^A_{\epsilon}$
with $\epsilon \in \bigl( 0,\sqrt{3}/3 \bigr)$,
for segregation and association, respectively.
For $\y \in \Y$,
let $e(\y)$ denote the edge of $T(\Y)$ opposite vertex $\y$,
and for $x \in T(\Y)$
let $\ell_\y(x)$ denote the line parallel to $e(\y)$ through $x$.
Then define
$T(\y,\epsilon) = \bigl\{x \in T(\Y): d(\y,\ell_\y(x)) \le \epsilon\bigr\}$.
Let $H^S_{\epsilon}$ be the model under which
$X_i \stackrel{iid}{\sim} \U \bigl(T(\Y) \setminus \cup_{\y \in \Y} T(\y,\epsilon)\bigr)$
and $H^A_{\epsilon}$ be the model under which
$X_i \stackrel{iid}{\sim} \U\bigl(\cup_{\y \in \Y} T(\y,\sqrt{3}/3 - \epsilon)\bigr)$.
Thus the segregation model excludes the possibility of
any $X_i$ occurring near a $\y_j$,
and the association model requires
that all $X_i$ occur near a $\y_j$.
The $\sqrt{3}/3 - \epsilon$ in the definition of the
association alternative is so that $\epsilon=0$
yields $H_0$ under both classes of alternatives.

{\bf Remark:}
These definitions of the alternatives
are given for the standard equilateral triangle.
The geometry invariance result of Theorem 1 from Section 3 still holds
under the alternatives, in the following sense.
If, in an arbitrary triangle,
a small percentage $\delta \cdot 100\%$ where $\delta \in (0,4/9)$ of the area is carved
away as forbidden from each vertex using line segments parallel
to the opposite edge, then
under the transformation to the standard equilateral triangle
this will result in the alternative $H^S_{\sqrt{3 \delta / 4}}$.
This argument is for
segregation with $\delta < 1/4$;
a similar construction is available for the other cases.

\section{Asymptotic Normality Under the Null and Alternative Hypotheses}
\label{sec:asy-norm}
First we present a ``geometry invariance" result which allows us to assume
$T(\Y)$ is the standard equilateral triangle,
$T\bigl((0,0),(1,0),\bigl( 1/2,\sqrt{3}/2 \bigr)\bigr)$, thereby simplifying our subsequent analysis.

{\bf Theorem 1:}
Let $\Y = \{\y_1,\y_2,\y_3\} \subset \R^2$
be three non-collinear points.
For $i=1,\cdots,n$
let $X_i \stackrel{iid}{\sim} F = \U(T(\Y))$,
the uniform distribution on the triangle $T(\Y)$.
Then for any $r \in [1,\infty]$
the distribution of $\rho(\X_n;h,\NY^r)$
is independent of $\Y$,
 hence the geometry of $T(\Y)$.

{\bf Proof:}
A composition of translation, rotation, reflections, and scaling
will transform any given triangle $T_o = T\bigl(\y_1,\y_2,\y_3\bigr)$
into the ``basic'' triangle $T_b = T\bigl((0,0),(1,0),(c_1,c_2) \bigr)$
with $0 < c_1 \le 1/2$, $c_2 > 0$ and $(1-c_1)^2+c_2^2 \le 1$,
preserving uniformity.
The transformation $\phi_e: \R^2 \rightarrow \R^2$
given by $\phi_e(u,v) = \left(u+\frac{1-2\,c_1}{\sqrt{3}}\,v,\frac{\sqrt{3}}{2\,c_2}\,v \right)$
takes $T_b$ to
the equilateral triangle
$T_e = T\bigl((0,0),(1,0),\bigl( 1/2,\sqrt{3}/2 \bigr)\bigr)$.
Investigation of the Jacobian shows that $\phi_e$
also preserves uniformity.
Furthermore, the composition of $\phi_e$ with the rigid motion transformations
maps
     the boundary of the original triangle $T_o$
  to the boundary of the equilateral triangle $T_e$,
     the median lines of $T_o$
  to the median lines of $T_e$,
and  lines parallel to the edges of $T_o$
  to lines parallel to the edges of $T_e$.
Since the joint distribution of any collection of the $h_{ij}$
involves only probability content of unions and intersections
of regions bounded by precisely such lines,
and the probability content of such regions is preserved since uniformity is preserved,
the desired result follows.
$\blacksquare$

Based on Theorem 1 and our uniform null hypothesis,
we may assume that
$T(\Y)$ is the standard equilateral triangle
with $\Y =\bigl\{(0,0),(1,0),\bigl( 1/2,\sqrt{3}/2 \bigr)\bigr\}$
henceforth.

For our $r$-factor proximity map and uniform null hypothesis,
the asymptotic null distribution of $\rho_n(r) = \rho(\X_n;h,\NY^r)$
can be derived as a function of $r$.  Let $\mu(r):=\E[\rho_n(r)]$ and
$\nu(r):=\Cov[h_{12},h_{13}]$. Notice that $\mu(r)=\E[h_{12}]/2=P(X_2 \in \NY^r(X_1))$ is
the probability of an arc occurring between any pair of vertices.

\subsection{Asymptotic Normality under the Null Hypothesis}
By detailed geometric  probability calculations, provided in Appendix 1,
the mean and the asymptotic variance of the relative density of the
$r$-factor proximity catch digraph can explicitly be computed.
The central limit theorem for $U$-statistics then establishes
the asymptotic normality under the uniform null hypothesis.
These results are summarized in the following theorem.

{\bf Theorem 2:}
For $r \in [1,\infty)$,
\begin{eqnarray}
 \frac{\sqrt{n}\,\bigl(\rho_n(r)-\mu(r)\bigr)}{\sqrt{\nu(r)}}
 \stackrel{\mathcal{L}}{\longrightarrow}
 \N(0,1)
\end{eqnarray}
where
\begin{eqnarray}
\label{eq:Asymean}
\mu(r) =
 \begin{cases}
  \frac{37}{216}r^2                                 &\text{for} \quad r \in [1,3/2), \\
  -\frac{1}{8}r^2 + 4 - 8r^{-1} + \frac{9}{2}r^{-2}  &\text{for} \quad r \in [3/2,2), \\
  1 - \frac{3}{2}r^{-2}                             &\text{for} \quad r \in [2,\infty), \\
 \end{cases}
\end{eqnarray}
and
\begin{equation}
\label{eq:Asyvar}
\nu(r) =\nu_1(r) \,\I(r \in [1,4/3)) + \nu_2(r) \,\I(r \in [4/3,3/2))+ \nu_3(r) \,\I(r \in [3/2,2)) + \nu_4(r) \,\I( r \in [2,\infty])
\end{equation}
with
{\small
\begin{align*}
  \nu_1(r) &=\frac{3007\,r^{10}-13824\,r^9+898\,r^8+77760\,r^7-117953\,r^6+48888\,r^5
-24246\,r^4+60480\,r^3-38880\,r^2+3888}{58320\,r^4},\\
  \nu_2(r) &=\frac{5467\,r^{10}-37800\,r^9+61912\,r^8+46588\,r^6-191520\,r^5+13608\,r^4
+241920\,r^3-155520\,r^2+15552}{233280\,r^4},  \\
  \nu_3(r) &=-[7\,r^{12}-72\,r^{11}+312\,r^{10}-5332\,r^8+15072\,r^7+13704\,r^6-
139264\,r^5+273600\,r^4-242176\,r^3\\
& +103232\,r^2-27648\,r+8640]/[960\,r^6],\\
  \nu_4(r) &=\frac{15\,r^4-11\,r^2-48\,r+25}{15\,r^6}.
\end{align*}
}
For $r=\infty$, $\rho_n(r)$ is degenerate.

See Appendix 1 for the proof.

\begin{figure}[ht]
\centering
\psfrag{mu(r)}{\scriptsize{$\mu(r)$}}
\psfrag{nu(r)}{\scriptsize{$\nu(r)$}}
\psfrag{r}{\scriptsize{$r$}}
\epsfig{figure=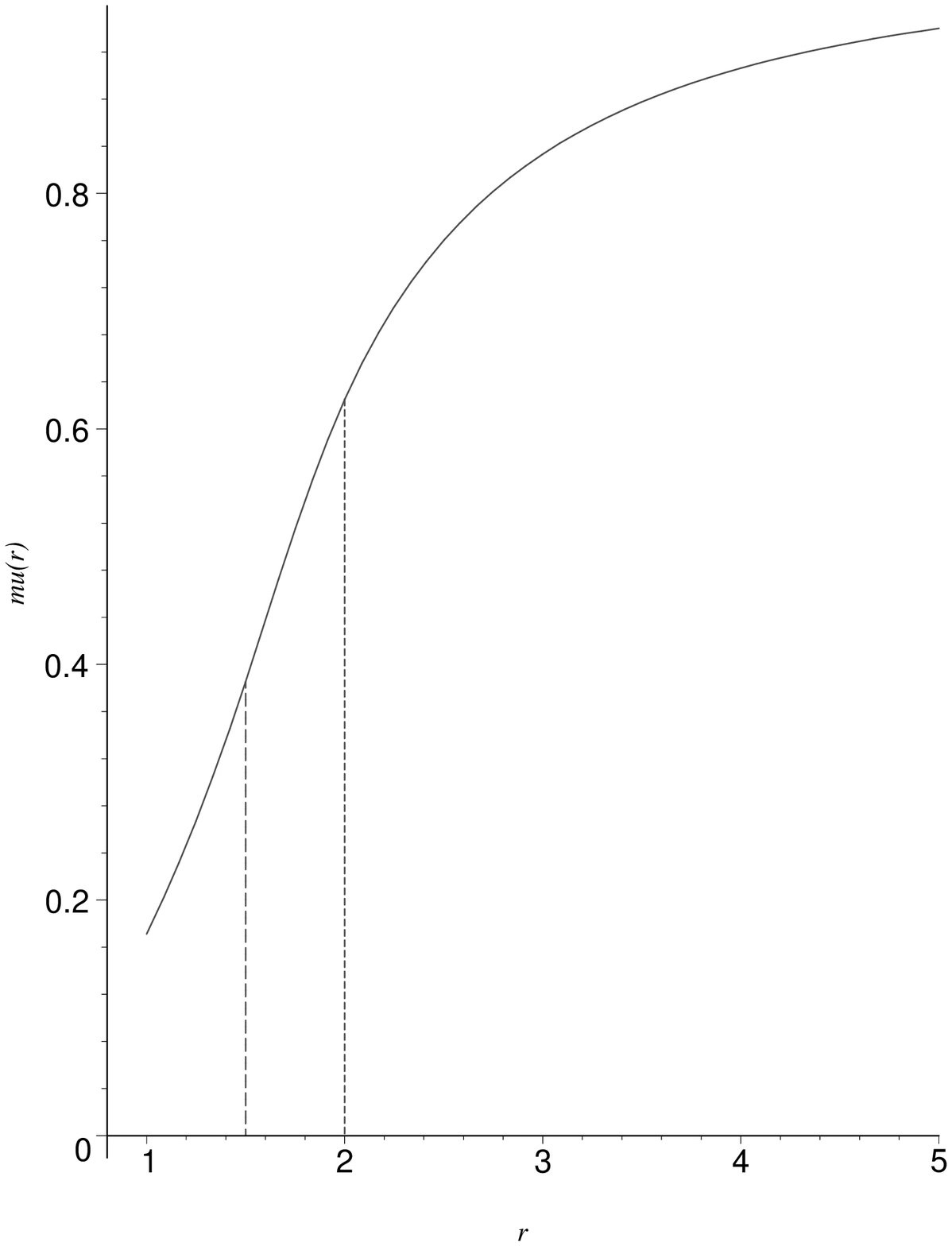, height=175pt, width=200pt}
\epsfig{figure=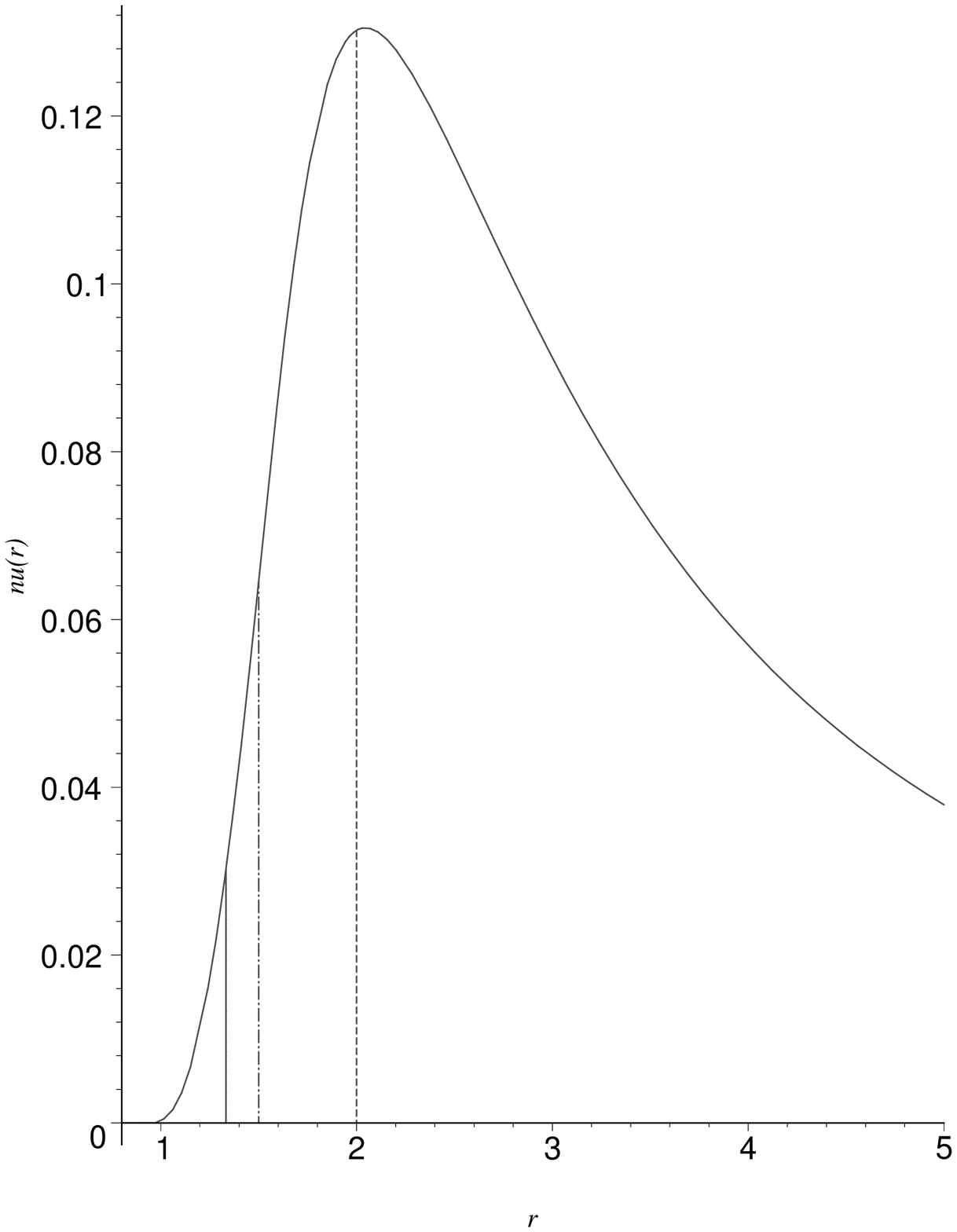, height=175pt, width=200pt}
\caption{
\label{fig:AsyNormCurves}
Asymptotic null mean $\mu(r)$ (left) and variance $\nu(r)$ (right),
from Equations (\ref{eq:Asymean}) and (\ref{eq:Asyvar}) in Theorem 2,
respectively. The vertical lines indicate the endpoints of the intervals
in the piecewise definition of the functions. Notice that the vertical axes are differently scaled.
}
\end{figure}

Consider the form of the mean and variance functions, which are depicted in
Figure \ref{fig:AsyNormCurves}. Note that $\mu(r)$ is monotonically increasing in $r$,
since the proximity region of any data point increases with $r$.
In addition, $\mu(r) \rightarrow 1$ as $r \rightarrow \infty$,
since the digraph becomes complete asymptotically,
which explains why $\rho_n(r)$ is degenerate, i.e. $\nu(r)=0$, when $r=\infty$.
Note also that $\mu(r)$ is continuous, with the value at $r=1$ $\mu(1) = 37/216$.

Regarding the asymptotic variance, note that $\nu(r)$ is continuous in $r$ with
$\lim_{r \rightarrow \infty} \nu(r) = 0$ and $\nu(1) = 34/58320\approx .000583$
and observe that $\sup_{r \ge 1} \nu(r) \approx .1305$ at $\argsup_{r \ge 1} \nu(r) \approx 2.045$.

To illustrate the limiting distribution, $r=2$ yields
$$
\frac{\sqrt{n}(\rho_n(2) - \mu(2))}{\sqrt{\nu(2)}}
=
\sqrt{\frac{192n}{25}} \left(\rho_n(2) - \frac{5}{8}\right)\stackrel{\mathcal{L}}{\longrightarrow} \N(0,1)$$
or equivalently
$$
\rho_n(2) \stackrel{\text{approx}}{\sim} \N\left(\frac{5}{8},\frac{25}{192n}\right)
.$$

\begin{figure}[ht]
\centering
\psfrag{Density}{ \Huge{\bf{Density}}}
\rotatebox{-90}{ \resizebox{2.1 in}{!}{ \includegraphics{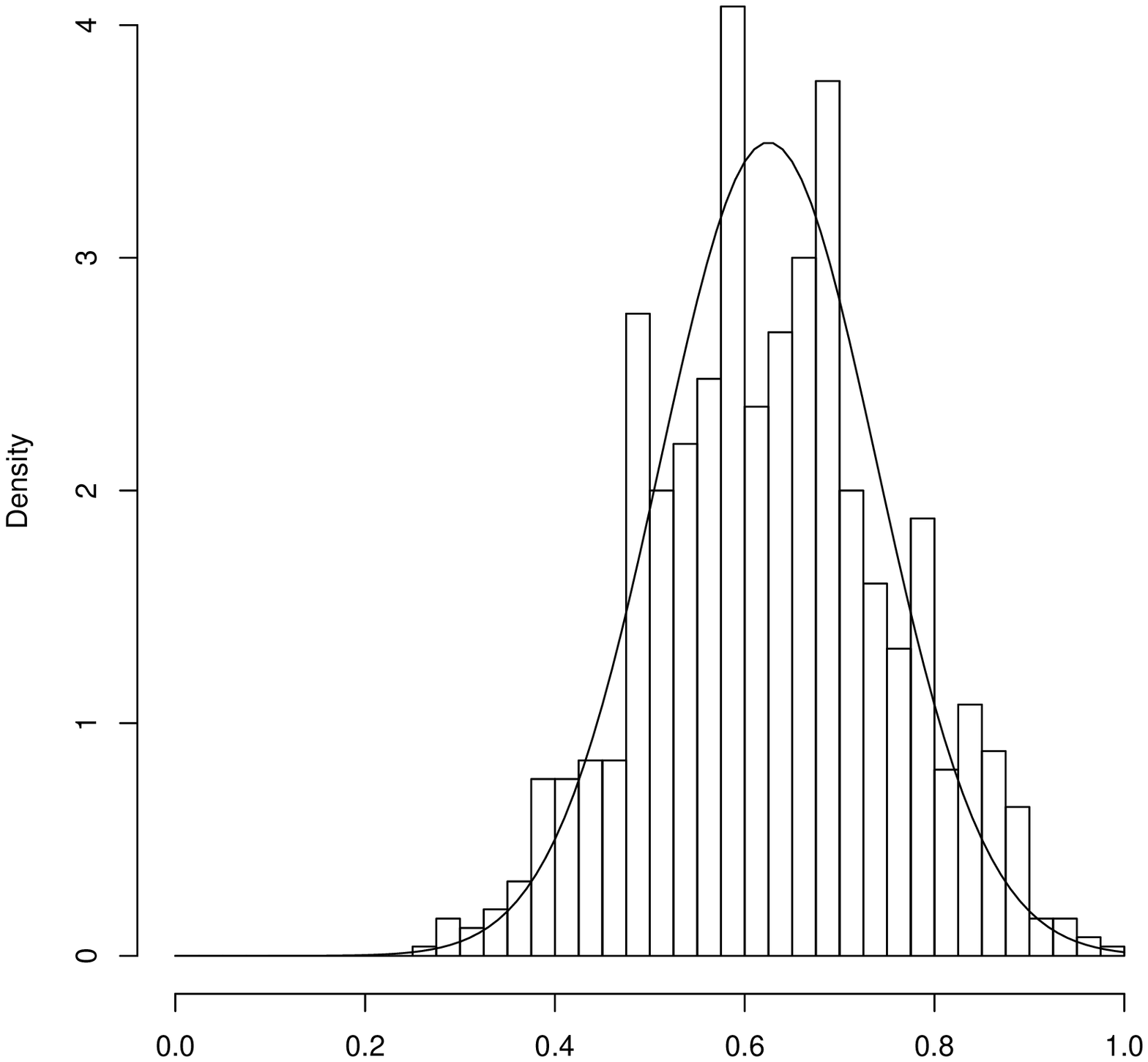} } }
\rotatebox{-90}{ \resizebox{2.1 in}{!}{ \includegraphics{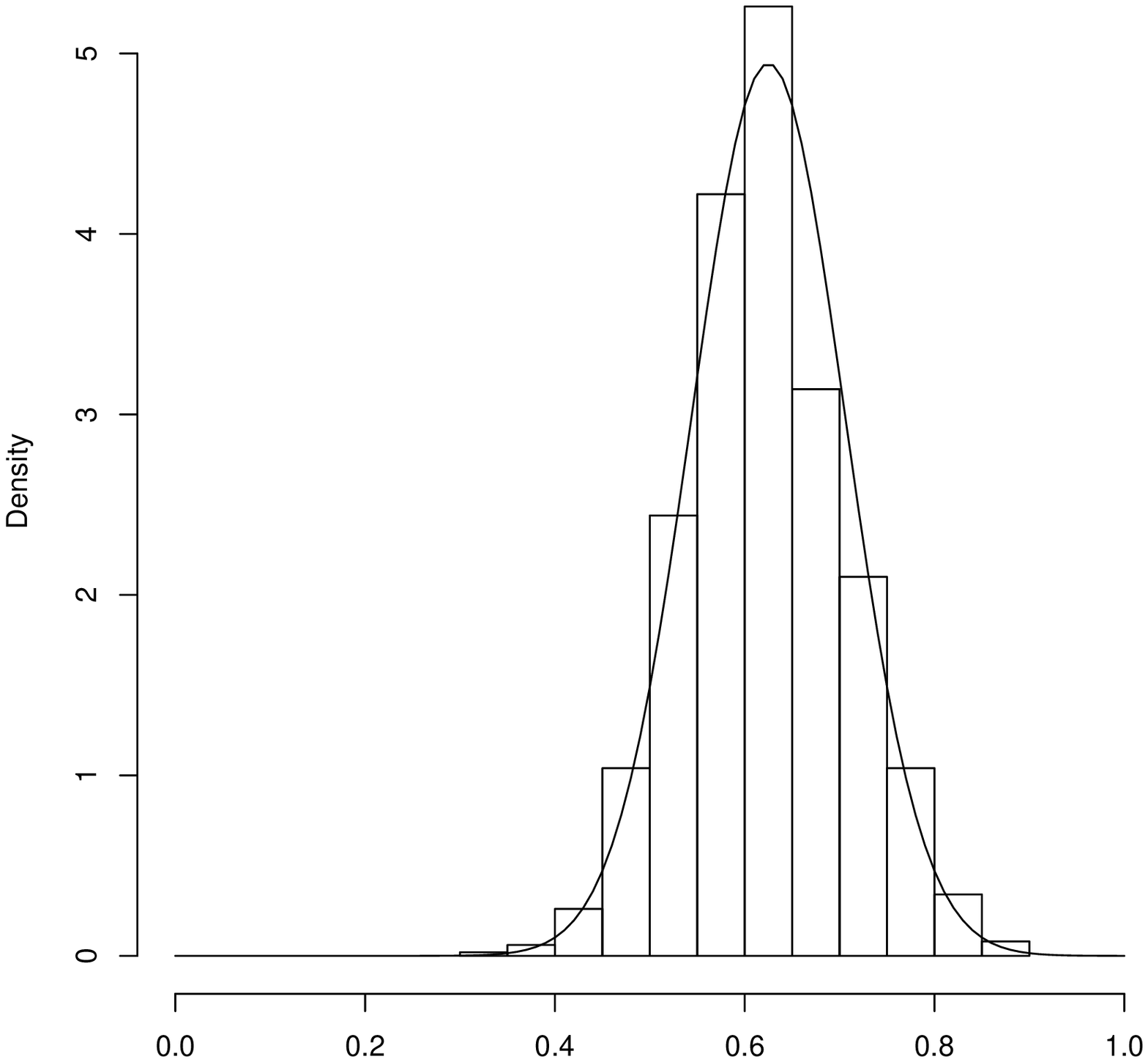} } }
\rotatebox{-90}{ \resizebox{2.1 in}{!}{ \includegraphics{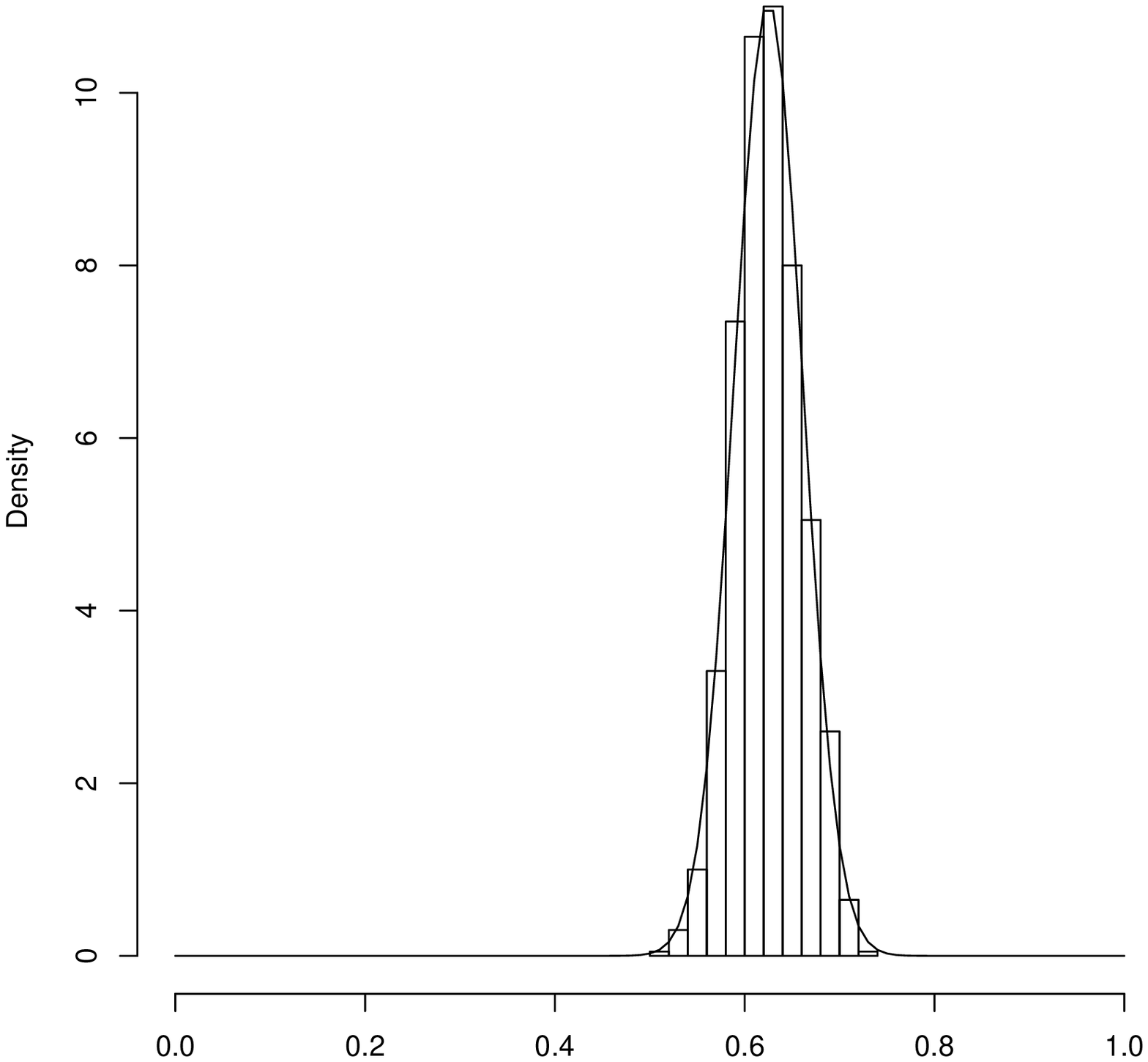} } }
\caption{
\label{fig:NormSkew}
Depicted are the distributions of
$\rho_n(2) \stackrel{\text{approx}}{\sim} \N \left(\frac{5}{8},\frac{25}{192n} \right)$
for $n=10,20,100$ (left to right).
Histograms are based on 1000 Monte Carlo replicates.
Solid curves represent the approximating normal densities given by Theorem 2.
Again, note that the vertical axes are differently scaled.
}
\end{figure}

\begin{figure}[ht]
\centering
\psfrag{Density}{ \Huge{\bf{Density}}}
\rotatebox{-90}{ \resizebox{2.1 in}{!}{ \includegraphics{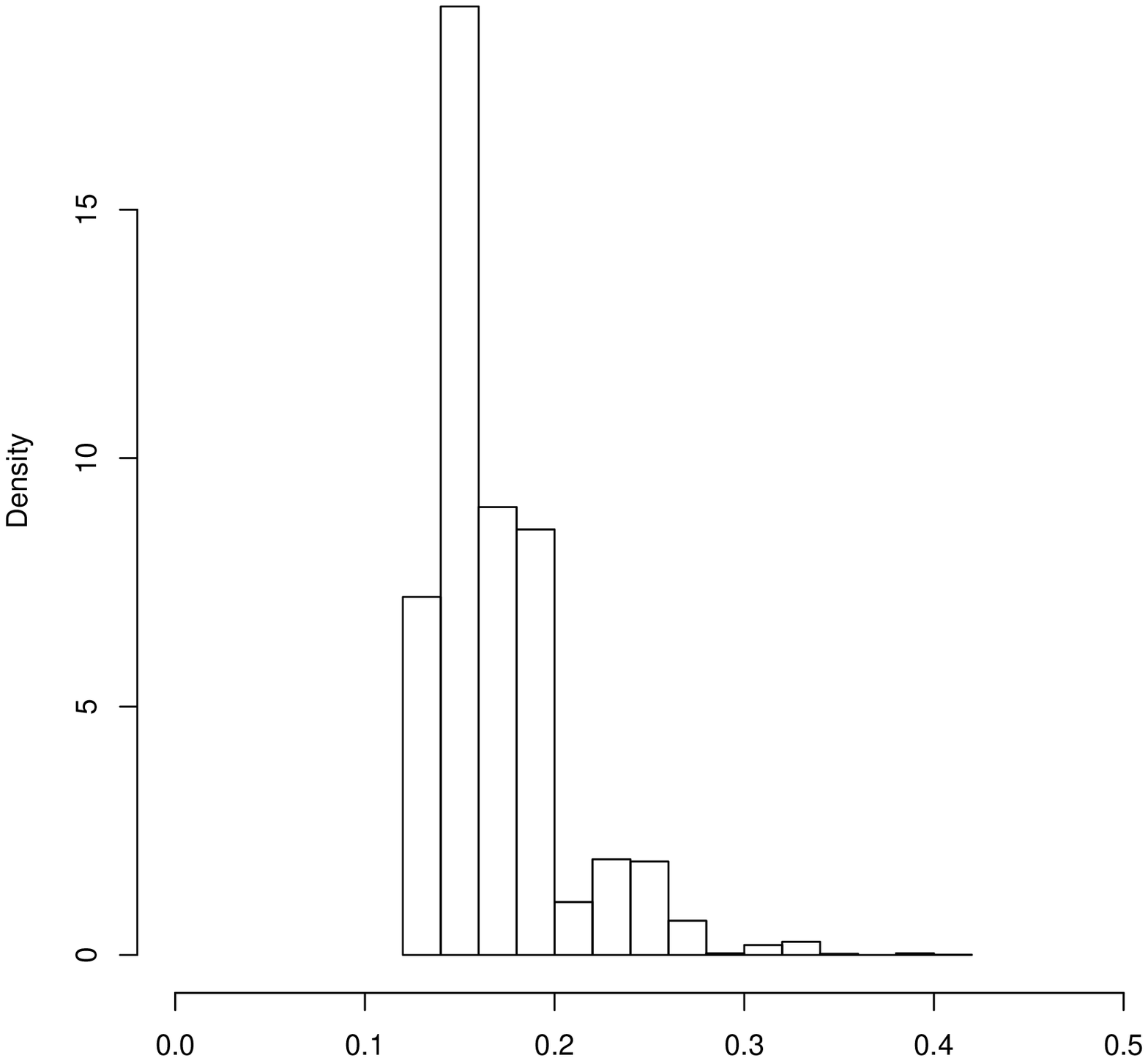} } }
\rotatebox{-90}{ \resizebox{2.1 in}{!}{ \includegraphics{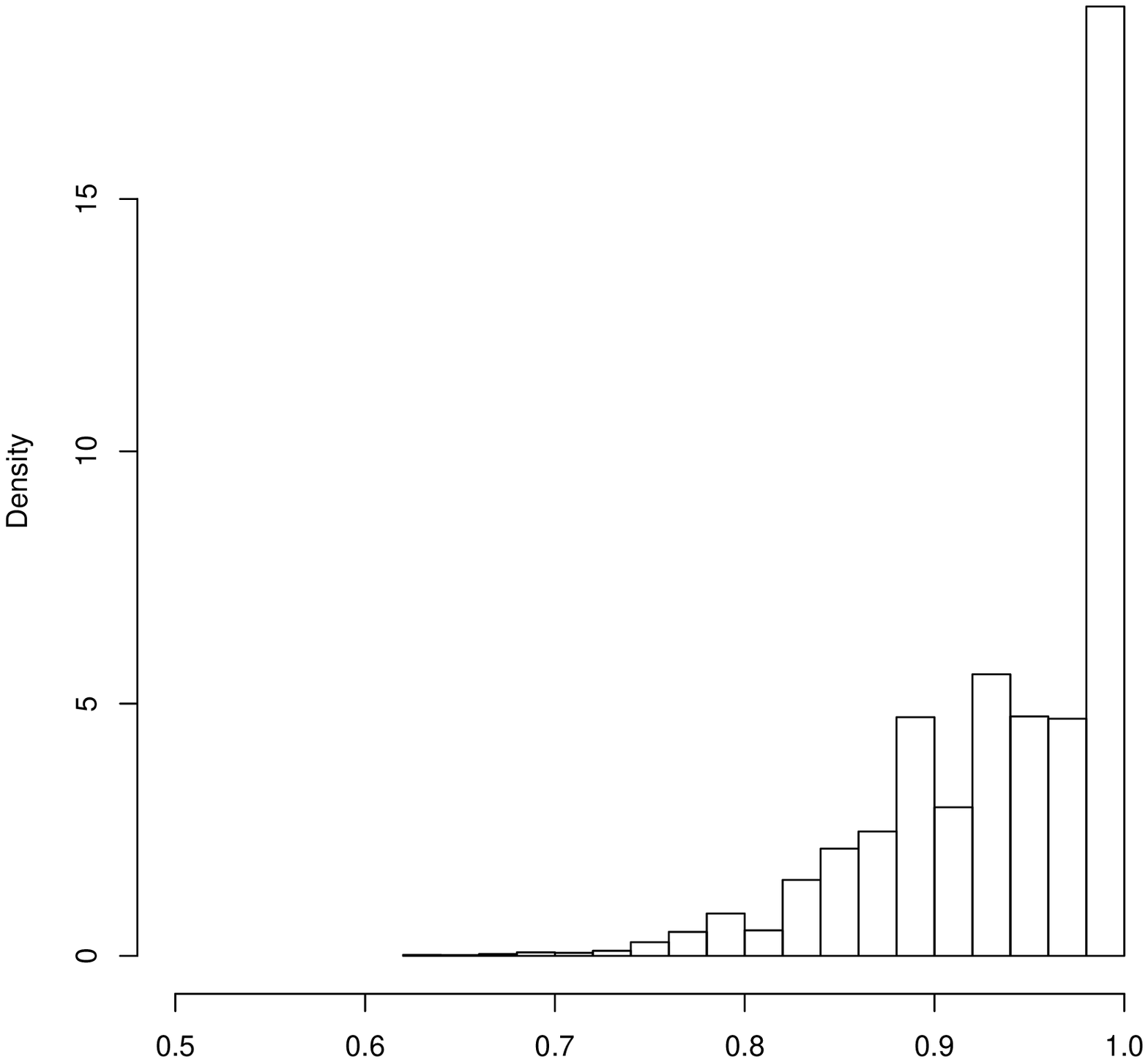} } }
\caption{
\label{fig:NormSkew1}
Depicted are the histograms for 10,000 Monte Carlo replicates of $\rho_{10}(1)$ (left)
and $\rho_{10}(5)$ (right) indicating severe small sample skewness for extreme values of $r$.
}
\end{figure}

Figure \ref{fig:NormSkew}
indicates that, for $r=2$,
the normal approximation is accurate even for small $n$
(although kurtosis may be indicated for $n=10$).
Figure \ref{fig:NormSkew1} demonstrates,
however, that severe skewness obtains for small values of $n$,
and extreme values of $r$.
The finite sample variance in Equation \ref{eq:var-rho} and
skewness may be derived analytically
in much the same way as was $\Cov[h_{12},h_{13}]$
for the asymptotic variance.
In fact,
the exact distribution of $\rho_n(r)$
is, in principle, available
by successively conditioning on the values of the $X_i$.
Alas,
while the joint distribution of $h_{12},h_{13}$ is available,
the joint distribution of $\{h_{ij}\}_{1 \leq i < j \leq n}$,
and hence the calculation for the exact distribution of $\rho_n(r)$,
is extraordinarily tedious and lengthy for even small values of $n$.

Letting $H_n(r) = \sum_{i=1}^n h(X_i,X_{n+1})$,
the exact distribution of $\rho_n(r)$
can be evaluated based on the recurrence
$$(n+1)n\rho_{n+1}(r) \stackrel{d}{=} n(n-1)\rho_n(r) + H_n(r)$$
by noting that the conditional random variable
$H_n(r)|X_{n+1}$
is the sum of $n$ independent and identically distributed random variables.
Alas, this calculation is also tedious for large $n$.

\subsection{Asymptotic Normality Under the Alternatives}
Asymptotic normality of relative density of the proximity catch digraphs under
the alternative hypotheses of segregation and association can be established
by the same method as under the null hypothesis.
Let $\E^S_{\epsilon}[\cdot]$ ( $\E^A_{\epsilon}[\cdot]$) be the expectation
with respect to the uniform distribution under the segregation
( association ) alternatives with $\epsilon \in \left( 0,\sqrt{3}/3 \right)$.

{\bf Theorem 3:}
Let $\mu_S(r,\epsilon)$ (and $\mu_A(r,\epsilon)$) be the mean and $\nu_S(r,\epsilon)$
(and $\nu_A(r,\epsilon)$) be the covariance, $\Cov[h_{12},h_{13}]$ for $r \in (0,1]$ and
$\epsilon \in \bigl( 0,\sqrt{3}/3 \bigr)$ under segregation (and association).
Then under $H^S_{\epsilon}$,
$\sqrt{n}(\rho_n(r)-\mu_S(r,\epsilon)) \stackrel {\mathcal L}{\longrightarrow} \N(0,\nu_S(r,\epsilon))$
for the values of the pair $(r,\epsilon)$ for which $\nu_S(r,\epsilon)>0$.
Likewise, under $H^A_{\epsilon}$,
$\sqrt{n}(\rho_n(r)-\mu_A(r,\epsilon)) \stackrel {\mathcal L}{\longrightarrow} \N(0,\nu_A(r,\epsilon))$
for the values of the pair $(r,\epsilon)$ for which $\nu_A(r,\epsilon)>0$.

{\bf Sketch of Proof:}
Under the alternatives, i.e. $\epsilon>0$ ,
$\rho_n(r)$ is a $U$-statistic
with the same symmetric kernel $h_{ij}$ as in the null case.
The mean
$\mu_S(r,\epsilon)=\E_{\epsilon}[\rho_n(r)] = \E_{\epsilon}[h_{12}]/2$ (and $\mu_A(r,\epsilon)$),
now a function of both $r$ and $\epsilon$, is again in $[0,1]$.
The asymptotic variance $\nu_S(r,\epsilon)=\Cov_{\epsilon}[h_{12},h_{13}]$
(and $\nu_A(r,\epsilon)$), also a function of both $r$ and $\epsilon$,
is bounded above by $1/4$, as before. The explicit forms of $\mu_S(r,\epsilon)$
and $\mu_A(r,\epsilon)$ is given, defined piecewise, in Appendix 2.
Sample values of $\mu_S(r,\epsilon)$, $\nu_S(r,\epsilon)$ and $\mu_A(r,\epsilon)$,
$\nu_A(r,\epsilon)$ are given in Appendix 3 for segregation with $\epsilon=\sqrt{3}/4$
and for association with $\epsilon=\sqrt{3}/12$.
Thus asymptotic normality obtains provided $\nu_S(r,\epsilon) > 0$ ($\nu_A(r,\epsilon) > 0$);
otherwise $\rho_n(r)$ is degenerate.
Note that under $H^S_{\epsilon}$,
$$\nu_S(r,\epsilon)>0 \text{ for } (r,\epsilon) \in \bigl[ 1,\sqrt{3}/(2 \epsilon) \bigr)
\times \bigl( 0,\sqrt{3}/4 \bigr] \cup \bigl[ 1,\sqrt{3}/\epsilon-2 \bigr) \times \bigl( \sqrt{3}/4,\sqrt{3}/3 \bigr),$$
and under $H^A_{\epsilon}$,
$$\nu_A(r,\epsilon)>0 \text{ for }(r,\epsilon)\in (1,\infty) \times \bigl( 0,\sqrt{3}/3 \bigr)
\cup \{1\} \times \bigl( 0,\sqrt{3}/12 \bigr). \;\;\blacksquare$$

Notice that for the association class of alternatives
any $r \in (1,\infty)$ yields asymptotic normality
for all $\epsilon \in \bigl( 0,\sqrt{3}/3 \bigr)$,
while for the segregation class of alternatives
only $r=1$ yields this universal asymptotic normality.


\section{The Test and Analysis}
The relative density of the proximity catch digraph
is a test statistic for the segregation/association alternative;
rejecting for extreme values of $\rho_n(r)$ is appropriate
since under segregation we expect $\rho_n(r)$ to be large,
while under association we expect $\rho_n(r)$ to be small.
Using the test statistic
\begin{equation}
R = \frac{\sqrt{n} (\rho_n(r) - \mu(r))}{\sqrt{\nu(r)}},
\end{equation}
the asymptotic critical value
for the one-sided level $\alpha$ test against segregation
is given by
\begin{equation}
z_{\alpha} = \Phi^{-1}(1-\alpha)
\end{equation}
where $\Phi(\cdot)$ is the standard normal distribution function.
Against segregation, the test rejects for $R>z_{1-\alpha}$ and against association,
the test rejects for $R<z_{\alpha}$.

\subsection{Consistency}
{\bf Theorem 4:}
The test against $H^S_{\epsilon}$ which rejects for $R>z_{1-\alpha}$
and
the test against $H^A_{\epsilon}$ which rejects for $R<z_{\alpha}$
are consistent for $r \in [1,\infty)$ and $\epsilon \in \bigl( 0,\sqrt{3}/3 \bigr)$.

{\bf Proof:}
Since the variance of the asymptotically normal test statistic,
under both the null and the alternatives,
converges to 0 as $n \rightarrow \infty$
(or is degenerate),
it remains to show that the mean under the null, $\mu(r)=\E[\rho_n(r)]$,
is less than (greater than) the mean under the alternative,
$\mu_S(r,\epsilon)=E_{\epsilon}[\rho_n(r)]$ ($\mu_A(r,\epsilon)$)
against segregation (association) for $\epsilon > 0$.
Whence it will follow that power converges to 1 as $n \rightarrow \infty$.

Detailed analysis of $\mu_S(r,\epsilon)$ and $\mu_A(r,\epsilon)$ in Appendix 2
indicates that under segregation $\mu_S(r,\epsilon)>\mu(r)$ for all $\epsilon >0$
and $r \in [1,\infty)$. Likewise, detailed analysis of $\mu_A(r,\epsilon)$ in Appendix 3
indicates that under association $\mu_A(r,\epsilon)<\mu(r)$ for all
$\epsilon >0$ and $r \in [1,\infty)$.
Hence the desired result follows for both alternatives.
$\blacksquare$

In fact, the analysis of $\mu(r,\epsilon)$ under the alternatives reveals more
than what is required for consistency.
Under segregation, the analysis indicates that $\mu_S(r,\epsilon_1) < \mu_S(r,\epsilon_2)$
for $\epsilon_1<\epsilon_2$.
Likewise, under association, the analysis indicates that $\mu_A(r,\epsilon_1) > \mu_A(r,\epsilon_2)$
for $\epsilon_1<\epsilon_2$.

\subsection{Monte Carlo Power Analysis}

\begin{figure}[ht]
\centering
\psfrag{kernel density estimate}{ \Huge{\bf{kernel density estimate}}}
\psfrag{relative density}{ \Huge{\bf{relative density}}}
\rotatebox{-90}{ \resizebox{2.1 in}{!}{ \includegraphics{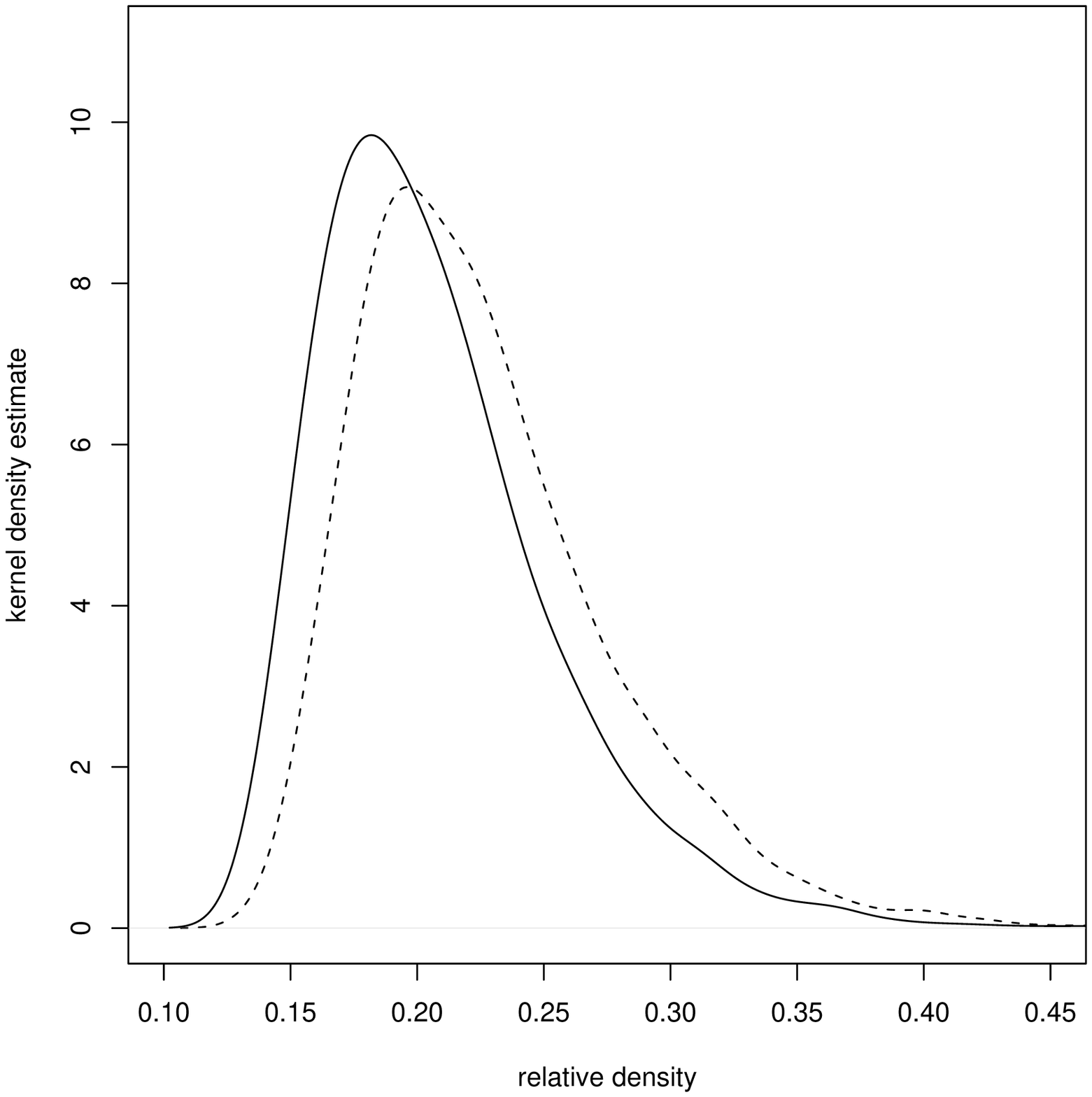} } }
\rotatebox{-90}{ \resizebox{2.1 in}{!}{ \includegraphics{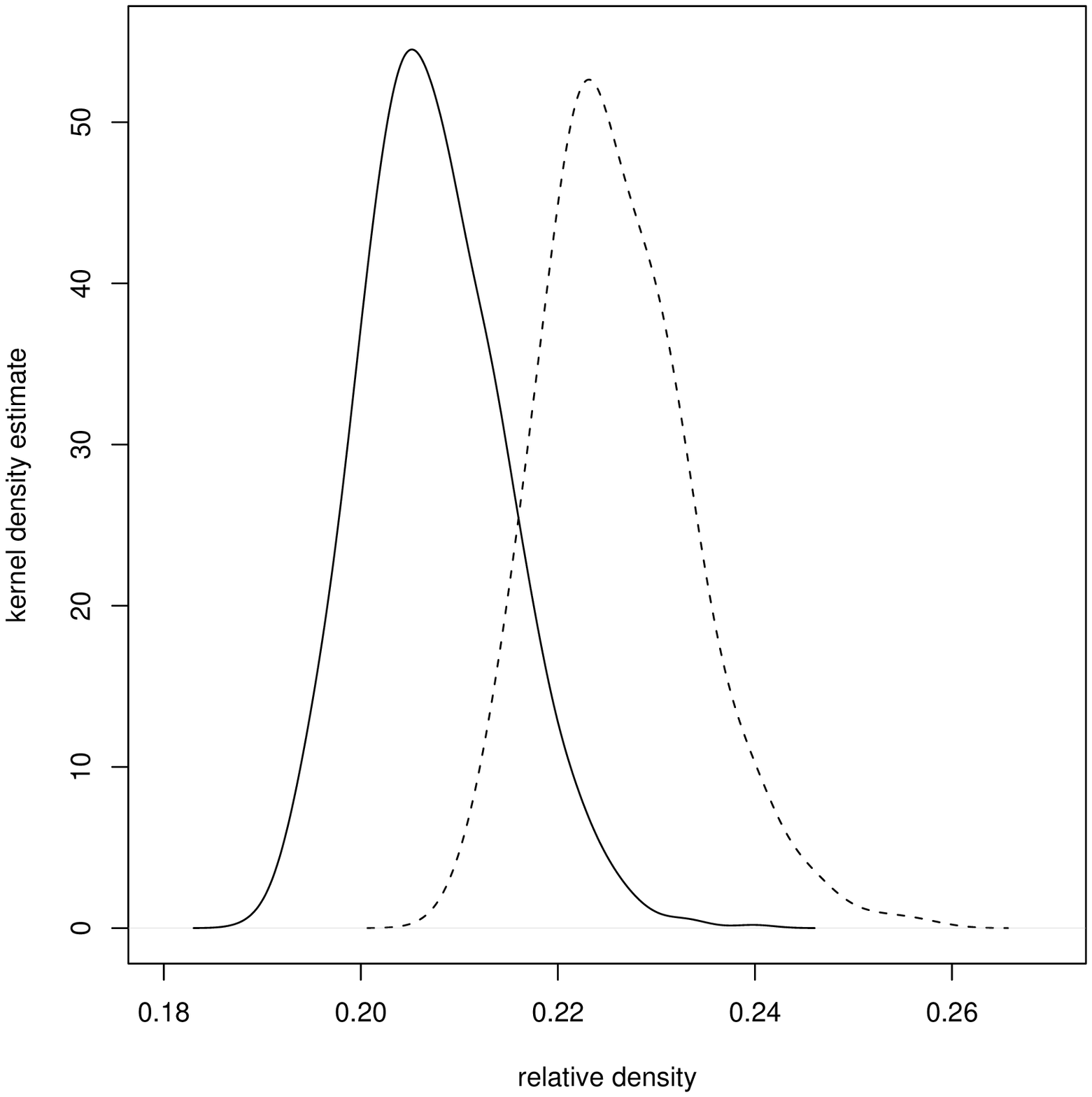} } }
\caption{ \label{fig:SegSimPowerPlots}
Two Monte Carlo experiments against the segregation alternative $H^S_{\sqrt{3}/8}$.
Depicted are kernel density estimates for $\rho_n(11/10)$ for
$n=10$ (left) and $n=100$ (right) under the null (solid) and alternative (dashed).
}
\end{figure}

In Figure \ref{fig:SegSimPowerPlots}, we present a Monte Carlo investigation
against the segregation alternative $H^S_{\sqrt{3}/8}$ for $r=11/10$ and $n=10,100$.
With $n=10$, the null and alternative probability density functions
for $\rho_{10}(1.1)$ are very similar, implying small power
(10,000 Monte Carlo replicates yield $\widehat{\beta}^S_{mc} = 0.0787$,
which is based on the empirical critical value).
With $n=100$,
there is more separation
between null and alternative probability density functions;
for this case, 1000 Monte Carlo replicates yield $\widehat{\beta}^S_{mc} = 0.77$.
Notice also that the probability density functions are more skewed for $n=10$,
while approximate normality holds for $n=100$.

\begin{figure}[]
\centering
\psfrag{power}{ \Huge{\bf{power}}}
\psfrag{r}{\Huge{$r$}}
\rotatebox{-90}{ \resizebox{2.1 in}{!}{ \includegraphics{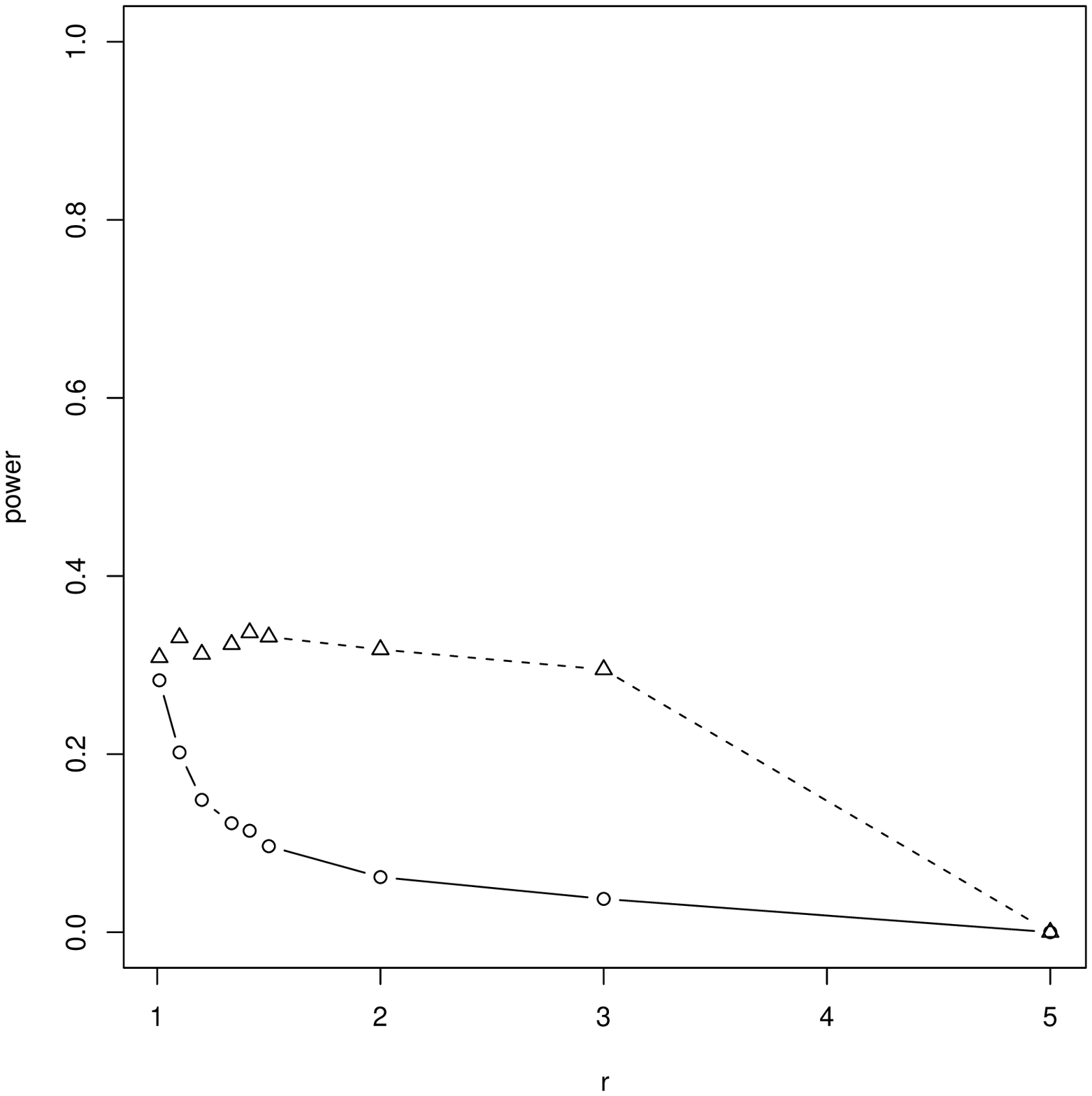} } }
\rotatebox{-90}{ \resizebox{2.1 in}{!}{ \includegraphics{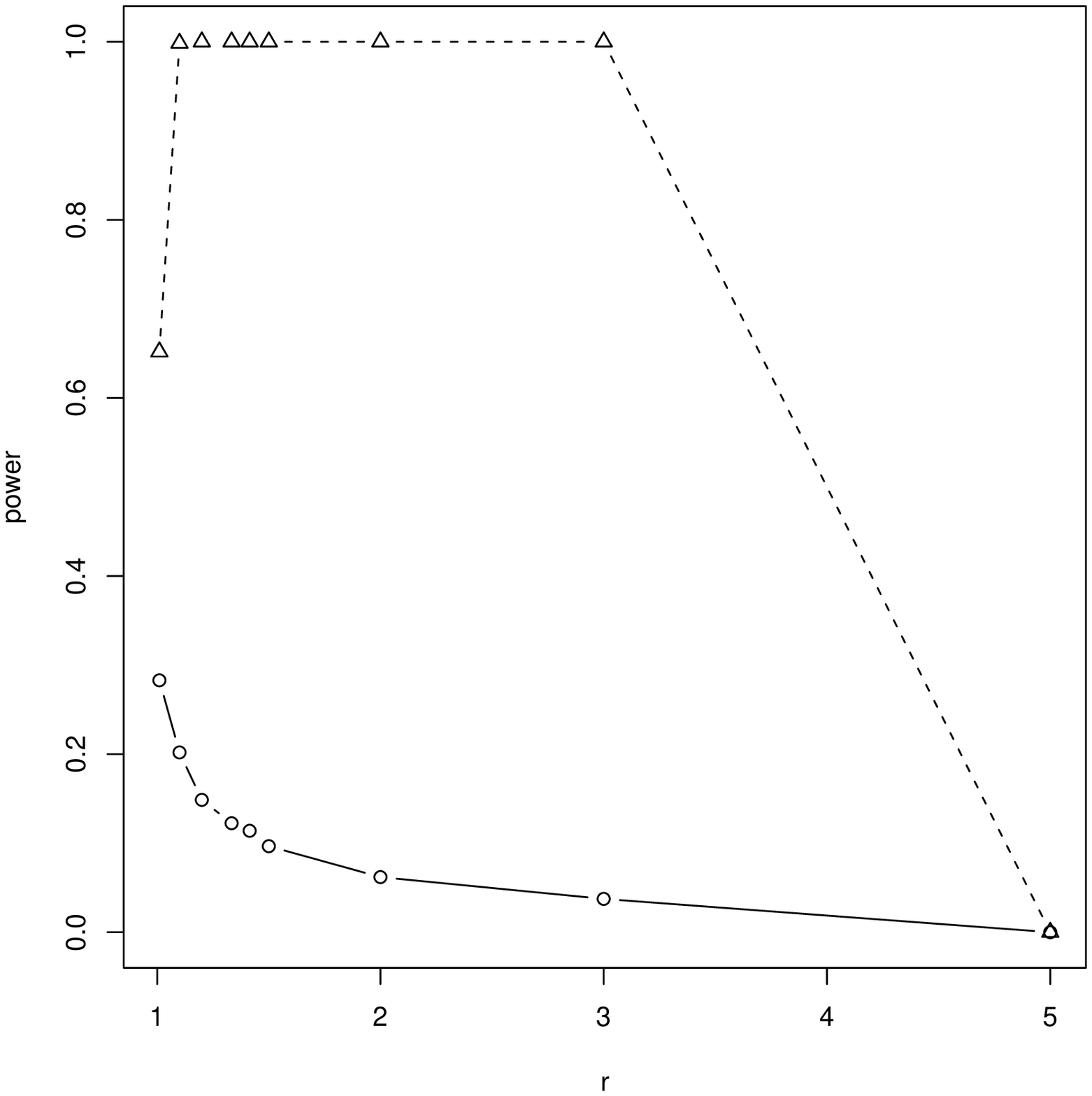} } }
\caption{ \label{fig:SegSimPowerCurve}
Monte Carlo power using the asymptotic critical value against segregation alternatives
$H^S_{\sqrt{3}/8}$ (left)
and
$H^S_{\sqrt{3}/4}$ (right)
as a function of $r$, for $n=10$.  The circles represent the empirical significance
levels while triangles represent the empirical power values.
The $r$ values plotted are $1,\,11/10,\,12/10,\,4/3,\,\sqrt{2},\,,\,2,\,3,\,5,\,10$.
}
\end{figure}
For a given alternative and sample size,
we may consider analyzing the power of the test --- using the asymptotic critical value---
as a function of the proximity factor $r$.
In Figure \ref{fig:SegSimPowerCurve}, we present
a Monte Carlo investigation of power against
$H^S_{\sqrt{3}/8}$
and
$H^S_{\sqrt{3}/4}$
as a function of $r$ for $n=10$.  The empirical significance level is about $.05$ for $r=2,\,3$
which have the empirical power $\widehat{\beta}^S_{10}(r,\sqrt{3}/8) \approx .35$,
and $\widehat{\beta}^S_{10}(r,\sqrt{3}/4) =1$.
So, for small sample sizes, moderate values of $r$ are more appropriate for normal approximation,
as they yield the desired significance level and the more severe the segregation,
the higher the power estimate.

\begin{figure}[]
\centering
\psfrag{kernel density estimate}{ \Huge{\bf{kernel density estimate}}}
\psfrag{relative density}{ \Huge{\bf{relative density}}}
\rotatebox{-90}{ \resizebox{3 in}{!}{ \includegraphics{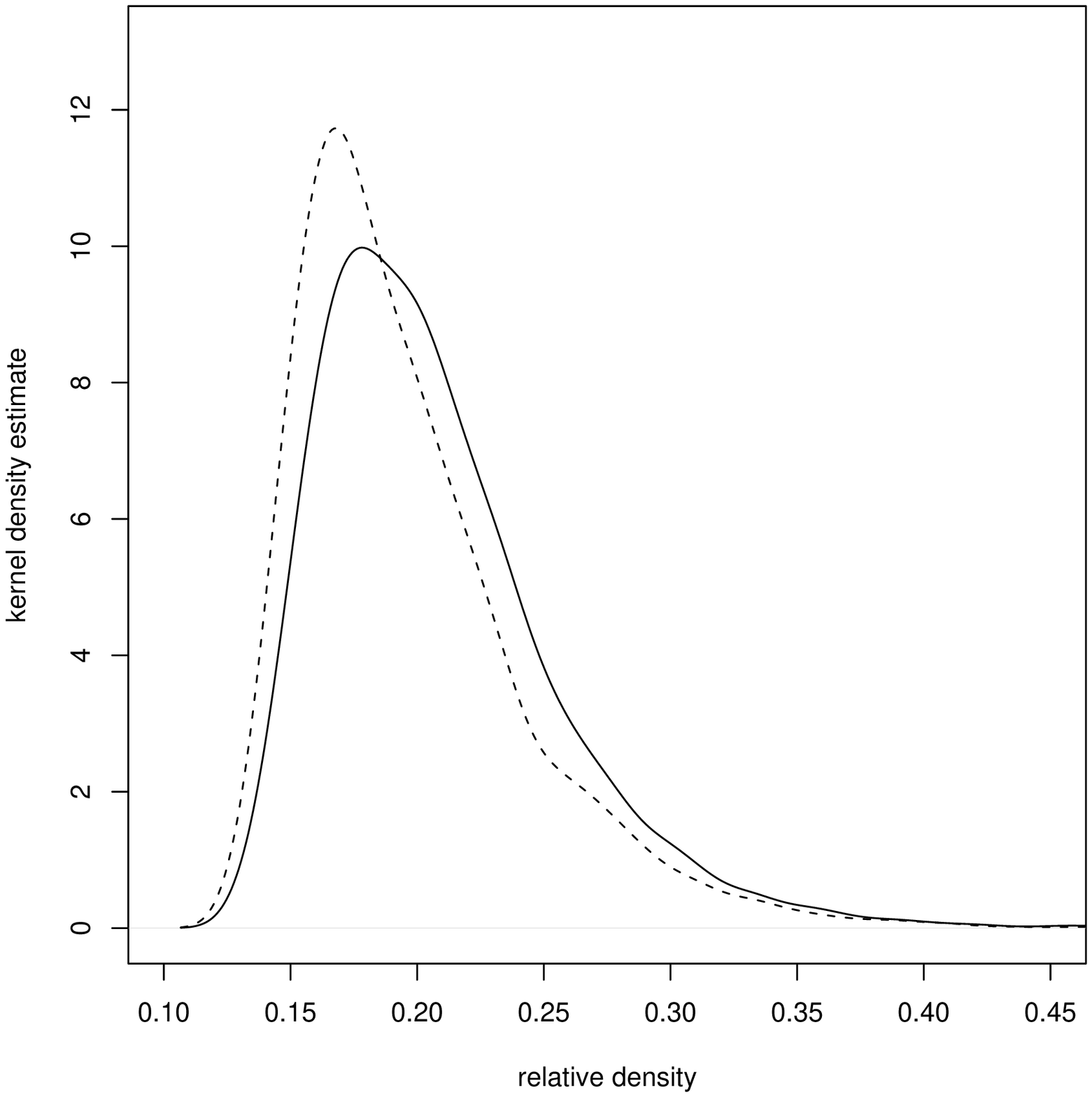} } }
\rotatebox{-90}{ \resizebox{3 in}{!}{ \includegraphics{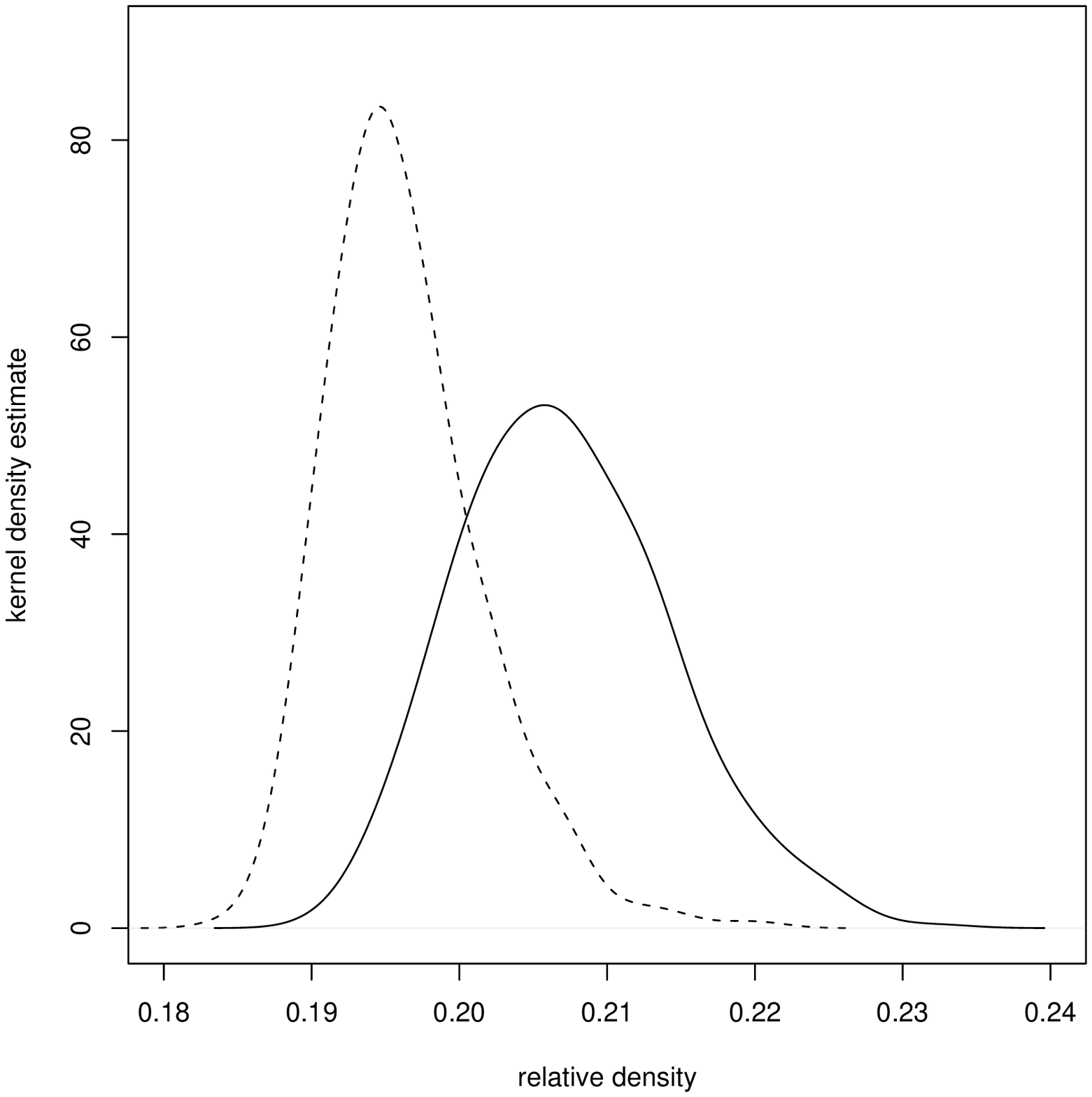} } }
\caption{ \label{fig:AggSimPowerPlots}
Two Monte Carlo experiments against the association alternative $H^A_{\sqrt{3}/12}$.
Depicted are kernel density estimates for $\rho_n(11/10)$ for
$n=10$ (left) and $n=100$ (right) under the null (solid) and alternative (dashed).
}
\end{figure}

\begin{figure}[]
\centering
\psfrag{power}{ \Huge{\bf{power}}}
\psfrag{r}{\Huge{$r$}}
\rotatebox{-90}{ \resizebox{3 in}{!}{ \includegraphics{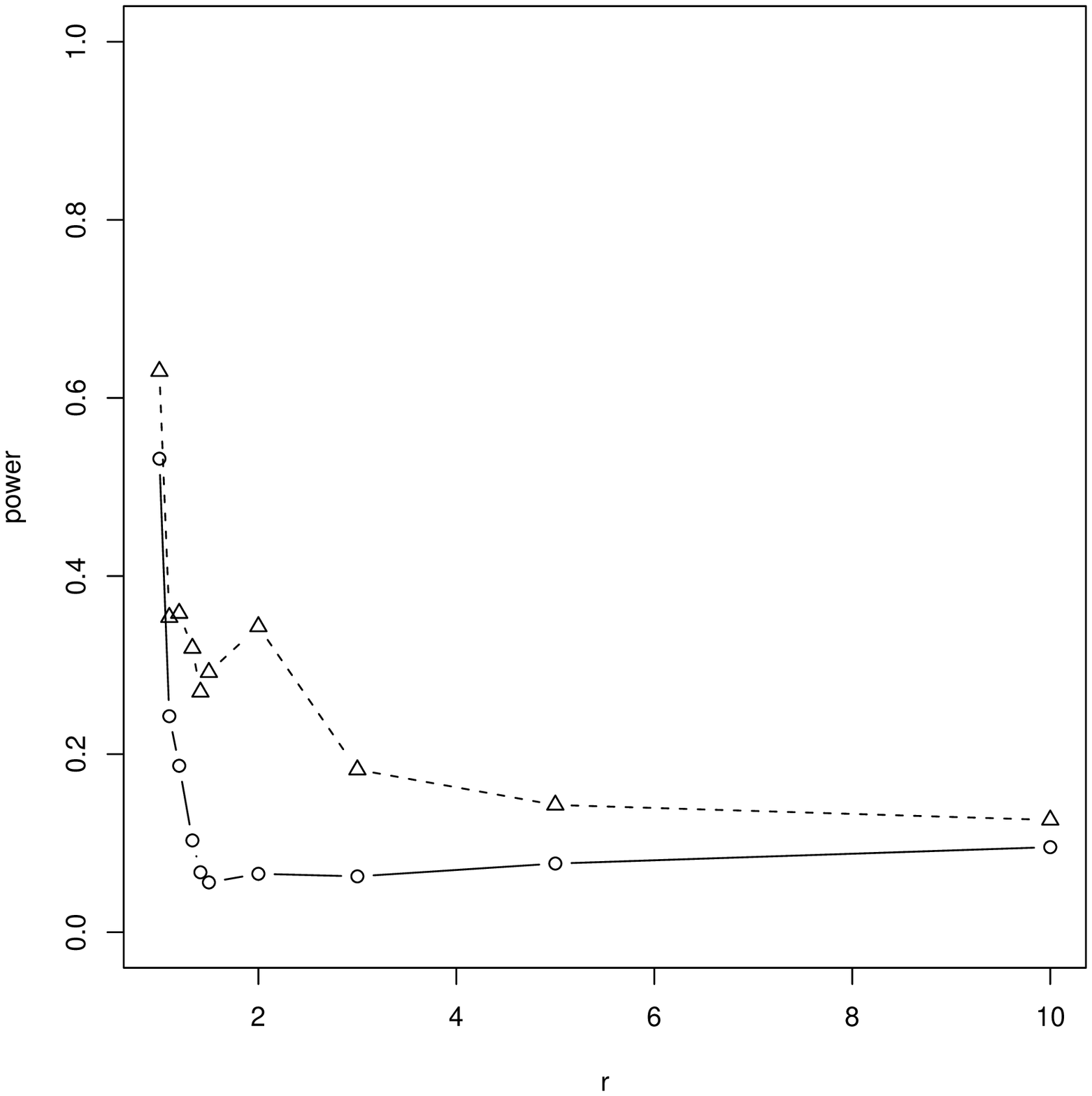} } }
\rotatebox{-90}{ \resizebox{3 in}{!}{ \includegraphics{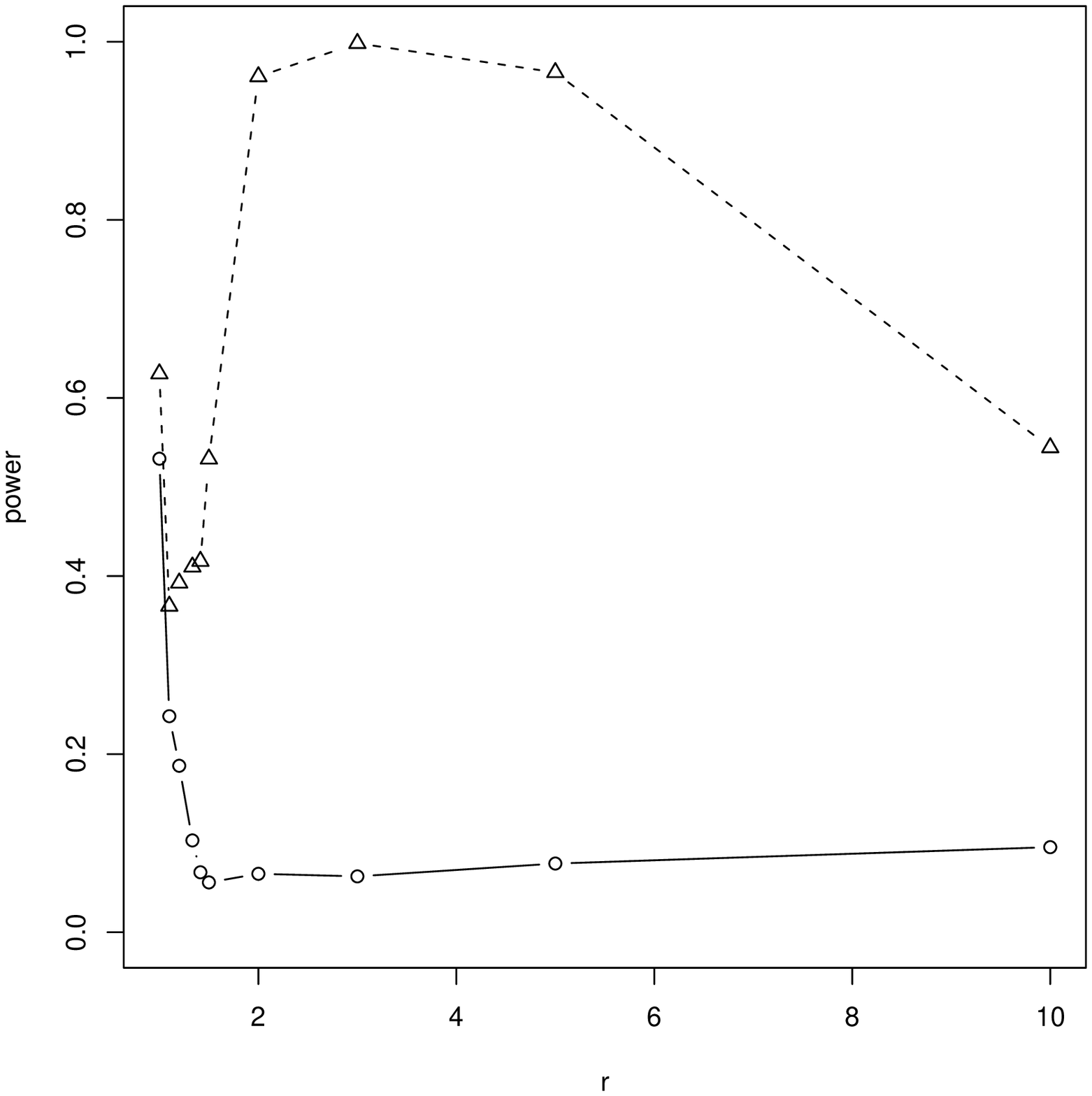} } }
\caption{ \label{fig:AggSimPowerCurve}
Monte Carlo power using the asymptotic critical value against association alternatives
$H^A_{\sqrt{3}/12}$ (left)
and
$H^A_{5\,\sqrt{3}/24}$ (right)
as a function of $r$, for $n=10$.
The $r$ values plotted are $1,\,11/10,\,12/10,\,4/3,\,\sqrt{2},\,,\,2,\,3,\,5,\,10$.
}
\end{figure}

In Figure \ref{fig:AggSimPowerPlots}, we present a Monte Carlo investigation
against the association alternative $H^A_{\sqrt{3}/12}$ for $r=11/10$ and $n=10$ and $100$.
The analysis is same as in the analysis of the Figure \ref{fig:SegSimPowerPlots}.
In Figure \ref{fig:AggSimPowerCurve}, we present
a Monte Carlo investigation of power against
$H^A_{\sqrt{3}/12}$
and
$H^A_{5\,\sqrt{3}/24}$
as a function of $r$ for $n=10$.  The empirical significance level is about $.05$
for $r=3/2,\,2,\,3,\,5$ which have the empirical power $\widehat{\beta}^A_{10}(r,\sqrt{3}/12) \le .35$
with maximum power at $r=2$, and $\widehat{\beta}^A_{10}(r,5\,\sqrt{3}/24) =1$ at $r=3$.
So, for small sample sizes, moderate values of $r$ are more appropriate for normal approximation,
as they yield the desired significance level, and the more severe the association,
the higher the power estimate.

\subsection{Pitman Asymptotic Efficacy}
\label{sec:Pitman}

Pitman asymptotic efficiency (PAE)
provides for an investigation of ``local asymptotic power''
--- local around $H_0$.
This involves the limit as $n \rightarrow \infty$ as well as
the limit as $\epsilon \rightarrow 0$.
A detailed discussion of PAE can be found in \cite{kendall:1979} and \cite{eeden:1963}.
For segregation or association alternatives
the PAE is given by $\PAE(\rho_n(r)) = \frac{\left( \mu^{(k)}(r,\epsilon=0) \right)^2}{\nu(r)}$
where $k$ is the minimum order of the derivative with respect to $\epsilon$ for which
$\mu^{(k)}(r,\epsilon=0) \not= 0$.  That is, $\mu^{(k)}(r,\epsilon=0) \not=0$ but
$\mu^{(l)}(r,\epsilon=0)=0$ for $l=1,2,\ldots,k-1$.
Then under segregation alternative $H^S_{\epsilon}$ and association alternative $H^A_{\epsilon}$,
the PAE of $\rho_n(r)$ is given by
$$
\PAE^S(r) =
   \frac{\left( \mu_S^{\prime\prime}(r,\epsilon=0) \right)^2}{\nu(r)} \text{ and }
\PAE^A(r) =
   \frac{\left( \mu_A^{\prime\prime}(r,\epsilon=0) \right)^2}{\nu(r)},
$$
respectively, since $\mu_S^{\prime}(r,\epsilon=0)=\mu_A^{\prime}(r,\epsilon=0) = 0$.
Equation (\ref{eq:Asyvar}) provides the denominator;
the numerator requires $\mu(r,\epsilon)$ which is provided in Appendix 2 for under both
segregation and association alternatives, where we only use the intervals of $r$ that
donot vanish as $\epsilon \rightarrow 0$.

\begin{figure}[]
\centering
\psfrag{r}{\scriptsize{$r$}}
\psfrag{pS}{\scriptsize{$\PAE^S(r)$}}
\epsfig{figure=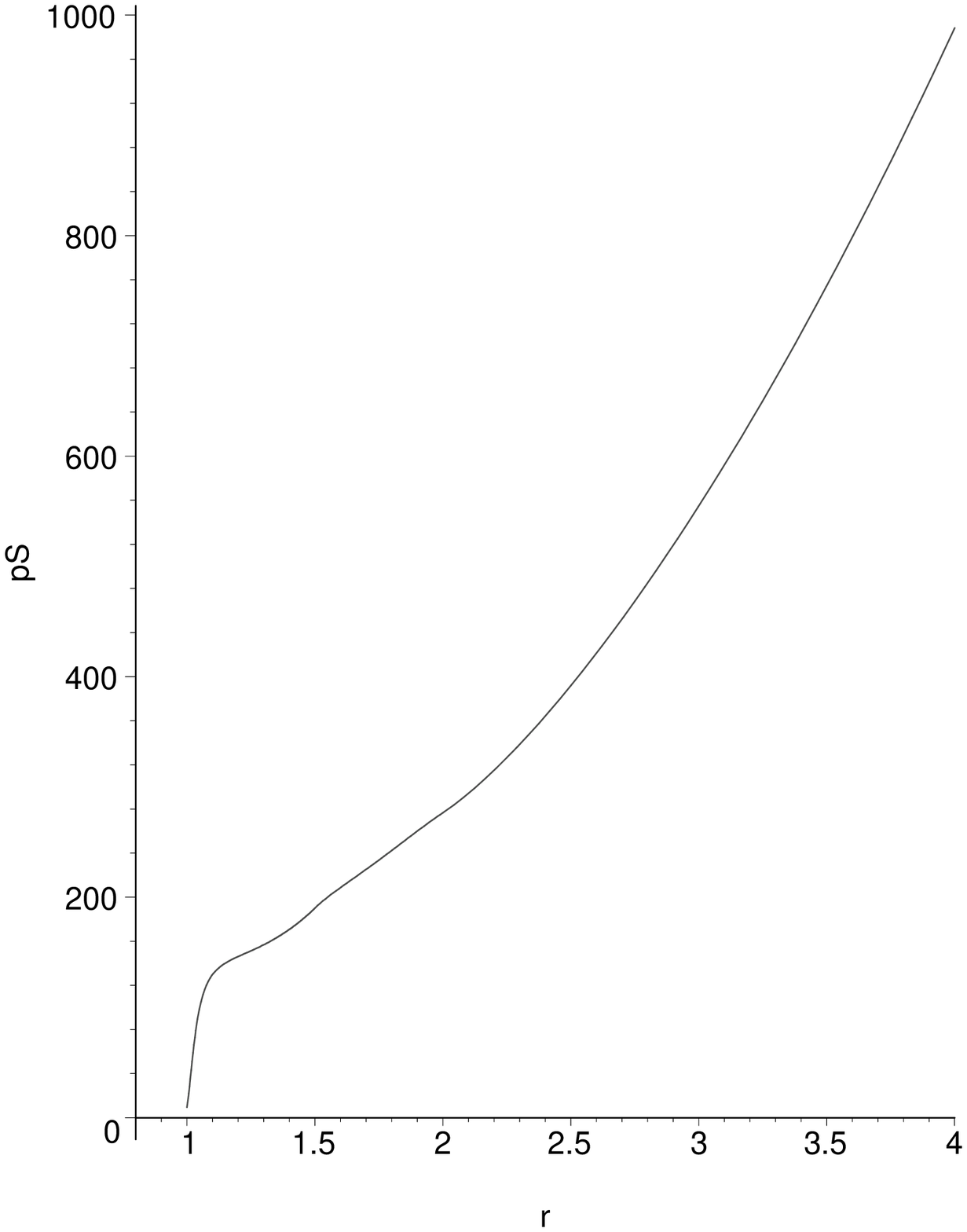, height=200pt, width=200pt}
\psfrag{pA}{\scriptsize{$\PAE^A(r)$}}
\epsfig{figure=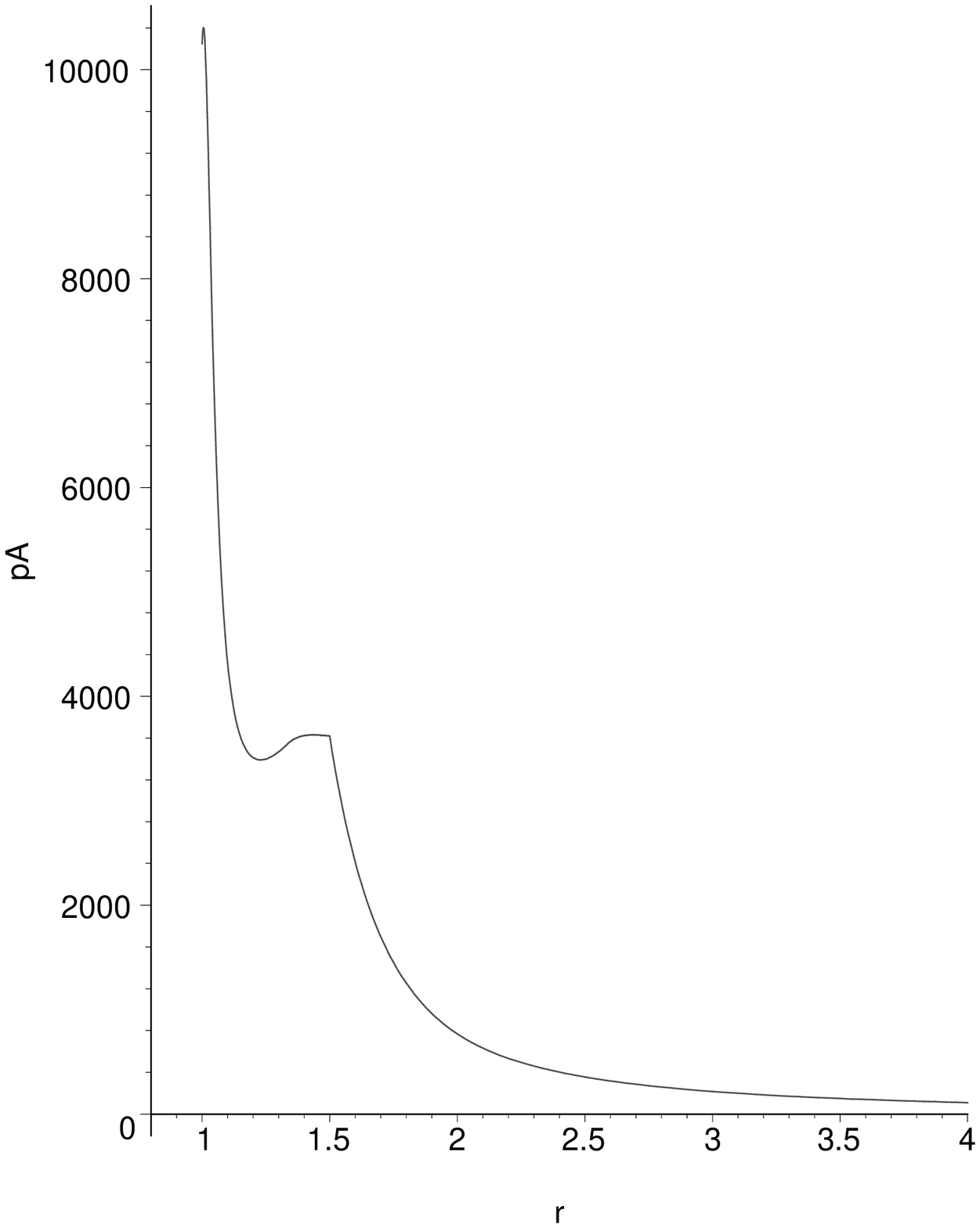, height=200pt, width=200pt}
\caption{ \label{fig:PAECurves}
Pitman asymptotic efficiency against segregation (left) and association (right)
as a function of $r$.
Notice that vertical axes are differently scaled.}
\end{figure}

In Figure \ref{fig:PAECurves}, we present the PAE as a function of $r$ for both segregation
and association.
Notice that
$\PAE^S(r=1) = 160/7 \approx 22.8571$, $\lim_{r \rightarrow \infty} \PAE^S(r) = \infty$,
$\PAE^A(r=1) = 174240/17 \approx 10249.4118$, $\lim_{r \rightarrow \infty} \PAE^A(r) = 0$,
$\argsup_{r \in [1,\infty)} \PAE^A(r) \approx 1.006$ with
$\sup _{r \in [1,\infty)} \PAE^A(r) \approx 10399.7726$. $\PAE^A(r)$
has also a local supremum at $r_l \approx 1.4356$ with $\PAE^A(r_l)\approx 3630.8932$.
Based on the asymptotic efficiency analysis, we suggest,
for large $n$ and small $\epsilon$,
choosing $r$ large for testing against segregation and
choosing $r$ small for testing against association.

\subsection{Hodges-Lehmann Asymptotic Efficacy}
\label{sec:Hodges-Lehmann}

\begin{figure}[]
\centering
\psfrag{r}{\scriptsize{$r$}}
\epsfig{figure=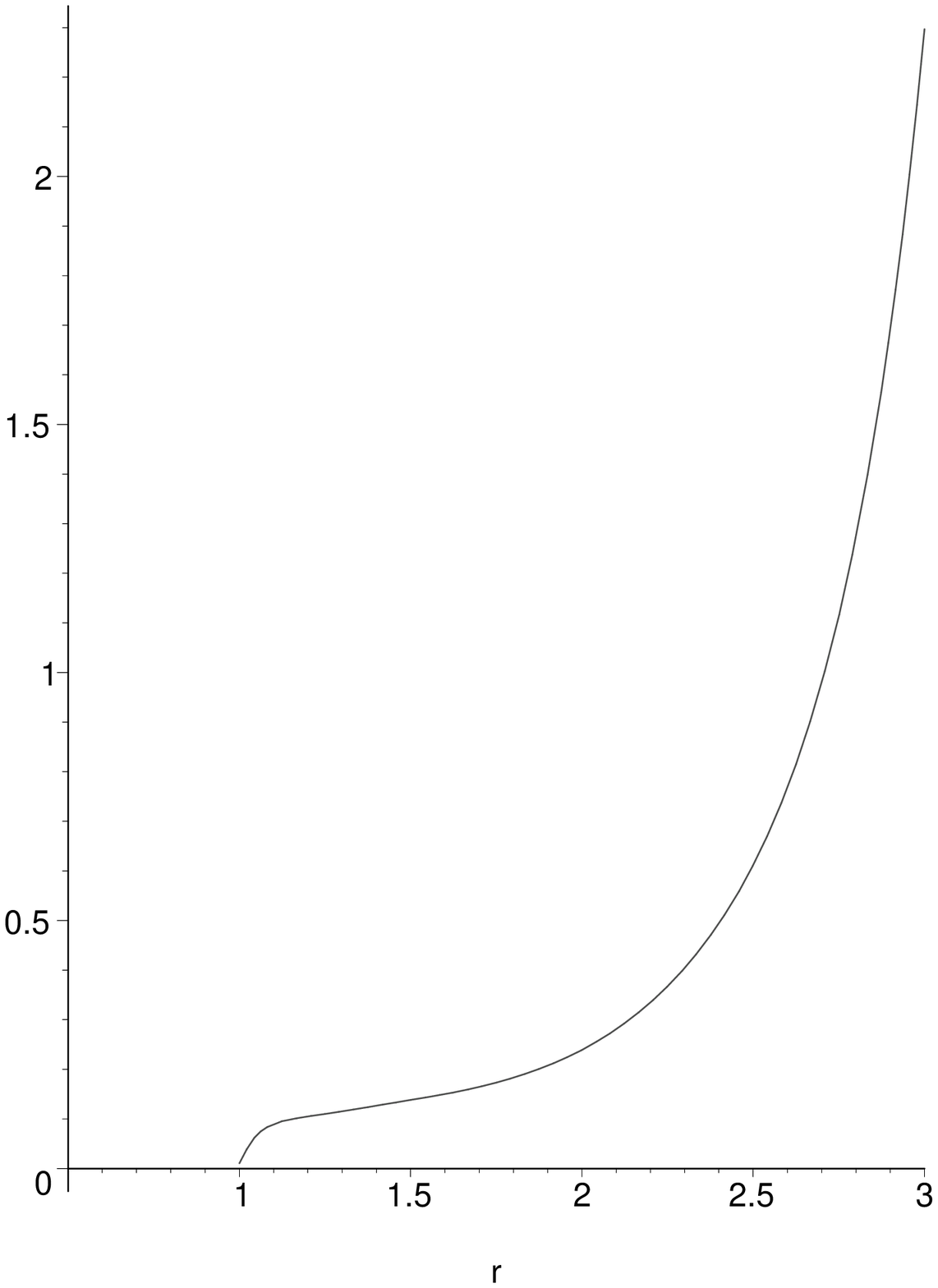, height=150pt, width=150pt}
\epsfig{figure=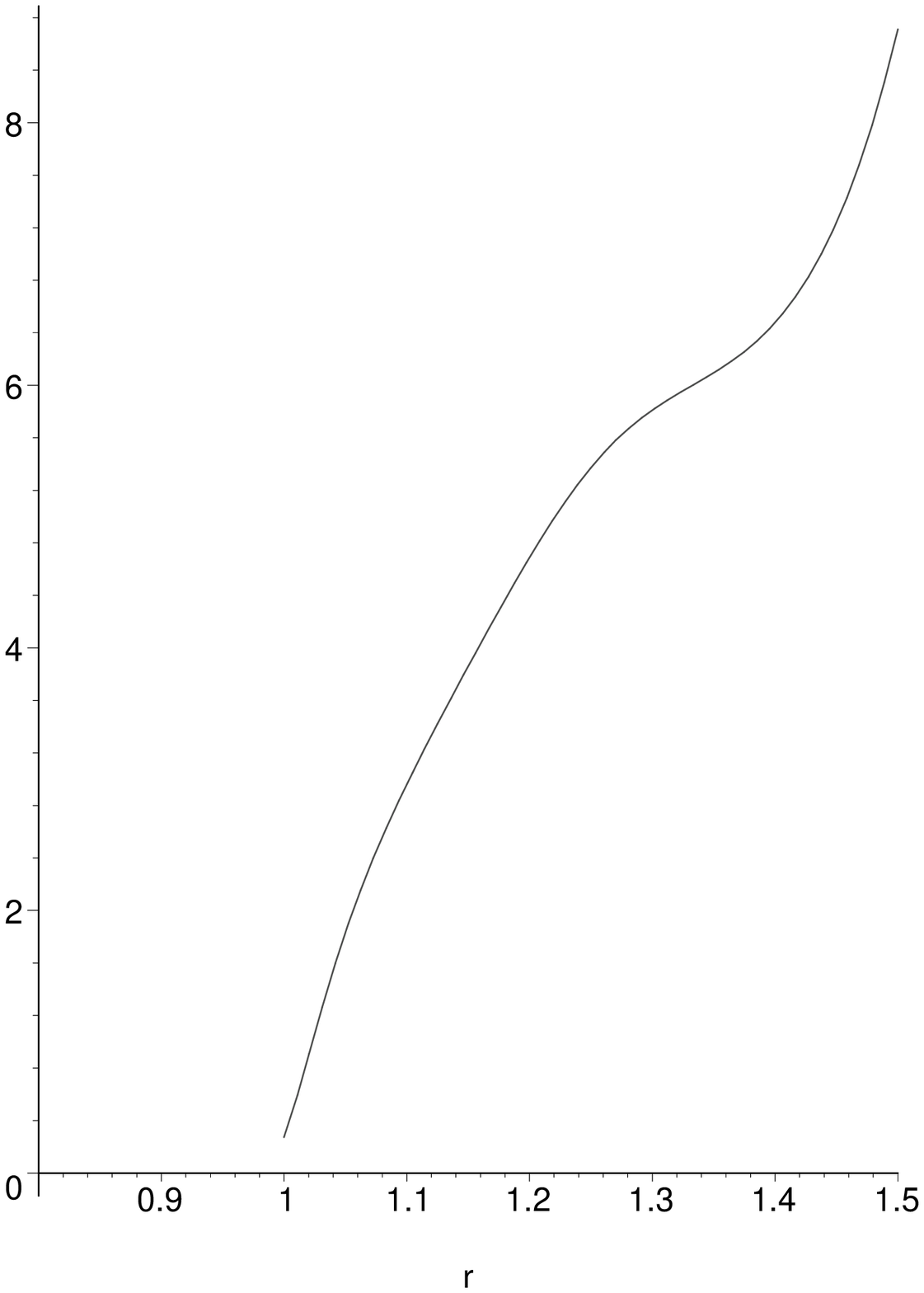, height=150pt, width=150pt}
\epsfig{figure=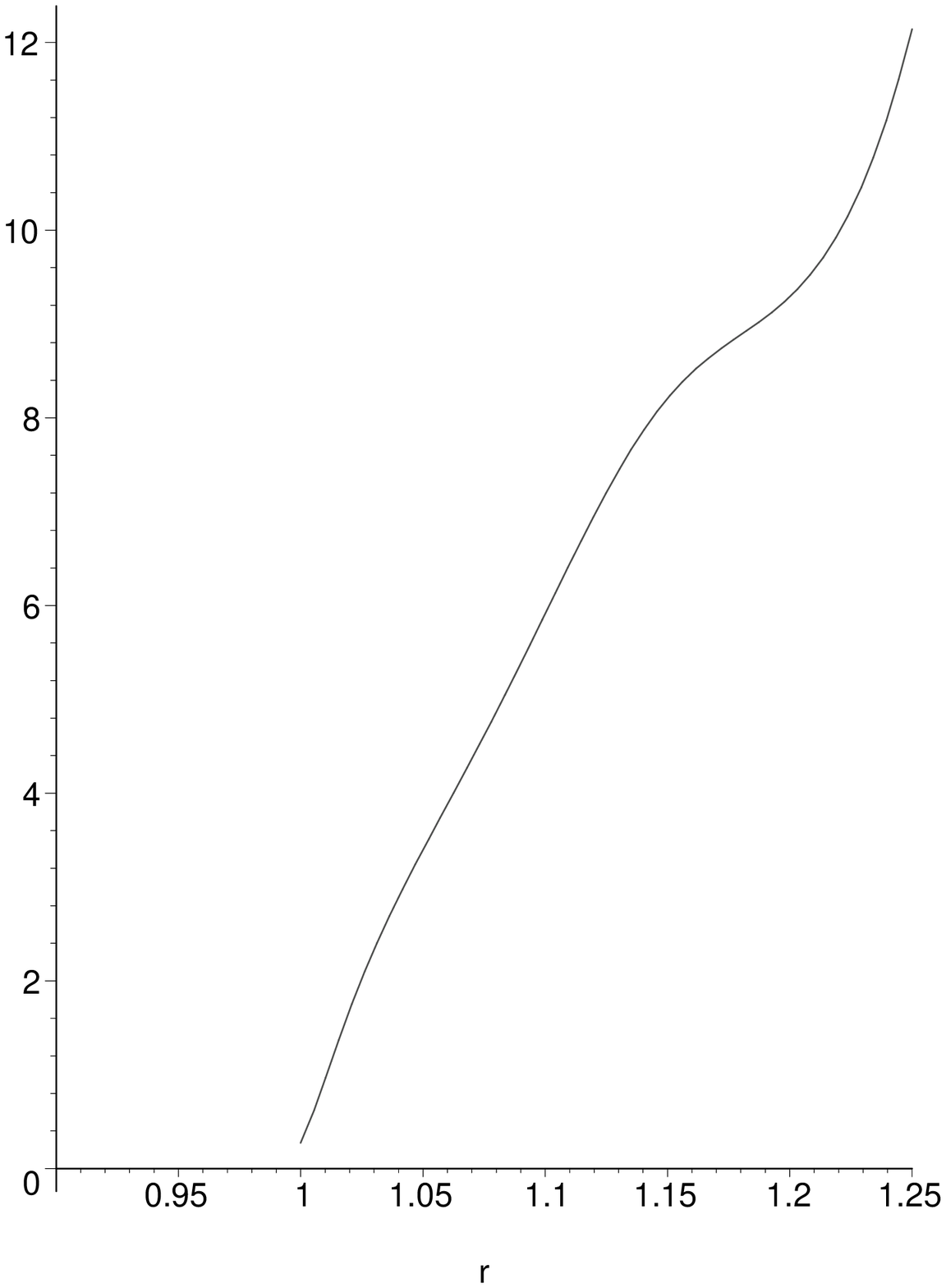, height=150pt, width=150pt}
\caption{ \label{fig:HLAEPlots for Seg}
Hodges-Lehmann asymptotic efficiency
against segregation alternative $H^S_{\epsilon}$
as a function of $r$
for $\epsilon= \sqrt{3}/8, \sqrt{3}/4, 2\sqrt{3}/7$ (left to right).}
\end{figure}

Hodges-Lehmann asymptotic efficiency (HLAE) of $\rho_n(r)$
(see e.g. \cite{hodges:1956}) under $H^S_{\epsilon}$
is given by
$$
\HLAE^S(r,\epsilon):=\frac{(\mu_S(r,\epsilon)-\mu(r))^2}{\nu_S(r,\epsilon)}.
$$
HLAE for association is defined similarly.
Unlike PAE, HLAE does not involve the limit as $\epsilon \rightarrow 0$.
Since this requires the mean and, especially,
the asymptotic variance of $\rho_n(r)$ {\it under an alternative},
we investigate HLAE for specific values of $\epsilon$.
Figure \ref{fig:HLAEPlots for Seg}
contains a graph of HLAE
against segregation as a function of $r$
for $\epsilon= \sqrt{3}/8,\, \sqrt{3}/4,\, 2\sqrt{3}/7$.
See Appendix 3 for explicit forms of $\mu_S(r,\epsilon)$ and $\nu_S(r,\epsilon)$ for $\epsilon=\sqrt{3}/4$.

From Figure \ref{fig:HLAEPlots for Seg}, we see that,
against $H^S_{\epsilon}$,
$\HLAE^S(r,\epsilon)$
appears to be an increasing function, dependent on $\epsilon$, of $r$.
Let $r_d(\epsilon)$ be the minimum $r$ such that
$\rho_n(r)$ becomes degenerate under the alternative $H^S_{\epsilon}$.
Then
$r_d(\sqrt{3}/8)=4$,
$r_d\bigl( \sqrt{3}/4\bigr)=2$, and
$r_d\bigl( 2\,\sqrt{3}/7 \bigr)=2$.  In fact, for $\epsilon \in (0,\sqrt{3}/4]$,
$r_d(\epsilon)=\sqrt{3}/(2\,\epsilon)$ and for $\epsilon \in (\sqrt{3}/4,\sqrt{3}/3)$,
$r_d(\epsilon)=\sqrt{3}/\epsilon -2$.
Notice that $\lim_{r \rightarrow r_d(\epsilon)}\HLAE^S(r,\epsilon)=\infty$,
which is in agreement with $\PAE^S$ as $\epsilon \rightarrow 0$;
since as $\epsilon \rightarrow 0$, HLAE becomes PAE and $r_d(\epsilon) \rightarrow \infty$
and under $H_0$, $\rho_n(r)$ is degenerate for $r=\infty$.
So HLAE suggests choosing $r$ large against segregation,
but in fact choosing $r$ too large will reduce power since
$r \geq r_d(\epsilon)$ guarantees the complete digraph
under the alternative and, as $r$ increases therefrom,
provides an ever greater probability of seeing the complete digraph
under the null.

\begin{figure}[ht]
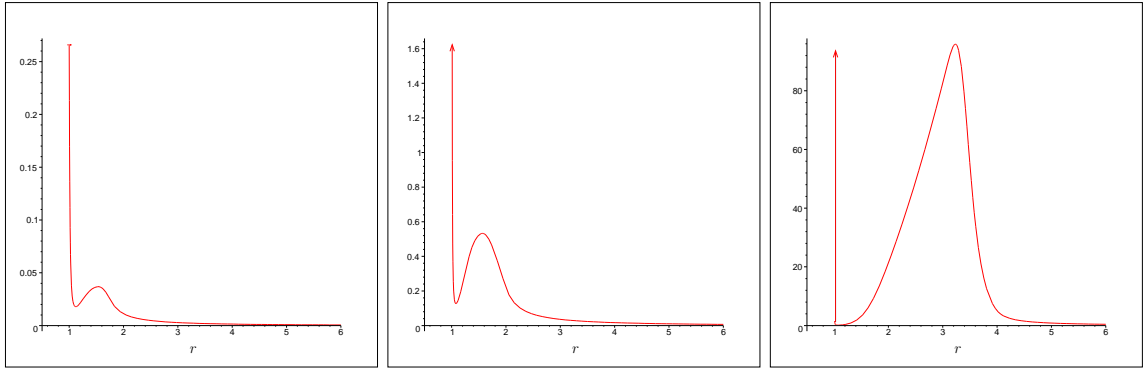

\centering
\scalebox{.23}{\input{hlae_agg3.pstex_t}}
\scalebox{.23}{\input{hlae_agg2.pstex_t}}
\scalebox{.23}{\input{hlae_agg1.pstex_t}}
\caption{ \label{fig:HLAEPlots for Agg}
Hodges-Lehmann asymptotic efficiency
against association alternative $H^A_{\epsilon}$
as a function of $r$
for $\epsilon= \sqrt{3}/21,\, \sqrt{3}/12,\, 5\,\sqrt{3}/24$ (left to right).}
\end{figure}

Figure \ref{fig:HLAEPlots for Agg}
contains a graph of HLAE
against association as a function of $r$
for $\epsilon= 5\,\sqrt{3}/24,\, \sqrt{3}/12, \, \sqrt{3}/21$.
See Appendix 3 for explicit forms of $\mu_A(r,\epsilon)$ and $\nu_A(r,\epsilon)$ for $\epsilon=\sqrt{3}/12$.
Notice that since $\nu(r,\epsilon)=0$ for $\epsilon \ge \sqrt{3}/12$,
$\HLAE^A(r=1,\epsilon)=\infty$ for $\epsilon \ge \sqrt{3}/12$
and $\lim_{r\rightarrow \infty}\HLAE^A(r,\epsilon)=0$.

In Figure \ref{fig:HLAEPlots for Agg} we see that,
against $H^A_{\epsilon}$,
$\HLAE^A(r,\epsilon)$
has a local supremum for $r$ sufficiently larger than 1.
Let $\tilde r$ be the value at which this local supremum is attained.  Then
$\tilde r(5\,\sqrt{3}/24)\approx 3.2323$,
$\tilde r(\sqrt{3}/12) \approx 1.5676$, and
$\tilde r(\sqrt{3}/21) \approx 1.533$.
Note that, as $\epsilon$ gets smaller, $\tilde r$ gets smaller.
Furthermore, $\HLAE^A(r=1,\sqrt{3}/21) < \infty$ and as $\epsilon \rightarrow 0$,
$\tilde r$ becomes the global supremum, and $\PAE^A(r=1)=0$ and
$\argsup_{r \ge 1}\PAE^A(r=1)\approx 1.006$.
So, when testing against association, HLAE suggests choosing moderate $r$,
whereas PAE suggests choosing small $r$.

\subsection{Asymptotic Power Function Analysis}
\label{sec:Asy-Power}
The asymptotic power function (see e.g. \cite{kendall:1979})
can also be investigated as a function of $r$, $n$, and $\epsilon$
using the asymptotic critical value and an appeal to normality.
Under a specific segregation alternative $H^S_{\epsilon}$,
the asymptotic power function is given by
\begin{eqnarray*}
\Pi^S(r,n,\epsilon) =
1-\Phi\left(\frac{z_{(1-\alpha)}\,\sqrt{\nu(r)}+\sqrt{n}\,(\mu(r)-\mu_S(r,\epsilon))}{\sqrt{\nu_S(r,\epsilon)}}\right),
\end{eqnarray*}
where $z_{1-\alpha}=\Phi^{-1}(1-\alpha)$.
  Under $H^A_{\epsilon}$,
we have
\begin{eqnarray*}
\Pi^A(r,n,\epsilon) =
\Phi\left(\frac{z_{\alpha}\,\sqrt{\nu(r)}+\sqrt{n}\,(\mu(r)-\mu_A(r,\epsilon))}{\sqrt{\nu_A(r,\epsilon)}}\right).
\end{eqnarray*}

\begin{figure}[ht]
\centering
\psfrag{r}{\scriptsize{$r$}}
\epsfig{figure=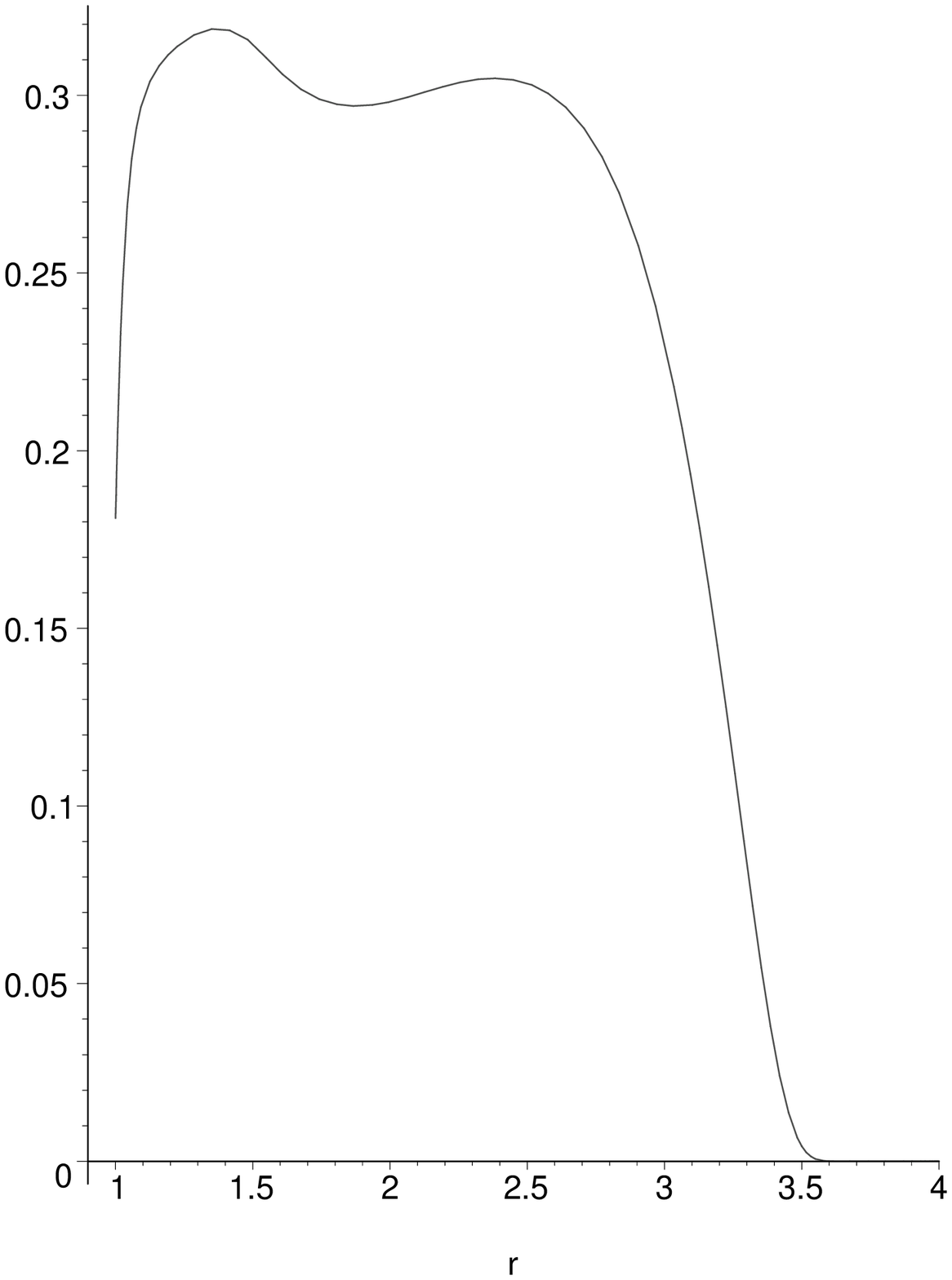,    height=115pt, width=115pt}
\epsfig{figure=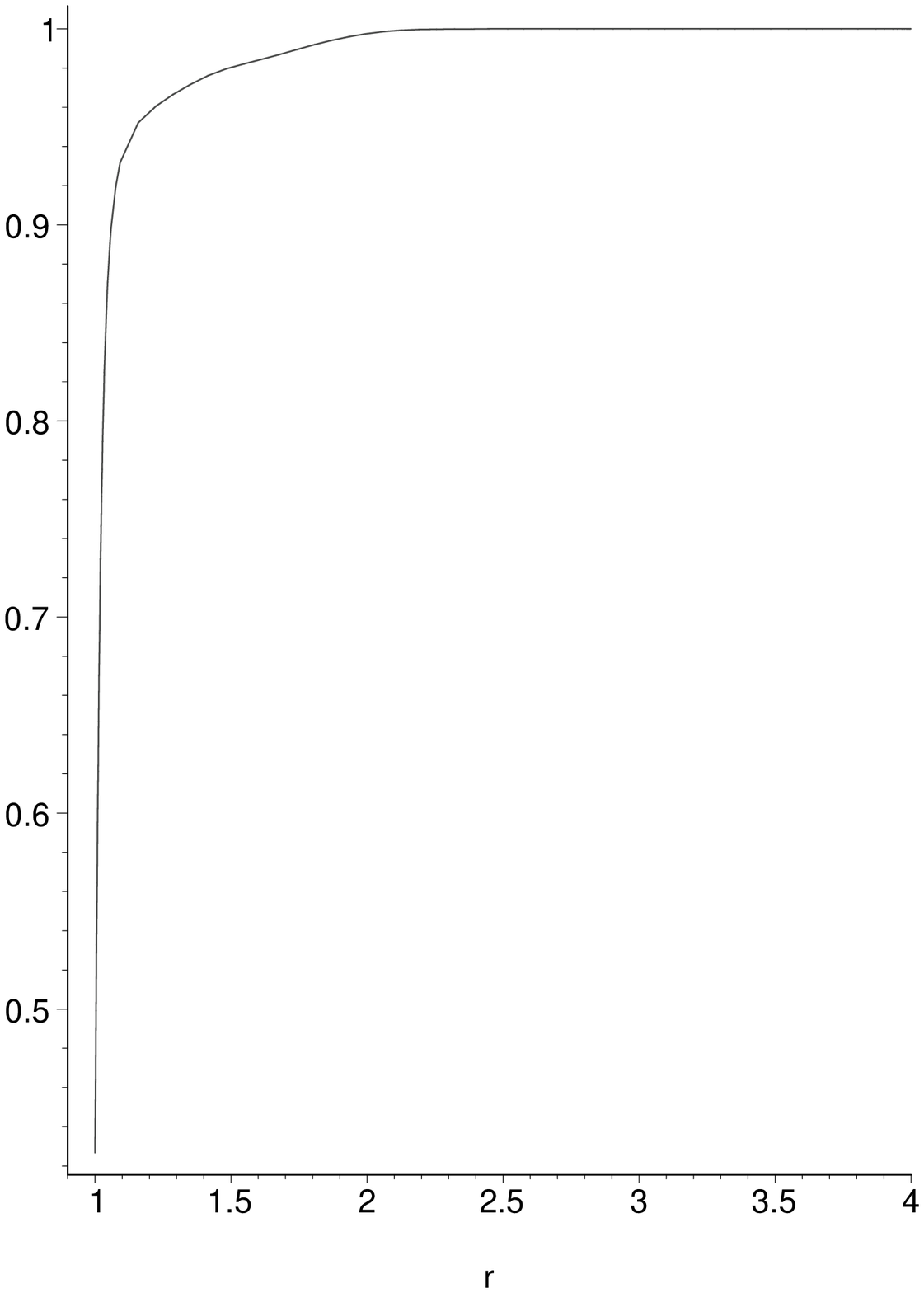,   height=115pt, width=115pt}
\epsfig{figure=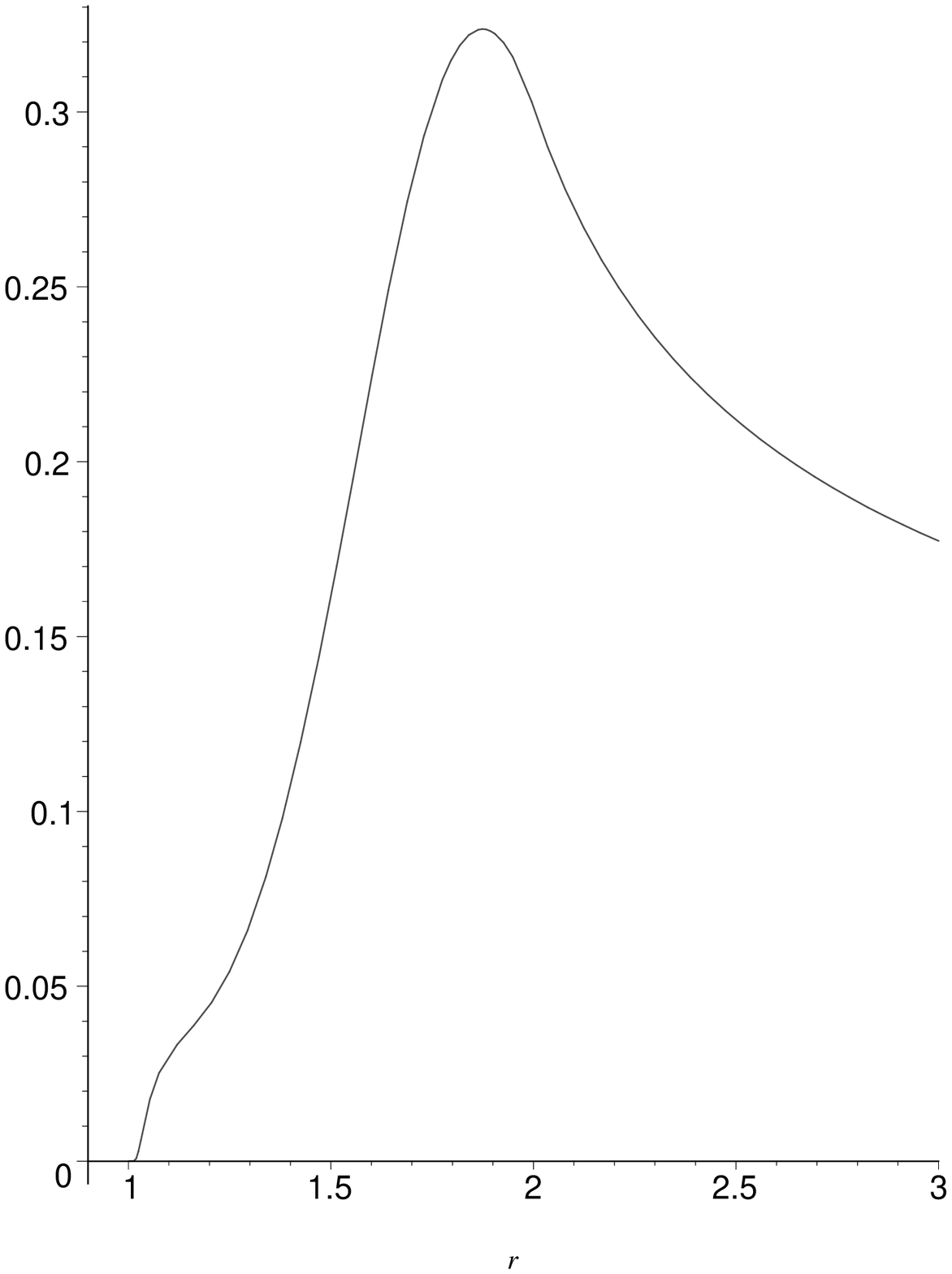,  height=115pt, width=115pt}
\epsfig{figure=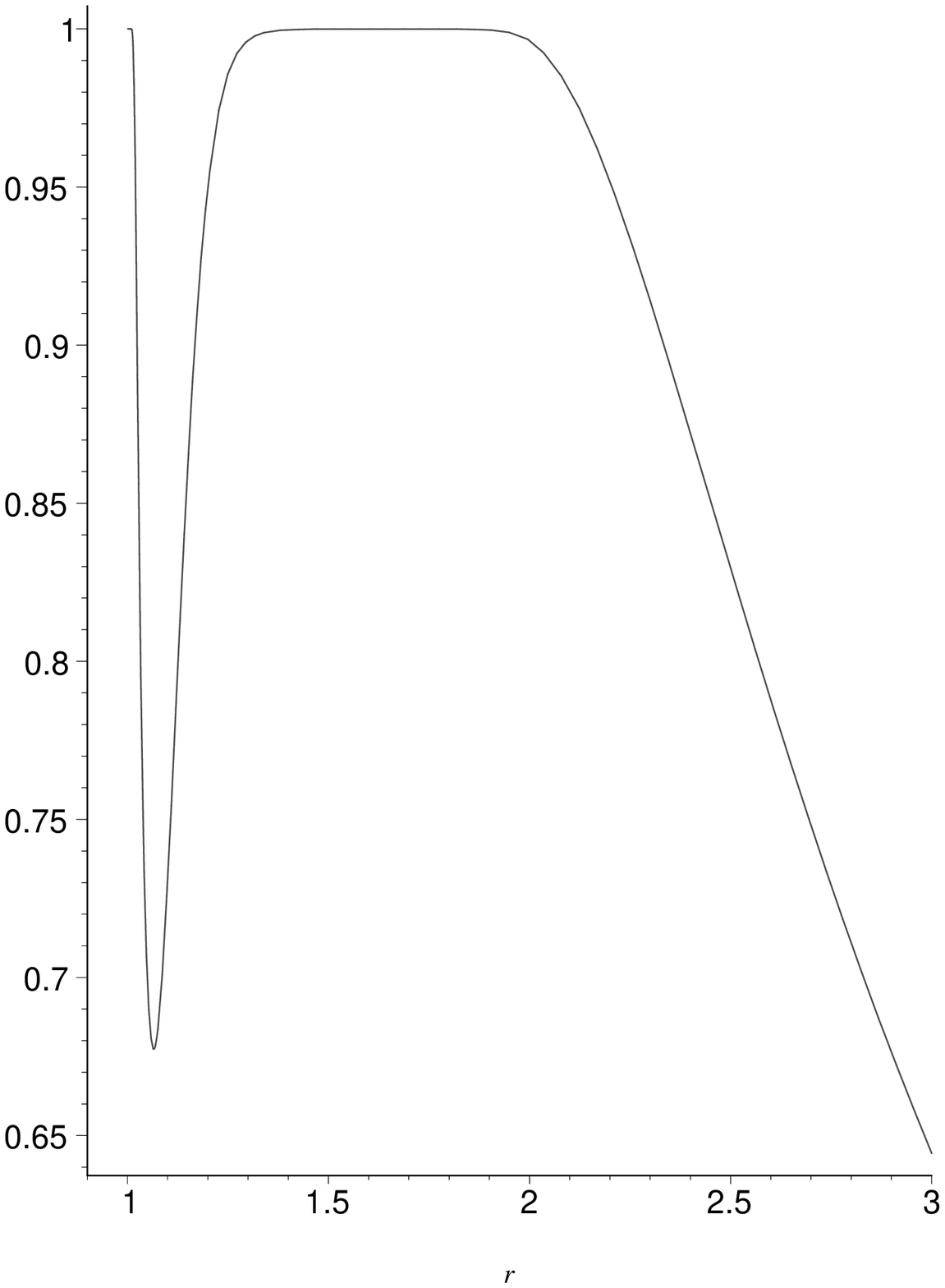, height=115pt, width=115pt}
\caption{
\label{fig:FSPPlots}
Asymptotic power function
against segregation alternative $H^S_{\sqrt{3}/8}$
as a function of $r$
for $n=10$ (first from left) and $n=100$ (second) and association alternative $H^A_{\sqrt{3}/12}$
as a function of $r$
for $n=10$ (third) and $n=100$ (fourth).
}
\end{figure}

Analysis of Figure \ref{fig:FSPPlots}
shows that, against $H^S_{\sqrt{3}/8}$,
a large choice of $r$ is warranted for $n=100$
but, for smaller sample size, a more moderate $r$ is recommended. Against $H^A_{\sqrt{3}/12}$,
a moderate choice of $r$ is recommended for both $n=10$ and $n=100$.
This is in agreement with Monte Carlo investigations.

\section{Multiple Triangle Case}

Suppose $\Y$ is a finite collection of points in $\R^2$ with $|\Y| \ge 3$.
Consider the Delaunay triangulation (assumed to exist) of $\Y$,
where $T_j$ denotes the $j^{th}$ Delaunay triangle,
$J$ denotes the number of triangles, and
$C_H(\Y)$ denotes the convex hull of $\Y$.
We wish to test
$H_0: X_i \stackrel{iid}{\sim} \U(C_H(\Y))$
against segregation and association alternatives.


The digraph $D$ is constructed using $N_{\Y_j}^r(\cdot)$ as described
in Section \ref{ref:data-random-PCD},
where for $X_i \in T_j$ the three points in $\Y$ defining the
Delaunay triangle $T_j$ are used as $\Y_j$.
Let $\rho_n(r,J)$ be the relative density of the digraph based on $\X_n$ and
$\Y$ which yields $J$ Delaunay triangles, and let $w_j := A(T_j) / A(C_H(\Y))$ for $j=1,\ldots,J$,
where $A(C_H(\Y))=\sum_{j=1}^{J}A(T_j)$ with $A(\cdot)$ being the area functional.
Then we obtain the following as a corollary to Theorem 2.

{\bf Corollary 1:}
The asymptotic null distribution for $\rho_n(r,J)$ conditional on $\mathcal W=\{w_1,\ldots,w_J\}$
for $r \in [1,\infty]$
is given by $\N(\mu(r,J),\nu(r,J)/n)$ provided that $\nu(r,J)>0$ with
\begin{equation}
\mu(r,J):=\mu(r) \,\sum_{j=1}^{J}w_j^2, \text{ and } \nu(r,J):= \nu(r) \,\sum_{j=1}^{J}w_j^3 +4\mu(r)^2\left[\sum_{j=1}^{J}w_j^3-\left(\sum_{j=1}^{J}w_j^2 \right)^2\right],
\end{equation}
where $\mu(r)$ and $\nu(r)$ are given by equations (\ref{eq:Asymean}) and (\ref{eq:Asyvar}),
respectively.

{\bf Proof:} See Appendix 4.
$\blacksquare$

By an appropriate application of Jensen's Inequality, we see that
$\sum_{j=1}^{J}w_j^3 \ge \bigl(\sum_{j=1}^{J}w_j^2 \bigr)^2.$
Therefore, $\nu(r,J)=0$ iff $\nu(r)=0$ and $\sum_{j=1}^{J}w_j^3=\bigl(\sum_{j=1}^{J}w_j^2 \bigr)^2$,
so asymptotic normality may hold even when $\nu(r)=0$.

Similarly, for the segregation (association) alternatives with $4\,\epsilon^2/3 \times 100 \%$ of
the triangles around the vertices of each triangle is forbidden (allowed),
we obtain the above asymptotic distribution of $\rho_n(r)$ with $\mu(r)$
being replaced by $\mu_S(r,\epsilon)$, $\nu(r)$ by $\nu_S(r,\epsilon)$,
$\mu(r,J)$, by $\mu_S(r,J,\epsilon)$, and $\nu(r,J)$ by $\nu_S(r,J,\epsilon)$.
Likewise for association.

Thus in the case of $J>1$, we have a (conditional) test of
$H_0: X_i \stackrel{iid}{\sim} \U(C_H(\Y))$
which once again
rejects against segregation for large values of $\rho_n(r,J)$ and
rejects against association for small values of $\rho_n(r,J)$.

\begin{figure}[ht]
\centering
\epsfig{figure=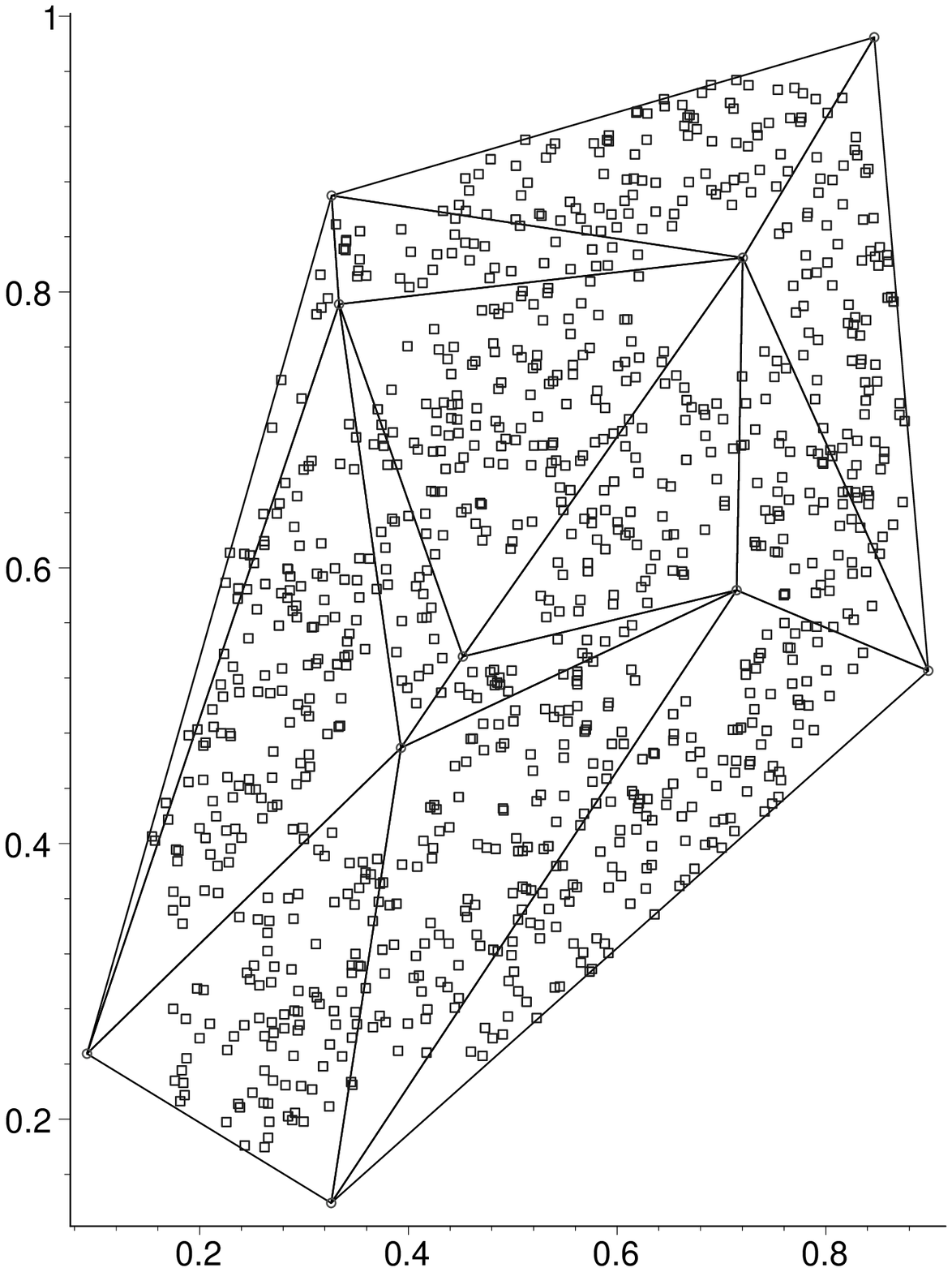, height=125pt, width=125pt}
\epsfig{figure=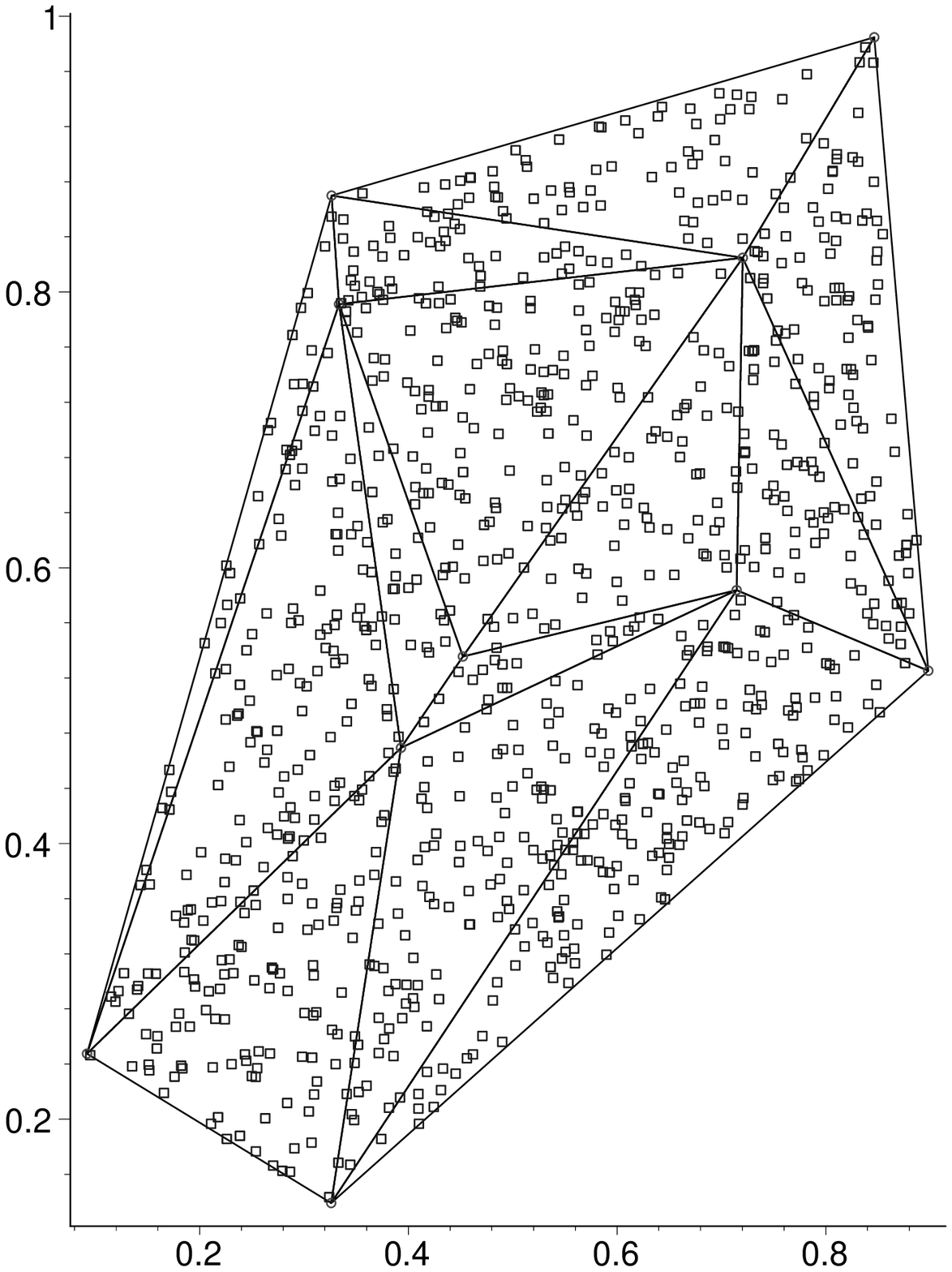,      height=125pt, width=125pt}
\epsfig{figure=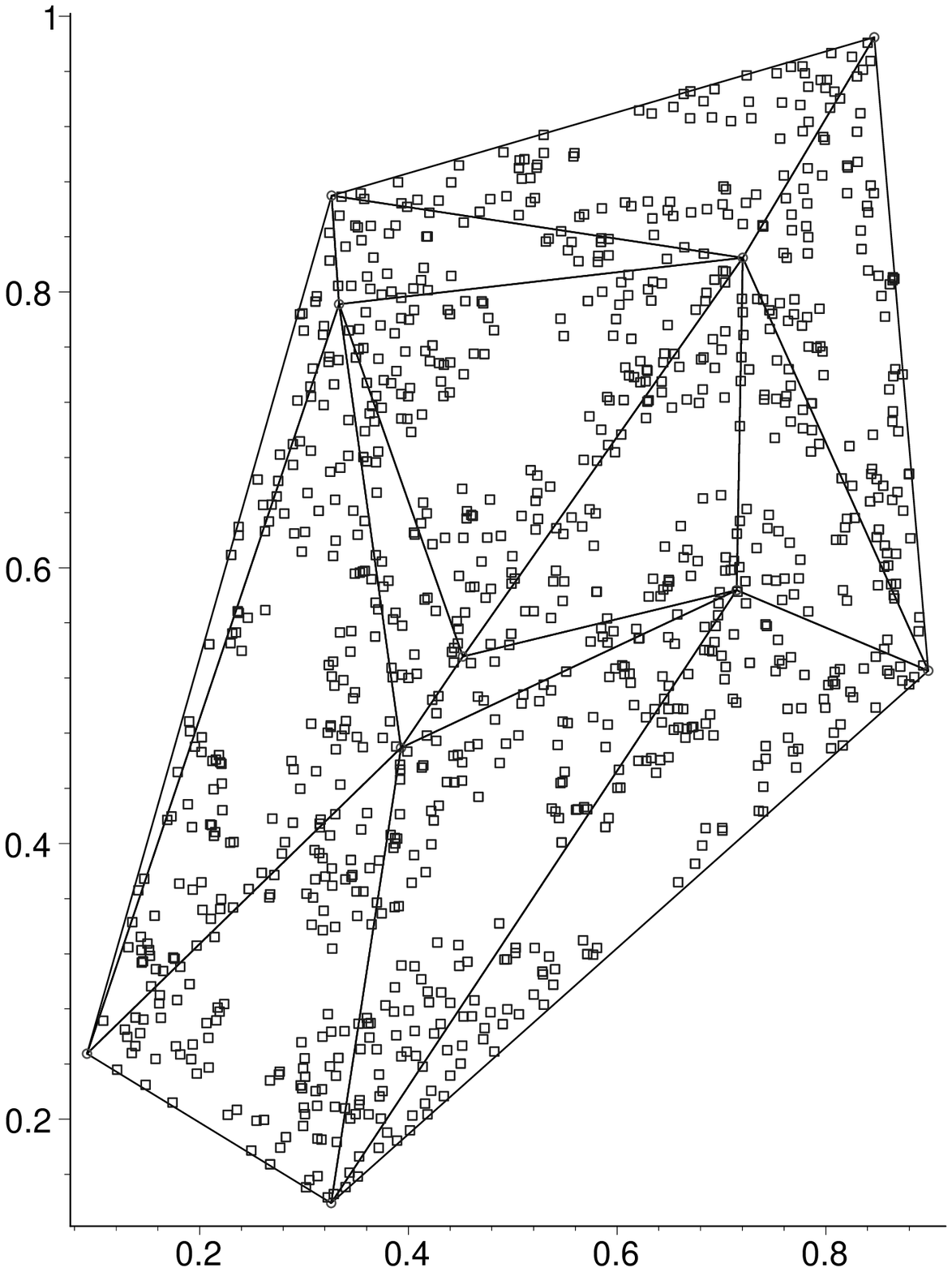, height=125pt, width=125pt}
\caption{\label{fig:deldata}
Realization of segregation (left), $H_0$ (middle), and association (right) for $|\Y|=10$, $J=13$, and $n=1000$.
}
\end{figure}

Depicted in Figure \ref{fig:deldata} are the segregation (with $\delta=1/16$ i.e.
$\epsilon=\sqrt{3}/8$), null, and association
(with $\delta=1/4$ i.e. $\epsilon=\sqrt{3}/12$) realizations (from left to right)
with $n=1000$, $|\Y|=10$, and $J=13$.  For the null realization,
the $p$-value is greater than 0.1 for all $r$ values and both alternatives.
For the segregation realization,
we obtain
$p < 0.0031$ for $ 1< r \le 5$
and
$p > 0.24$ for $r=1$ and $r \ge 10$.
For the association realization,
we obtain
$p < 0.0135$ for $1<r \le 3$,
$p=.14$ for $r=1$, and
$p > 0.25$ for for $r \ge 5$. Note that this is only for one realization of $\X_n$.

\begin{figure}[]
\centering
\psfrag{power}{ \Huge{\bf{power}}}
\psfrag{r}{ \Huge{$r$}}
\rotatebox{-90}{ \resizebox{2. in}{!}{ \includegraphics{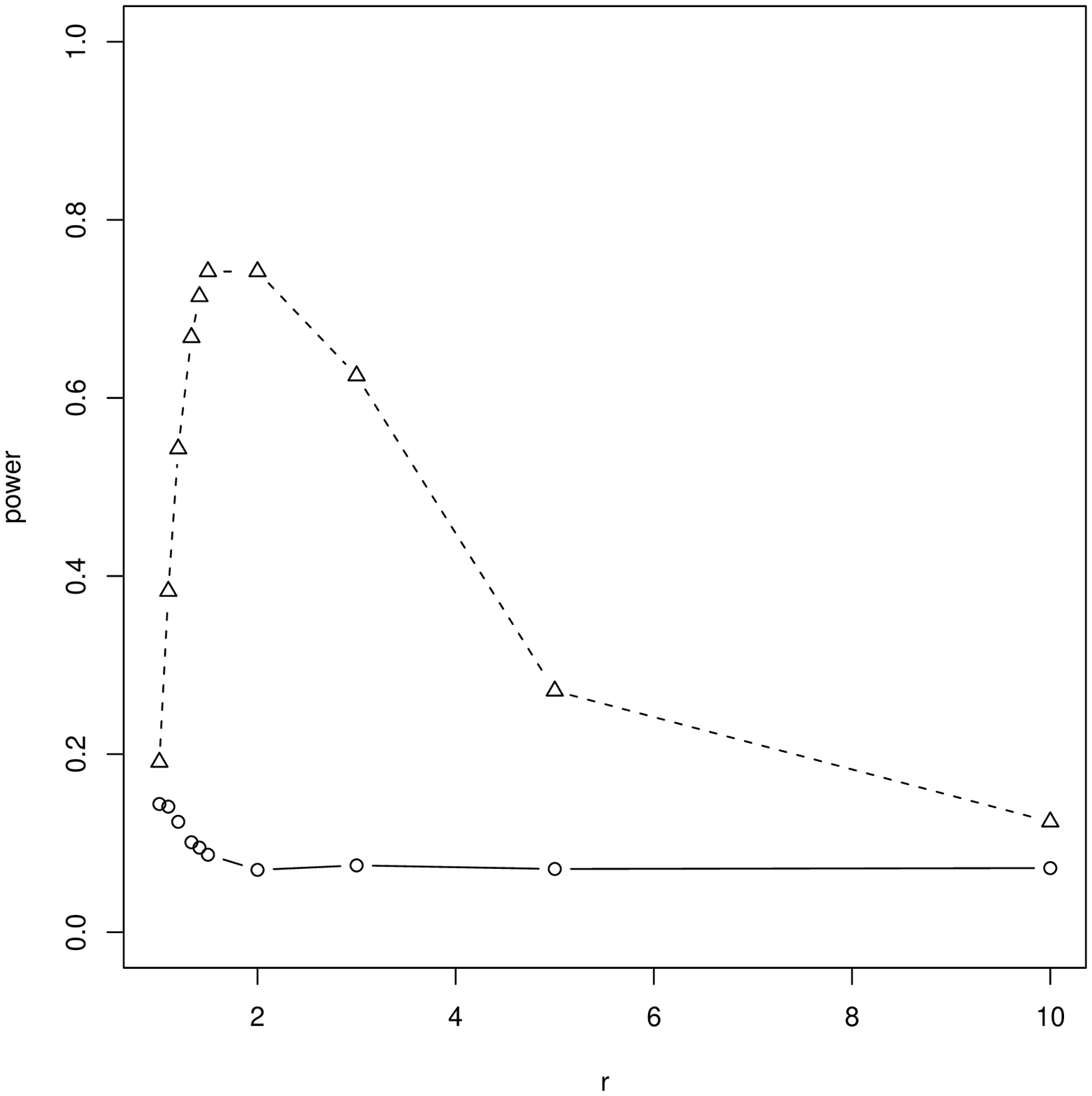}}}
\rotatebox{-90}{ \resizebox{2. in}{!}{ \includegraphics{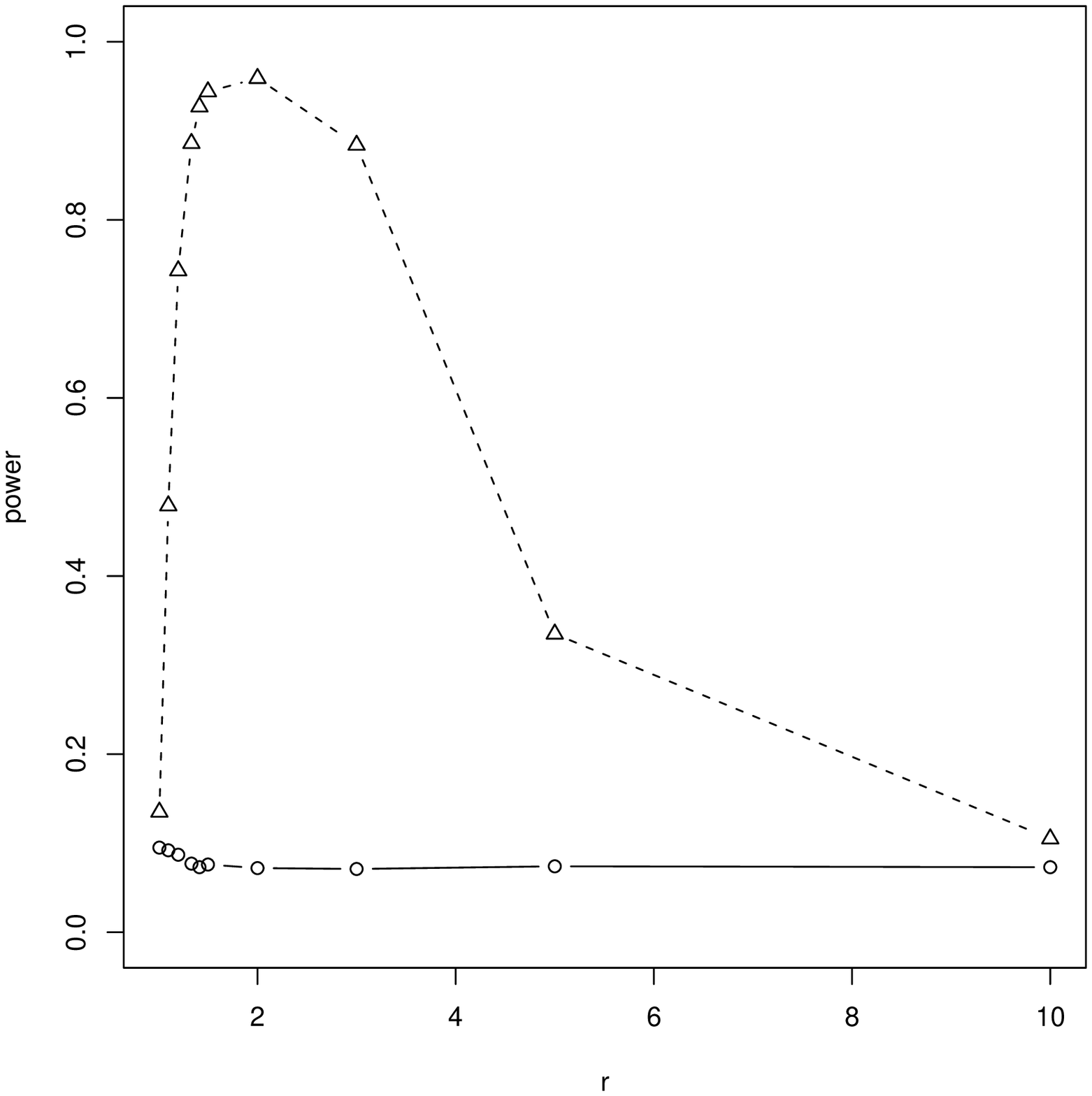}}}
\rotatebox{-90}{ \resizebox{2. in}{!}{ \includegraphics{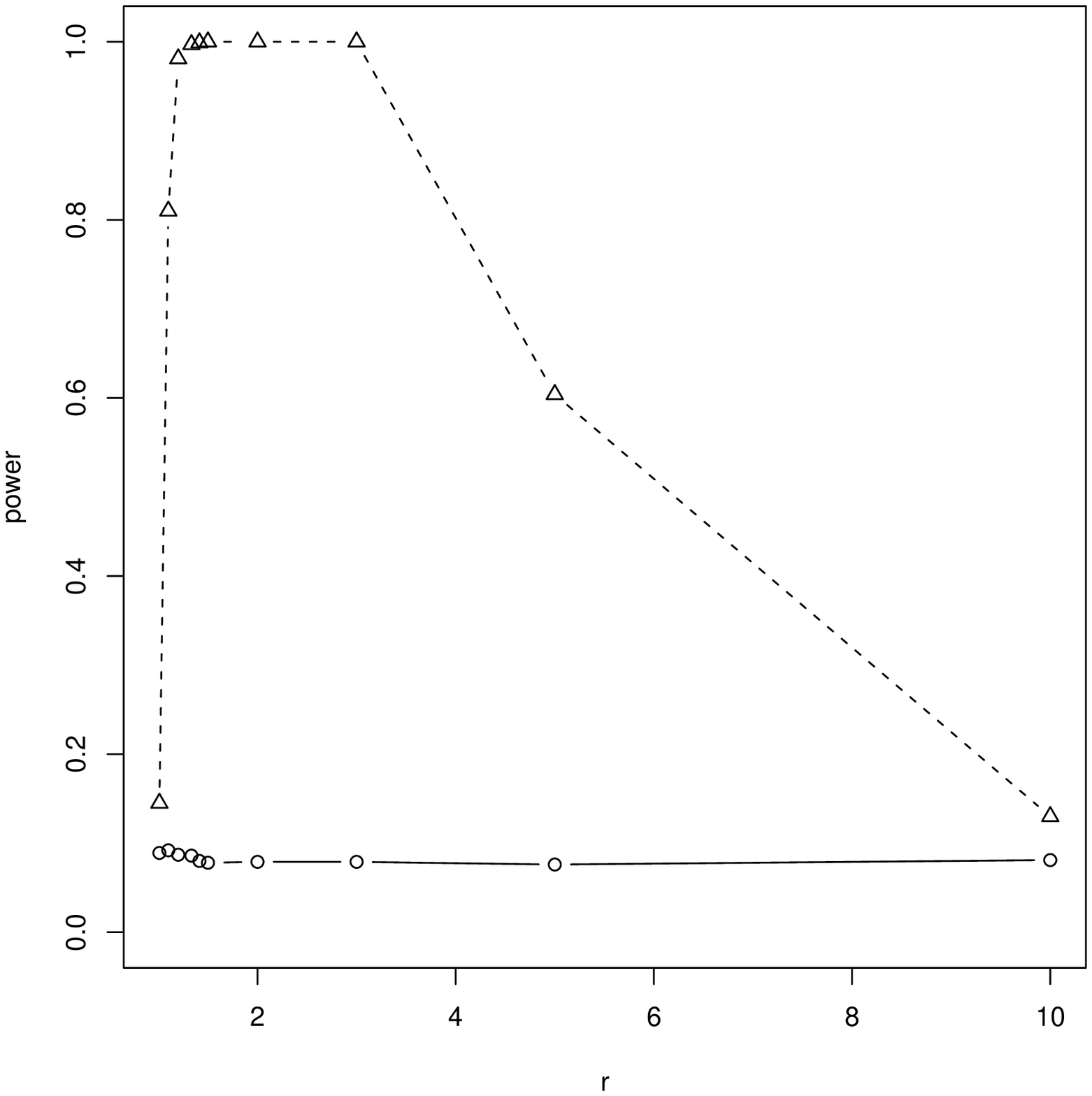}}}
\caption{
 \label{fig:MTSegSim1}
Monte Carlo power using the asymptotic critical value against $H^S_{\sqrt{3}/8}$,
as a function of $r$, for $n=100$ (left), $n=200$ (middle), and $n=500$ (right)
conditional on the realization of $\Y$ in Figure \ref{fig:deldata}.
The circles represent the empirical significance levels while triangles
represent the empirical power values.}
\end{figure}

\begin{figure}[]
\centering
\psfrag{power}{ \Huge{\bf{power}}}
\psfrag{r}{ \Huge{$r$}}
\rotatebox{-90}{ \resizebox{2. in}{!}{ \includegraphics{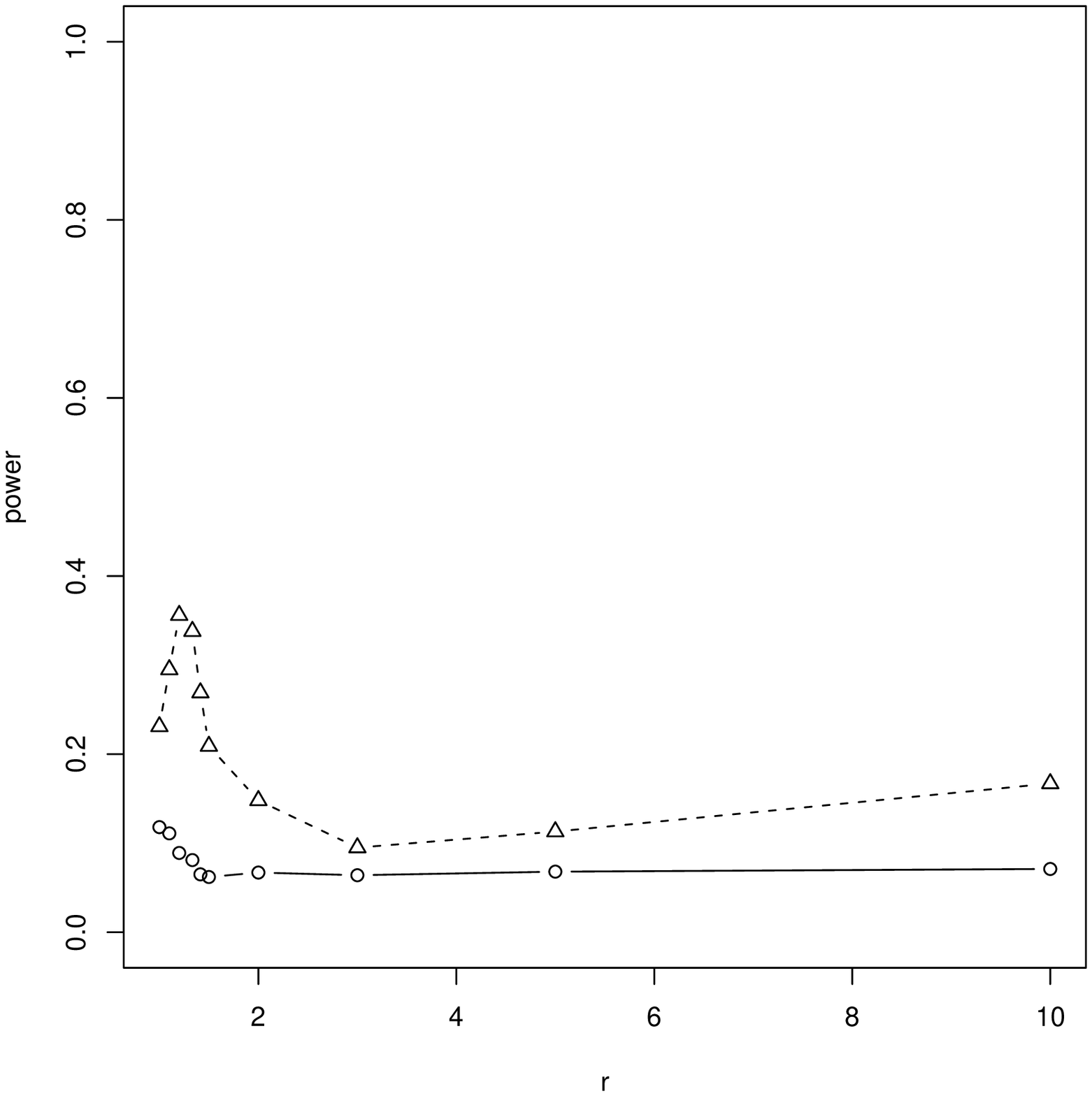}}}
\rotatebox{-90}{ \resizebox{2. in}{!}{ \includegraphics{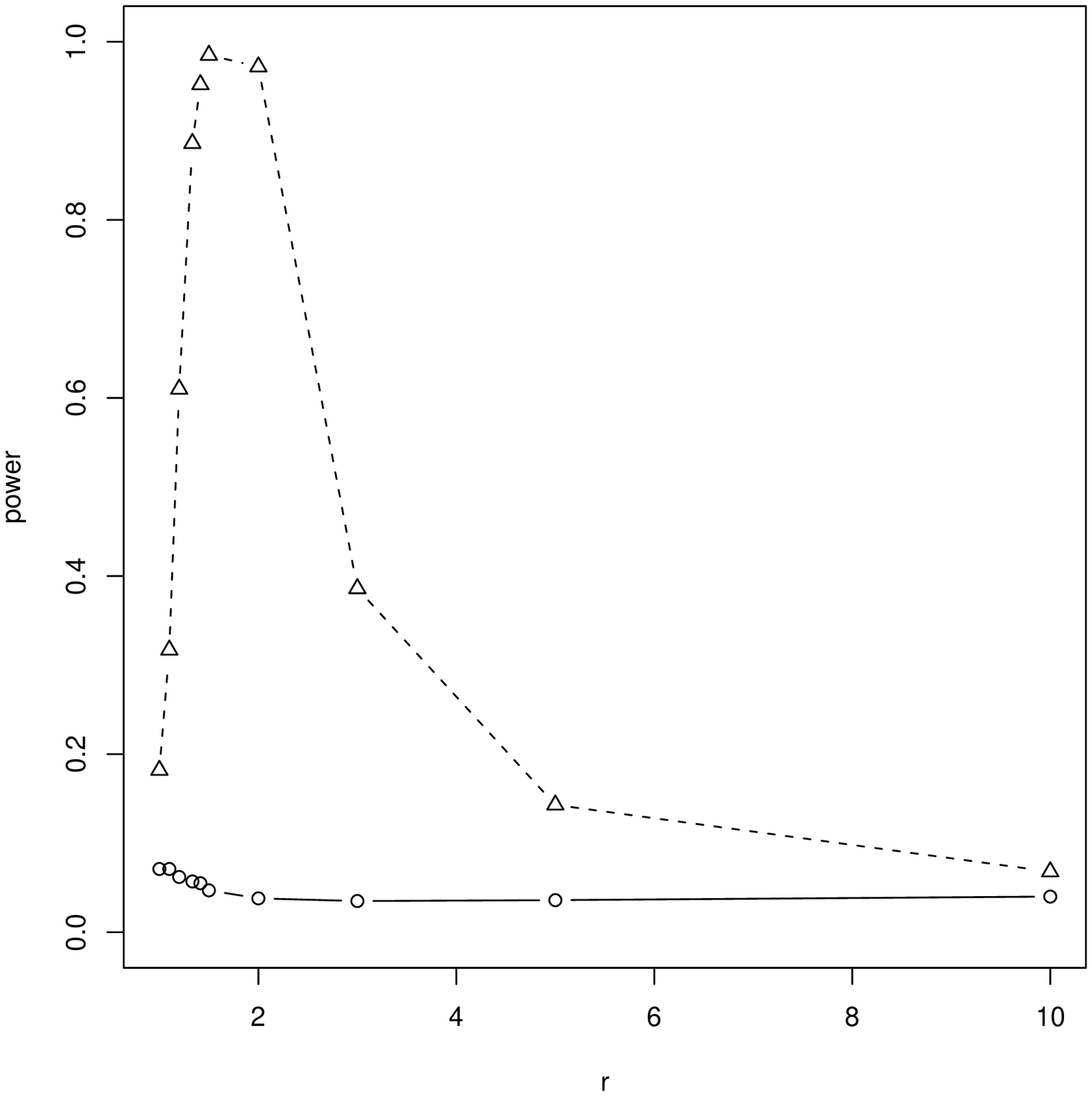}}}
\rotatebox{-90}{ \resizebox{2. in}{!}{ \includegraphics{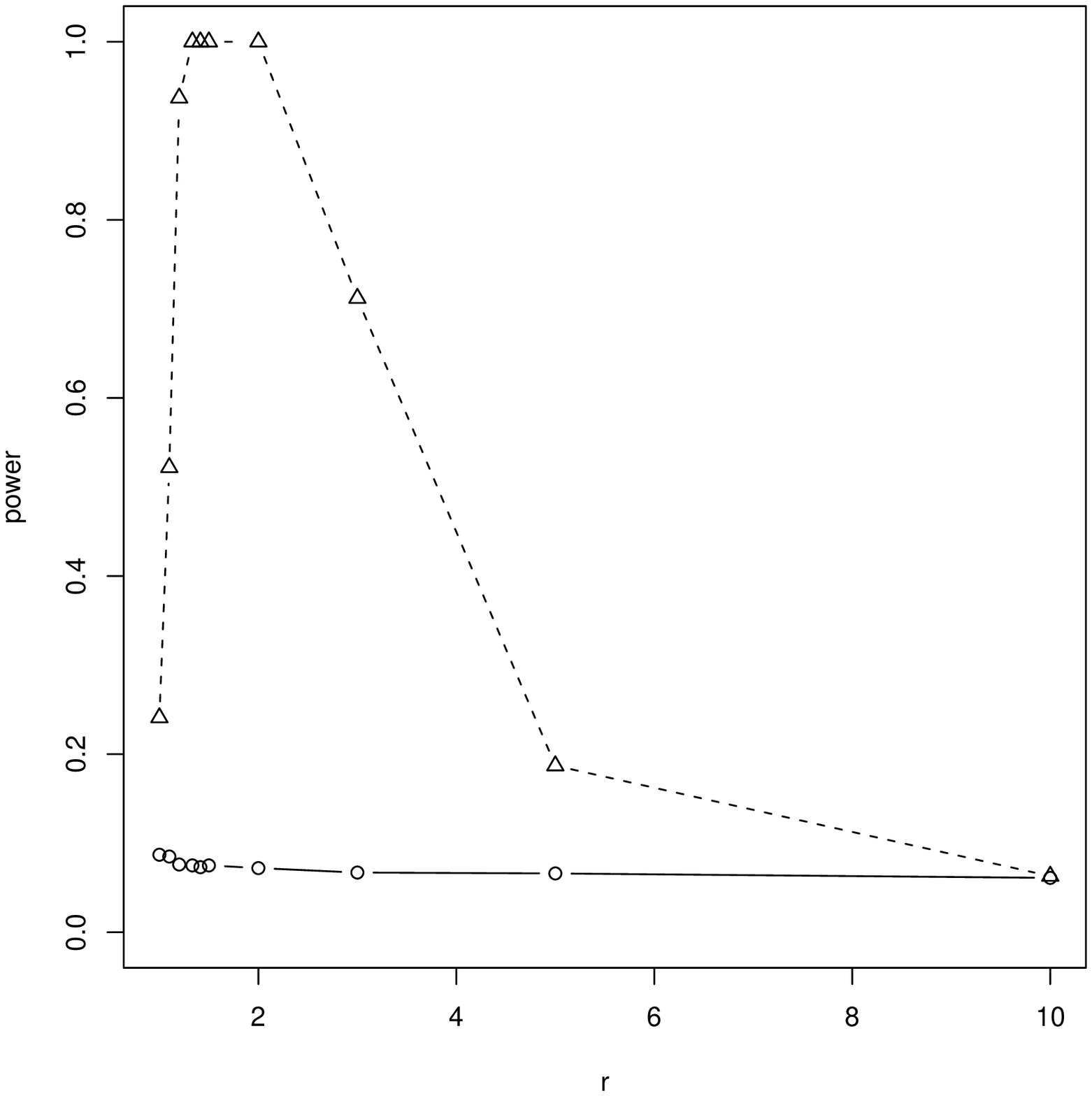}}}
\caption{
 \label{fig:MTAggSim1}
Monte Carlo power using the asymptotic critical value against $H^A_{\sqrt{3}/12}$
as a function of $r$, for $n=100$ (left), $n=200$ (middle), and $n=500$ (right)
conditional on the realization of $\Y$ in Figure \ref{fig:deldata}.
The circles represent the empirical significance levels while
triangles represent the empirical power values.}
\end{figure}

We implement the above described Monte Carlo experiment $1000$ times with $n=100$,
$n=200$, and $n=500$ and find the empirical significance levels $\widehat{\alpha}_S(n,J)$
and $\widehat{\alpha}_A(n,J)$ and the empirical powers $\widehat{\beta}^S_{n}(r,\sqrt{3}/8,J)$
and $\widehat{\beta}^A_{n}(r,\sqrt{3}/12,J)$.
These empirical estimates are presented in Table \ref{tab:MT-asy-emp-val} and plotted
in Figures \ref{fig:MTSegSim1} and \ref{fig:MTAggSim1}.
Notice that the empirical significance levels are all larger than .05 for both alternatives,
so this test is liberal in rejecting $H_0$ against both alternatives for the given
realization of $\Y$ and $n$ values.  The smallest empirical significance levels and
highest empirical power estimates occur at moderate $r$ values ($r=3/2,\,2,\,3$)
against segregation and at smaller $r$ values ($r=\sqrt{2},\,3/2$) against association.
Based on this analysis, for the given realization of $\Y$, we suggest the use of
moderate $r$ values for segregation and slightly smaller for association.
Notice also that as $n$ increases, the empirical power estimates gets larger
for both alternatives.

\begin{table}[t]
\centering
\begin{tabular}{|c|c|c|c|c|c|c|c|c|c|c|}
\hline
$r$  & 1 & 11/10 &6/5 & 4/3 & $\sqrt{2}$ &3/2 & 2 & 3 & 5 & 10\\
\hline
\multicolumn{11}{|c|}{$n=100$, $N=1000$} \\
\hline
$\widehat{\alpha}_S(n,J)$ & 0.144 & 0.141 & 0.124 & 0.101 & 0.095 & 0.087 & 0.070 & 0.075 & 0.071 & 0.072 \\
\hline
$\widehat{\beta}^S_{n}(r,\sqrt{3}/8,J)$ & 0.191 & 0.383 & 0.543 & 0.668 & 0.714 & 0.742 & 0.742 & 0.625 & 0.271 & 0.124 \\
\hline
\hline
$\widehat{\alpha}_A(n,J)$ & 0.118 & 0.111 & 0.089 & 0.081 & 0.065 & 0.062 & 0.067 & 0.064 & 0.068 & 0.071  \\
\hline
$\widehat{\beta}^A_{n}(r,\sqrt{3}/12,J)$ & 0.231 & 0.295 & 0.356 & 0.338 & 0.269 & 0.209 & 0.148 & 0.095 & 0.113 & 0.167 \\
\hline
\multicolumn{11}{|c|}{$n=200$, $N=1000$} \\
\hline
$\widehat{\alpha}_S(n,J)$ & 0.095 & 0.092 & 0.087 & 0.077 & 0.073 & 0.076 & 0.072 & 0.071 & 0.074 & 0.073 \\
\hline
$\widehat{\beta}^S_{n}(r,\sqrt{3}/8,J)$ & 0.135 & 0.479 & 0.743 & 0.886 & 0.927 & 0.944 & 0.959 & 0.884 & 0.335 & 0.105 \\
\hline
\hline
$\widehat{\alpha}_A(n,J)$ & 0.071 & 0.071 & 0.062 & 0.057 & 0.055 & 0.047 & 0.038 & 0.035 & 0.036 & 0.040  \\
\hline
$\widehat{\beta}^A_{n}(r,\sqrt{3}/12,J)$ & 0.182 & 0.317 & 0.610 & 0.886 & 0.952 & 0.985 & 0.972 & 0.386 & 0.143 & 0.068 \\
\hline
\multicolumn{11}{|c|}{$n=500$, $N=1000$} \\
\hline
$\widehat{\alpha}_S(n,J)$ & 0.089 & 0.092 & 0.087 & 0.086 & 0.080 & 0.078 & 0.079 & 0.079 & 0.076 & 0.081 \\
\hline
$\widehat{\beta}^S_{n}(r,\sqrt{3}/8,J)$ & 0.145 & 0.810 & 0.981 & 0.997 & 0.999 & 1.000 & 1.000 & 1.000 & 0.604 & 0.130 \\
\hline
\hline
$\widehat{\alpha}_A(n,J)$ & 0.087 & 0.085 & 0.076 & 0.075 & 0.073 & 0.075 & 0.072 & 0.067 & 0.066 & 0.061  \\
\hline
$\widehat{\beta}^A_{n}(r,\sqrt{3}/12,J)$ & 0.241 & 0.522 & 0.937 & 1.000 & 1.000 & 1.000 & 1.000 & 0.712 & 0.187 & 0.063 \\
\hline
\end{tabular}
\caption{
\label{tab:MT-asy-emp-val}
The empirical significance level and empirical power values under $H^S_{\sqrt{3}/8}$ and
$H^A_{\sqrt{3}/12}$, $N=1000$, $n=100$, and $J=13$, at $\alpha=.05$ for the realization
of $\Y$ in Figure \ref{fig:deldata}.}
\end{table}

The conditional test presented here is appropriate
when the $\mathcal W$ are fixed, not random.
An unconditional version requires the joint distribution
of the number and relative size of Delaunay triangles
when $\Y$ is, for instance, a Poisson point pattern.
Alas, this joint distribution is not available
\cite{okabe:2000}.

\subsection{Related Test Statistics in Multiple Triangle Case}
For $J>1$, we have derived the asymptotic distribution of $\rho_n(r,J)=|\A|/(n\,(n-1))$.
Let $\A_j$ be the number of arcs, $n_j:=|\X_n \cap T_j|$, and $\rho_{n_j}(r)$ be
the arc density for triangle $T_j$ for $j=1,\ldots,J$.
So $\sum_{j=1}^{J}\frac{n_j\,(n_j-1)}{n\,(n-1)} \rho_{n_j}(r)= \rho_n(r,J)$,
since $\sum_{j=1}^{J}\frac{n_j\,(n_j-1)}{n\,(n-1)} \rho_{n_j}(r)
=\frac{\sum_{j=1}^{J} |\A_j|}{n\,(n-1)}=\frac{|\A|}{n\,(n-1)}=\rho_n(r,J)$.

Let $\widehat{U}_n:=\sum_{j=1}^{J}w_j^2\,\rho_{n_j}(r)$ where $w_j=A(T_j)/A(C_H(\Y))$.
Since $\rho_{n_j}(r)$ are asymptotically independent, $\sqrt{n}(\widehat{U}_n-\mu(r,J))$
 and $\sqrt{n}(\rho_n(r,J)-\mu(r,J))$ both converge in distribution to  $\N(0,\nu(r,J))$.

In the denominator of $\rho_n(r,J)$, we use $n(n-1)$ as the maximum number of arcs possible.
However, by definition, we can at most have a digraph with $J$ complete
symmetric components of order $n_j$, for $j=1,\ldots,J$.
Then the maximum number possible is $n_t:=\sum_{j=1}^{J}n_j\,(n_j-1)$.
Then the (adjusted) arc density is $\rho^{adj}_{n,J}:=\frac{|\A|}{n_t}$.
Then $\rho^{adj}_{n,J}(r)=\frac{\sum_{j=1}^{J} |\A_j|}{n_t}=\sum_{j=1}^{J}\frac{n_j\,(n_j-1)}{n_t}\,\rho_{n_j}(r)$.
Since $\frac{n_j\,(n_j-1)}{n_t} \ge 0$ for each $j$,
and $\sum_{j=1}^{J}\frac{n_j\,(n_j-1)}{n_t}=1$, $\rho^{adj}_{n,J}(r)$ is
a mixture of $\rho_{n_j}(r)$'s. Then $\rho^{adj}_{n,J}(r)$ is asymptotically
normal with mean $\E[\rho^{adj}_{n,J}(r)]=\mu(r,J)$ and the variance of $\rho^{adj}_{n,J}(r)$ is
$$\frac{1}{n} \left[\nu(r) \Biggl(\sum_{j=1}^{J}w_j^3/(\sum_{j=1}^{J}w_j^2)^2 \Biggr)+
4\mu(r)^2\Biggl(\sum_{j=1}^{J}w_j^3/(\sum_{j=1}^{J}w_j^2)^2-1\Biggr) \right].$$

\subsection{Asymptotic Efficacy Analysis for $J>1$}
The PAE, HLAE, and asymptotic power function analysis are given for $J=1$ in
Sections \ref{sec:Pitman}, \ref{sec:Hodges-Lehmann}, and \ref{sec:Asy-Power},
respectively.
For $J>1$, the analysis will depend on both the number of triangles as well
as the size of the triangles.  So the optimal $r$ values with respect to
these efficiency criteria for $J=1$ do not necessarily hold for $J>1$,
hence the analyses need to be updated,
given the values of $J$ and $\mathcal W$.

Under segregation alternative $H^S_{\epsilon}$, the PAE of $\rho_n(r,J)$ is given by
$$
\PAE_J^S(r) = \frac{\bigl( \mu_S^{\prime\prime}(r,J,\epsilon=0)\bigr)^2}{\nu(r,J)}=
   \frac{\left( \mu_S^{\prime\prime}(r,\epsilon=0)\,\sum_{j=1}^{J}w_j^2 \right)^2}{\nu(r) \,\sum_{j=1}^{J}w_j^3 +4\mu_S(r,\epsilon=0)^2\Bigl(\sum_{j=1}^{J}w_j^3-\bigl(\sum_{j=1}^{J}w_j^2 \bigr)^2\Bigr)}.
$$
Under association alternative $H^A_{\epsilon}$ the PAE of $\rho_n(r,J)$ is similar.
\begin{figure}[]
\centering
\psfrag{r}{\scriptsize{$r$}}
\epsfig{figure=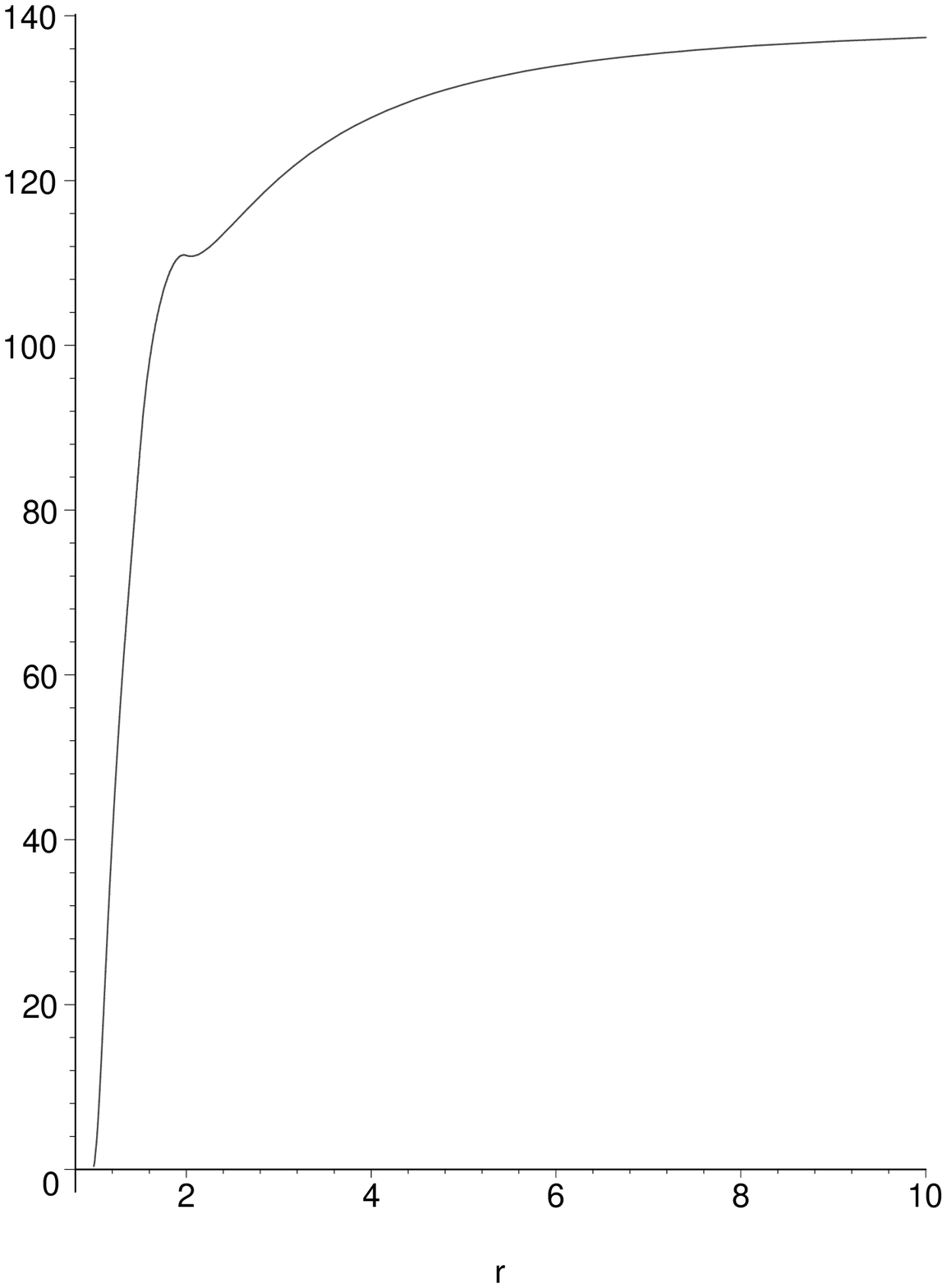, height=150pt, width=150pt}
\epsfig{figure=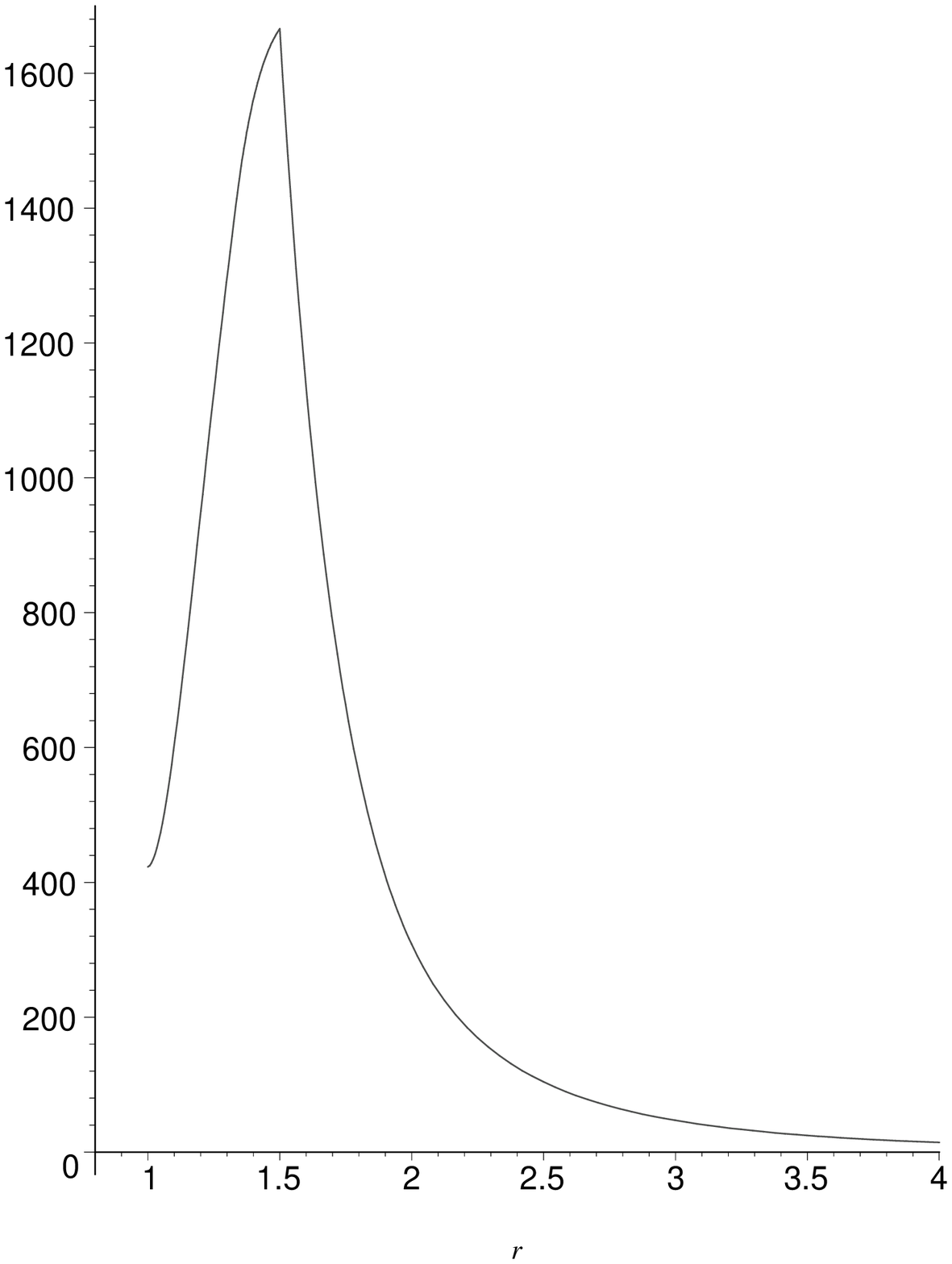, height=150pt, width=150pt}
\caption{\label{fig:MT-PAECurves}
Pitman asymptotic efficiency against segregation (left) and association (right)
as a function of $r$ with $J=13$.
Notice that vertical axes are differently scaled.}
\end{figure}
In Figure \ref{fig:MT-PAECurves}, we present the PAE as a function of $r$ for
both segregation and association conditional on the realization of $\Y$ in Figure \ref{fig:deldata}.
Notice that, unlike $J=1$ case, $\PAE_J^S(r)$ is bounded. Some values of note are
$\PAE_J^S(\rho_n(1)) = .3884$, $\lim_{r \rightarrow \infty} \PAE_J^S(r) =
\frac{8\,\sum_{j=1}^{J}w_j^2}{256\,\Bigl(\sum_{j=1}^{J}w_j^3-
\bigl(\sum_{j=1}^{J}w_j^2 \bigr)^2\Bigr)} \approx 139.34$, $\argsup_{r \in [1,2]} \PAE_J^S(r) \approx 1.974$. As for association, $\PAE_J^A(r=1) = 422.9551$, $ \lim_{r \rightarrow \infty} \PAE_J^A(r) = 0$,
$\argsup_{r \ge 1} \PAE_J^A(r) = 1.5$ with $\PAE_J^A(r=1.5) \approx 1855.9672$.
Based on the asymptotic efficiency analysis, we suggest,
for large $n$ and small $\epsilon$,
choosing moderate $r$ for testing against segregation and association.

Under segregation, the HLAE of $\rho_n(r,J)$ is given by
$$
\HLAE_J^S(r,\epsilon):=\frac{(\mu_S(r,J,\epsilon)-\mu(r,J))^2}{\nu_S(r,J,\epsilon)}=
\frac{\Bigl(\mu_S(r,\epsilon)\,\bigl(\sum_{j=1}^{J}w_j^2 \bigr)-
\mu(r)\,\bigl(\sum_{j=1}^{J}w_j^2 \bigr)\Bigr)^2}{\nu_S(r,\epsilon) \,
\sum_{j=1}^{J}w_j^3 +4\mu_S(r,\epsilon)^2\Bigl(\sum_{j=1}^{J}w_j^3-
\bigl(\sum_{j=1}^{J}w_j^2 \bigr)^2\Bigr)}.
$$
 Notice that $\HLAE_J^S(r,\epsilon=0)=0$ and $\lim_{\rightarrow \infty}
\HLAE_J^S(r,\epsilon)=0$ and HLAE is bounded provided that $\nu(r,J)>0$.

\begin{figure}[]
\centering
\psfrag{r}{\scriptsize{$r$}}
\epsfig{figure=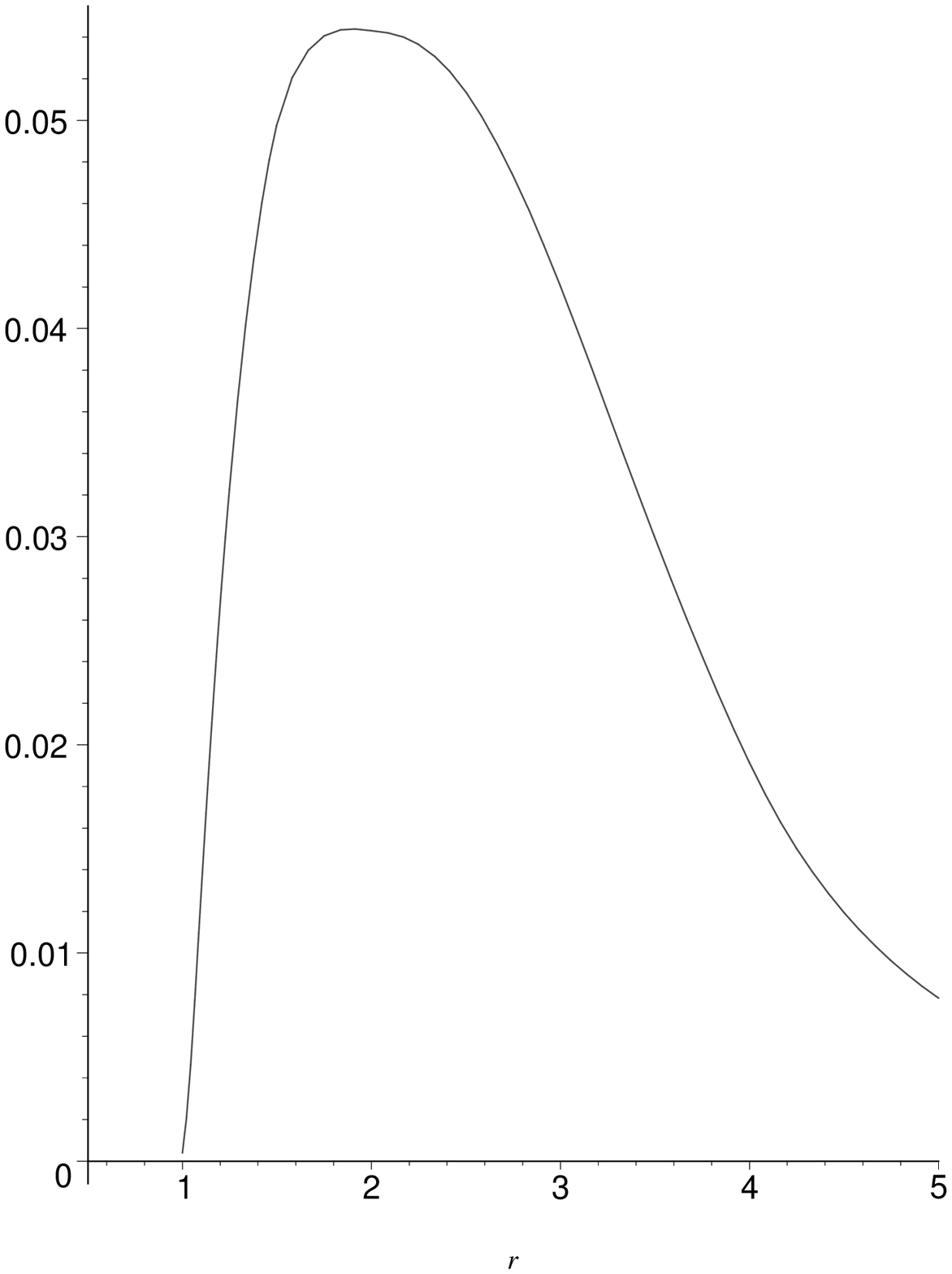, height=125pt, width=125pt}
\epsfig{figure=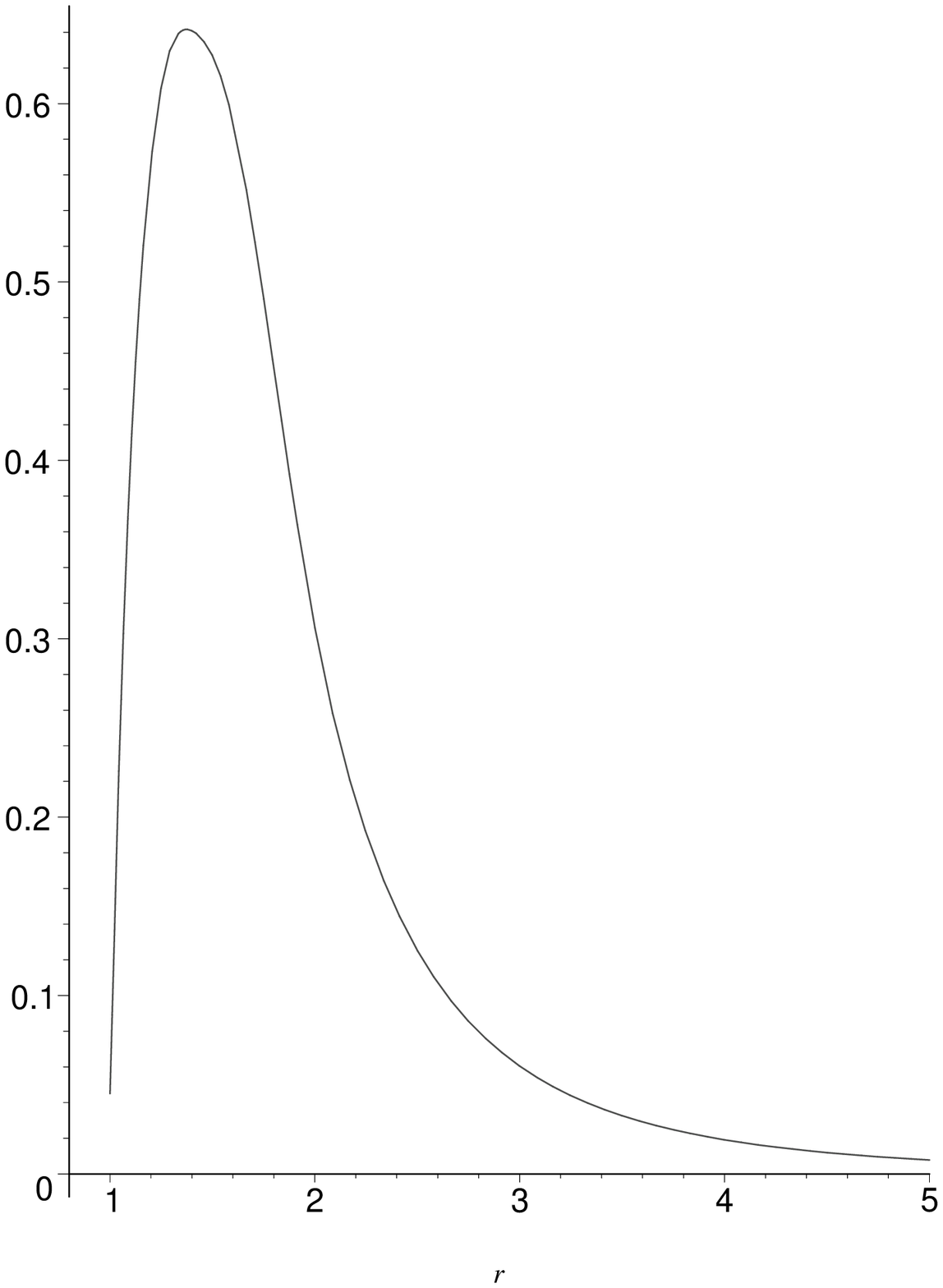, height=125pt, width=125pt}
\epsfig{figure=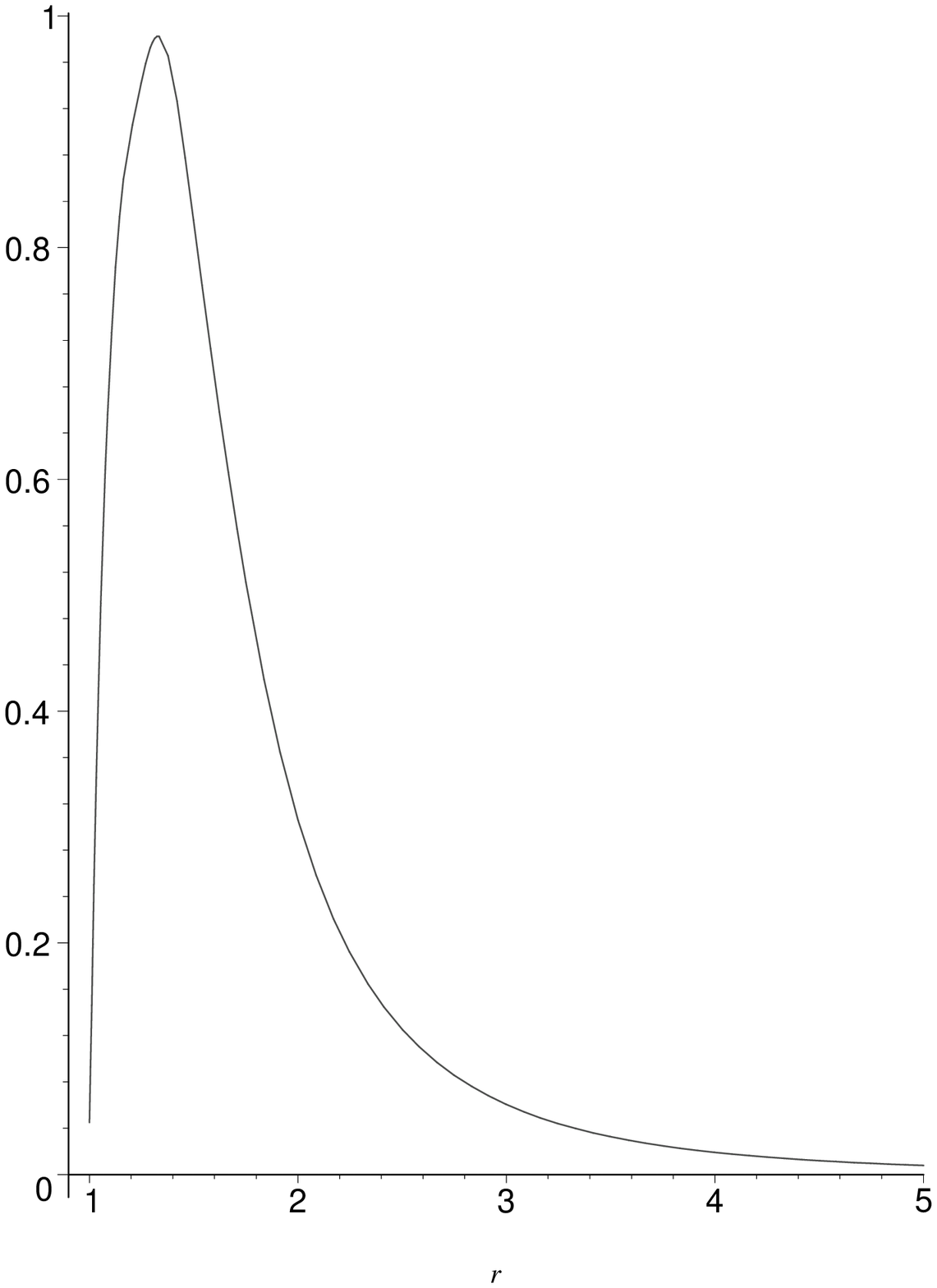, height=125pt, width=125pt}
\caption{
\label{fig:MT-HLAE-Seg}
Hodges-Lehmann asymptotic efficiency
against segregation alternative $H^S_{\epsilon}$
as a function of $r$
for $\epsilon= \sqrt{3}/8, \sqrt{3}/4, 2\,\sqrt{3}/7$ (left to right) and $J=13$.}
\end{figure}

We calculate HLAE of $\rho_n(r,J)$ under $H^S_{\epsilon}$ for $\epsilon=\sqrt{3}/8$,
$\epsilon=\sqrt{3}/4$, and $\epsilon=2\,\sqrt{3}/7$. In Figure \ref{fig:MT-HLAE-Seg}
we present $\HLAE_J^S(r,\epsilon)$ for these $\epsilon$ values conditional on
the realization of $\Y$ in Figure \ref{fig:deldata}.
Note that with $\epsilon=\sqrt{3}/8$, $\HLAE_J^S(r=1,\sqrt{3}/8) \approx .0004$
and $\argsup_{r \in [1,\infty]} \HLAE_J^S(r,\sqrt{3}/8) \approx 1.8928$ with the
supremum $\approx .0544$.
With $\epsilon=\sqrt{3}/4$,
$\HLAE_J^S(r=1,\sqrt{3}/4) \approx .0450$ and
$\argsup_{r \in [1,\infty]} \HLAE_J^S(r,\sqrt{3}/4) \approx 1.3746$
with the supremum $\approx .6416$.
With $\epsilon=2\,\sqrt{3}/7$, $\HLAE_J^S(r=1,2\,\sqrt{3}/7) \approx .045$
and $\argsup_{r \in [1,\infty]} \HLAE_J^S(r,2\,\sqrt{3}/7) \approx 1.3288$
with the supremum $ \approx .9844$.
Furthermore, we observe that $\HLAE_J^S(r,2\,\sqrt{3}/7)>\HLAE_J^S(r,\sqrt{3}/4)>\HLAE_J^S(r,\sqrt{3}/8)$.
Based on the HLAE analysis for the given $\Y$ we suggest moderate $r$
values for moderate segregation and small $r$ values for severe segregation.

The explicit form of $\HLAE_J^A(r,\epsilon)$ is similar to $\HLAE_J^S(r,\epsilon)$
which implies $\HLAE_J^A(r,\epsilon=0)=0$ and $\lim_{\rightarrow \infty}\HLAE_J^A(r,\epsilon)=0$.

\begin{figure}[ht]
\centering
\psfrag{r}{\scriptsize{$r$}}
\epsfig{figure=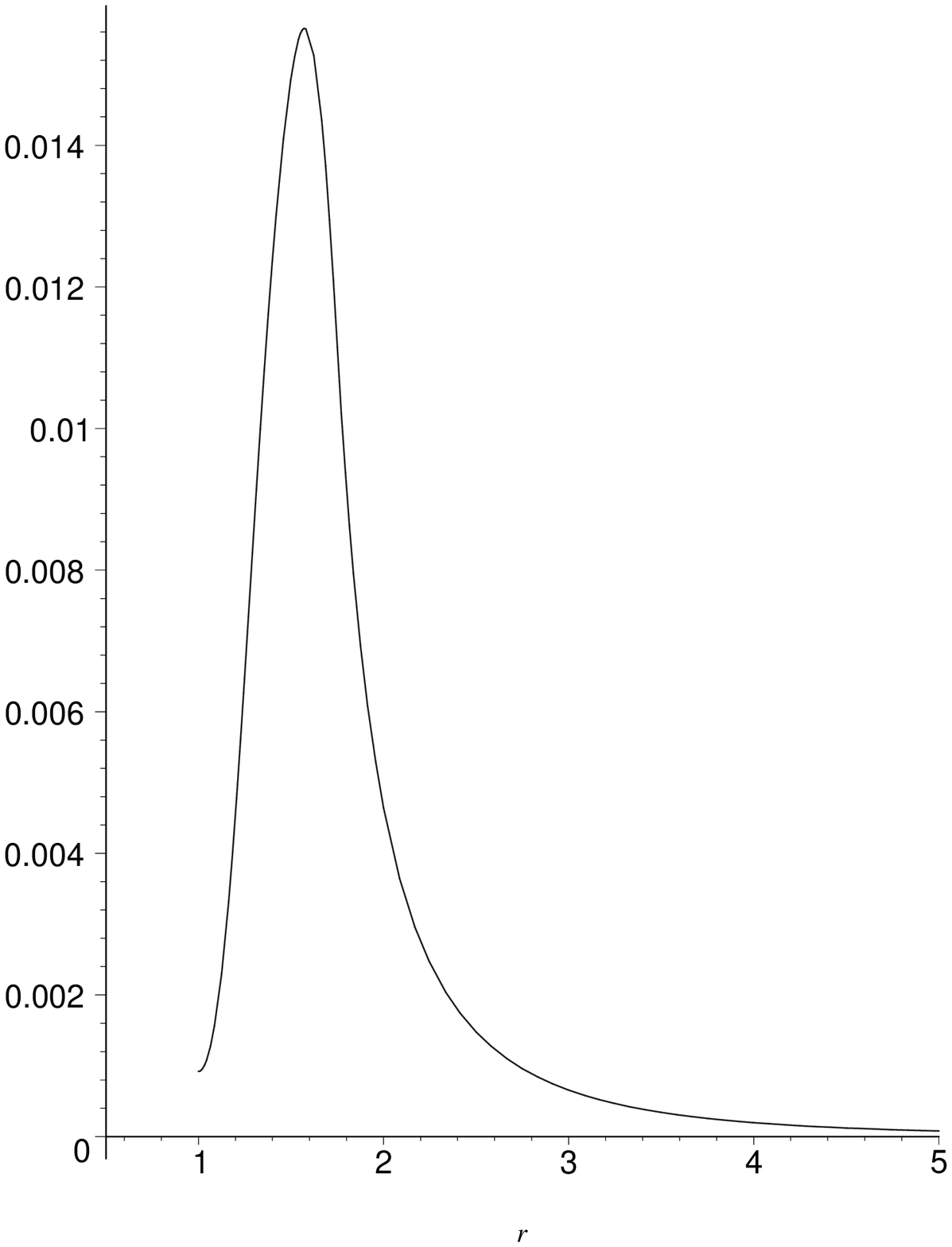, height=125pt, width=125pt}
\epsfig{figure=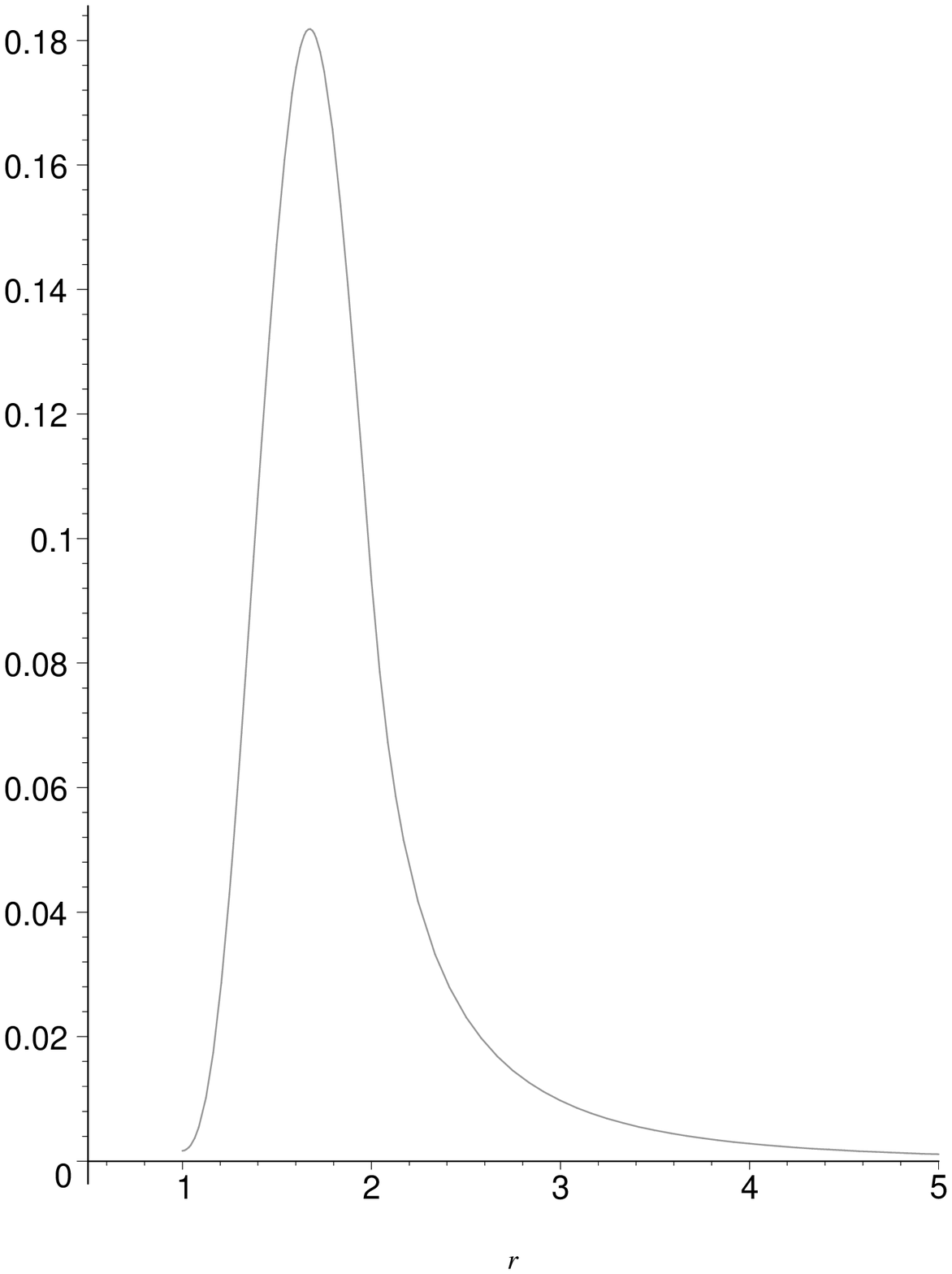, height=125pt, width=125pt}
\epsfig{figure=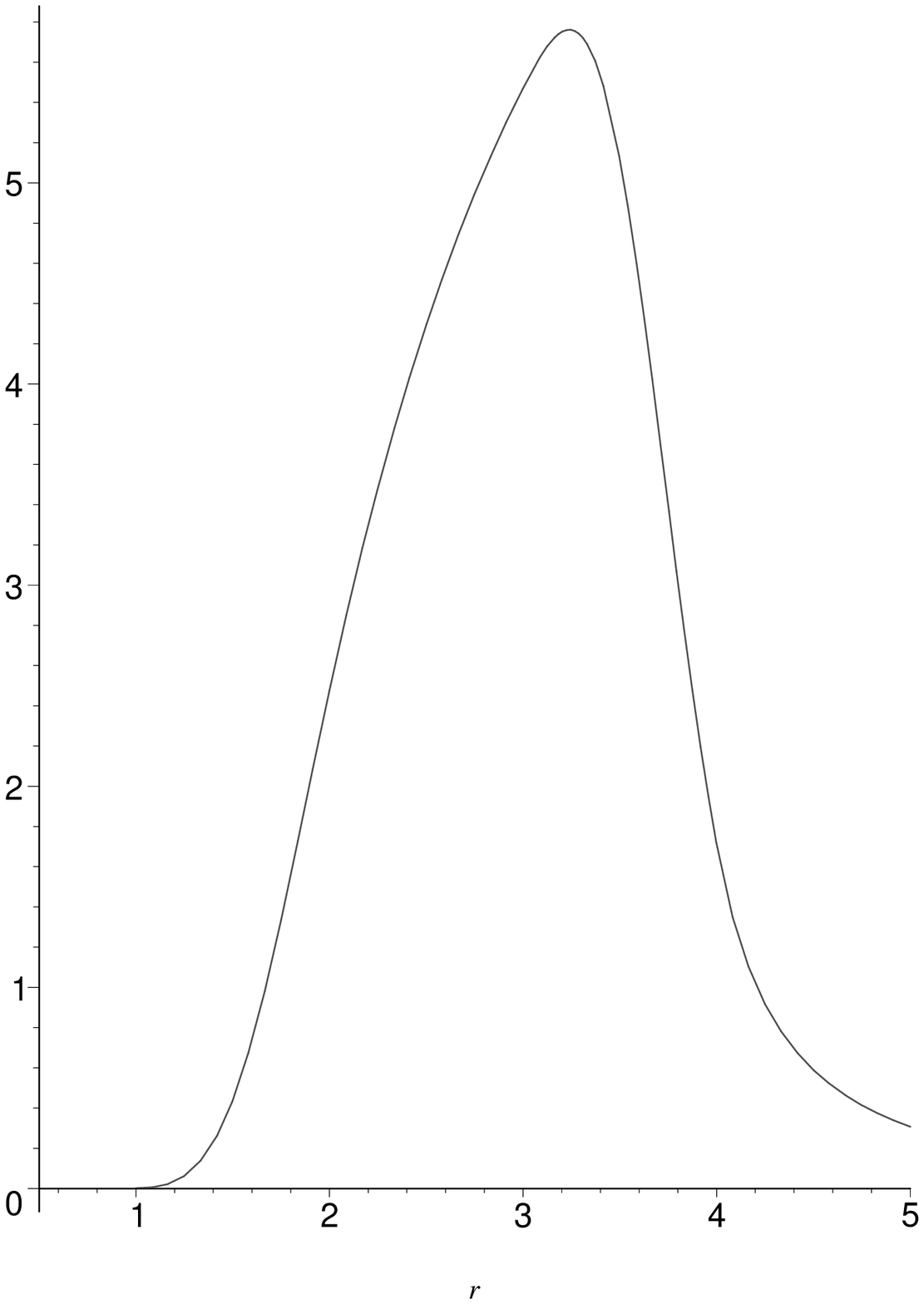, height=125pt, width=125pt}
\caption{
\label{fig:MT-HLAE-Agg}
Hodges-Lehmann asymptotic efficiency
against association alternative $H^A_{\epsilon}$
as a function of $r$
for $\epsilon= \sqrt{3}/21, \sqrt{3}/12, 5\,\sqrt{3}/24$ (left to right) and $J=13$.}
\end{figure}

We calculate HLAE of $\rho_n(r,J)$ under $H^A_{\epsilon}$ for $\epsilon=\sqrt{3}/21$,
$\epsilon=\sqrt{3}/12$, and $\epsilon=5\,\sqrt{3}/24$. In Figure \ref{fig:MT-HLAE-Agg}
we present $\HLAE_J^S(r,\epsilon)$ for these $\epsilon$ values conditional on the
realization of $\Y$ in Figure \ref{fig:deldata}
Note that with $\epsilon=\sqrt{3}/21$, $\HLAE_J^A(r=1,\sqrt{3}/21) \approx.0009$ and
$\argsup_{r \in [1,\infty]} \HLAE_J^A(r,\sqrt{3}/21) \approx 1.5734$ with the
supremum $\approx .0157$.  With $\epsilon=\sqrt{3}/12$, $\HLAE_J^A(r=1,\sqrt{3}/12) \approx .0168$
and $\argsup_{r \in [1,\infty]} \HLAE_J^A(r,\sqrt{3}/12) \approx 1.6732$ with the supremum
$\approx .1818$. With $\epsilon=5\,\sqrt{3}/24$, $\HLAE_J^A(r=1,5\,\sqrt{3}/24) \approx .0017$ and \\
$\argsup_{r \in [1,\infty]} \HLAE_J^A(r,5\,\sqrt{3}/24) \approx 3.2396$ with the supremum $\approx 5.7616$.
Furthermore, we observe that $\HLAE_J^A(r,5\,\sqrt{3}/24)>\HLAE_J^A(r,\sqrt{3}/12)>\HLAE_J^A(r,\sqrt{3}/21)$.
Based on the HLAE analysis for the given $\Y$ we suggest
moderate $r$ values for moderate association and large $r$ values for severe association.

\section{Discussion and Conclusions}
\label{sec:discussion}
In this article we investigate the mathematical properties
of a random digraph method for the analysis of spatial point patterns.

The first proximity map similar to the $r$-factor proximity map $\NY^r$ in
literature is the spherical proximity map $N_S(x):=B(x,r(x))$,
(see the references for CCCD in the Introduction).
A slight variation of $N_S$ is the arc-slice proximity map $N_{AS}(x):=B(x,r(x)) \cap T(x)$
where $T(x)$ is the Delaunay cell that contains $x$ (see \cite{ceyhan:CS-JSM-2003}).
Furthermore, Ceyhan and Priebe introduced the central similarity proximity map
$N_{CS}$ in \cite{ceyhan:CS-JSM-2003} and $\NY^r$ in \cite{ceyhan:TR-dom-num-NPE-spatial}.
The $r$-factor proximity map, when compared to the others,
has the advantages that the asymptotic distribution of the domination number
$\gamma_n(\NY^r)$ is tractable (see \cite{ceyhan:TR-dom-num-NPE-spatial}),
an exact minimum dominating set can be found in polynomial time.
Moreover $\NY^r$ and $N_{CS}$ are geometry invariant for uniform data over triangles.
Additionally, the mean and variance of relative density $\rho_n$ is not
analytically tractable for $N_S$ and $N_{AS}$.
While $\NY^r(x)$, $N_{CS}(x)$, and $N_{AS}(x)$ are well defined only for $x \in C_H(\Y)$,
the convex hull of $\Y$, $N_S(x)$ is well defined for all $x \in \R^d$.
The proximity maps $N_S$ and $N_{AS}$ require no effort to extend to higher dimensions.

The $N_S$ (the proximity map associated with CCCD) is used in classification in the literature,
but not for testing spatial patterns between two or more classes.
We develop a technique to test the patterns of segregation or association.
There are many tests available for segregation and association in ecology literature.
See \cite{dixon:1994} for a survey on these tests and relevant references.
Two of the most commonly used tests are Pielou's $\chi^2$ test of independence
and Ripley's test based on $K(t)$ and $L(t)$ functions.
However, the test we introduce here is not comparable to either of them.
Our test is a conditional test --- conditional on a realization of $J$
(number of Delaunay triangles) and $\mathcal W$ (the set of relative areas of the Delaunay triangles)
and we require the number of triangles $J$ is fixed and relatively
small compared to $n=|\X_n|$.
Furthermore, our method deals with a slightly different type of data than most methods to
examine spatial patterns.
The sample size for one
type of point (type $\X$ points) is much larger compared to the the other (type $\Y$ points).
This implies that in practice,
$\Y$ could be stationary or have much longer life span than members of $\X$.
For example, a special type of fungi might constitute $\X$ points, while the
tree species around which the fungi grow might be viewed as the $\Y$ points.

There are two major types of asymptotic structures for spatial data \cite{lahiri:1996}.
In the first, any two observations are required to be at least a fixed distance apart,
hence as the number of observations increase, the region on which the process
is observed eventually becomes unbounded.
This type of sampling structure is called ``increasing domain asymptotics".
In the second type, the region of interest is a fixed
bounded region and more or more points are observed in this region.
Hence the minimum distance between data points tends to zero
as the sample size tends to infinity.
This type of structure is called ``infill asymptotics", due to Cressie \cite{cressie:1991}.
The sampling structure for our asymptotic analysis is infill,
as only the size of the type $X$ process tends to infinity,
while the support, the convex hull of a given set of points from
type $Y$ process, $C_H(\Y)$ is a fixed bounded region.

Moreover, our statistic that can be written as a
$U$-statistic based on the locations of type $X$ points
with respect to type $Y$ points.
This is one advantage of the proposed method: most statistics for spatial patterns
can not be written as $U$-statistics.
The $U$-statistic form avails us the asymptotic normality,
once the mean and variance is obtained by tedious detailed geometric calculations.

The null hypothesis we consider is
considerably more restrictive than current approaches,
which can be used much more generally.
The null hypothesis for testing segregation or association can be described in two slightly
different forms \cite{dixon:1994}:
\begin{itemize}
\item[(i)]complete spatial randomness,
that is, each class is distributed randomly throughout the area of interest.
It describes both the arrangement of the locations
and the association between classes.
\item[(ii)] random labeling of locations, which is less restrictive than spatial randomness,
in the sense that arrangement of the locations can either be random or non-random.
\end{itemize}
Our conditional test is closer to the former in this regard.
Pielou's test provide insight only on the association between classes,
hence there is no assumption on the allocation of the observations, which
makes it more appropriate for testing the null hypothesis of random labeling.
Ripley's test can be used for both types of null hypotheses,
in particular,
it can be used to test a type of spatial randomness against another type of spatial randomness.

The test based on the mean domination number in \cite{ceyhan:TR-dom-num-NPE-spatial}
is not a conditional test,
but requires both $n$ and number of Delaunay triangles $J$ to be large.
The comparison for a large but fixed $J$ is possible.
Furthermore, under segregation alternatives, the Pitman asymptotic efficiency is
not applicable to the mean domination number case, however,
for large $n$ and $J$ we suggest the use of it over arc density since for each $\epsilon>0$,
Hodges-Lehmann asymptotic efficiency is unbounded for the mean domination number case,
while it is bounded for arc density case with $J>1$.
As for the association alternative, HLAE suggests moderate $r$ values which
has finite Hodges-Lehmann asymptotic efficiency.
So again, for large $J$ and $n$ mean domination number is preferable.
The basic advantage of $\rho_n(r)$ is that, it does not require $J$ to be large,
so for small $J$ it is preferable.

Although the statistical analysis and the mathematical properties
related to the $r$-factor proximity catch digraph are done in $\R^2$,
the extension to $\R^d$ with $d > 2$ is straightforward.
See Ceyhan and Priebe \cite{ceyhan:TR-dom-num-NPE-spatial} for more detail on the construction
of the associated proximity region in higher dimensions.
Moreover, the geometry invariance, asymptotic normality of the $U$-statistic
and consistency of the tests hold for $d>2$.

%

\section*{Appendix 1: Derivation of $\mu(r)$ and $\nu(r)$}
In the standard equilateral triangle, let $\y_1=(0,0)$, $\y_2=(1,0)$, $\y_3=\bigl( 1/2,\sqrt{3}/2 \bigr)$,
$M_C$ be the center of mass, $M_j$ be the midpoints of the edges $e_j$ for $j=1,2,3$.
Then $M_C=\bigl(1/2,\sqrt{3}/6\bigr)$, $M_1=\bigl(3/4,\sqrt{3}/4 \bigr)$,
$M_2=\bigl(1/4,\sqrt{3}/4\bigr)$, $M_3=(1/2,0)$.

Recall that $\E[\rho_n(r)]=\frac{1}{n\,(n-1)}\sum \sum_{i < j } \,\E[h_{ij}]=
\frac{1}{2}\E[h_{12}]=\mu(r)=P\bigl(X_j \in \NY^r(X_i)\bigr)$.

Let $\X_n$ be a random sample of size $n$ from $\U(T(\Y))$.
For $x_1=(u,v)$, $\ell_r(x_1)=r\,v+r\,\sqrt{3}\,u-\sqrt{3}\,x.$
Next, let $N_1:=\ell_r(x_1)\cap e_3$ and $N_2:=\ell_r(x_1)\cap e_2$.
Then for $z_1 \in T_s:=T(\y_1,M_3,M_C)$, $N_{\Y}^r(z_1)=T(\y_1,N_1,N_2)$
provided that $\ell_r(x_1)$ is not outside of $T(\Y)$, where
 $$N_1=\bigl(r\,\bigl(y_1+\sqrt{3}\,x_1\bigr)\sqrt{3}/3,0\bigr)\text{ and }
N_2=\bigl(r\,\bigl(y_1+\sqrt{3}\,x_1\bigr)\sqrt{3}/6,\bigl(y_1+\sqrt{3}\,x_1\bigr) r/2\bigr).$$
Now we find $\mu(r)$ for $r \in [1,\infty)$.

First, observe that, by symmetry,
 $$\mu(r)=P\bigl(X_2 \in \NY^r(X_1)\bigr)=6\,P\bigl(X_2 \in \NY^r(X_1), X_1 \in T_s\bigr).$$
 Let $\ell_s(r,x)$ be the line such that $r\,d(\y_1,\ell_s(r,x))=d(\y_1,e_1)$ and
$\ell_s(r,x)\cap T(\Y)\not=\emptyset$, so $\ell_s(r,x)=\sqrt{3}(\frac{1}{r}-x)$.
Then if $x_1 \in T_s$ is above $\ell_s(r,x)$ then $\NY^r(x_1)=T(\Y)$,
otherwise, $\NY^r(x_1)=T_r(x_1)\subsetneq T(\Y)$.

For $r \in [1,3/2)$, $\ell_s(r,x)\cap T_s=\emptyset$, so $\NY^r(x_1)=
T_r(x_1)\subsetneq T(\Y)$ for all $x \in T_s$.  Then
$$
P(X_2 \in \NY^r(X_1), X_1 \in T_s)=\int_0^{1/2}\int_0^{x/\sqrt{3}}
\frac{A(\NY^r(x_1))}{A(T(\Y))^2}dydx = \frac{37}{1296}\,r^2.
$$
where $A(\NY^r(x_1))=\frac{\sqrt{3}}{12}\,r^2(y+\sqrt{3}\,x)^2 $ and $A(T(\Y))=\sqrt{3}/4$.
Hence for $r \in [1,3/2)$, $\mu(r)=\frac{37}{216}\,r^2$.

\begin{figure} [ht]
    \centering
   \scalebox{.4}{\input{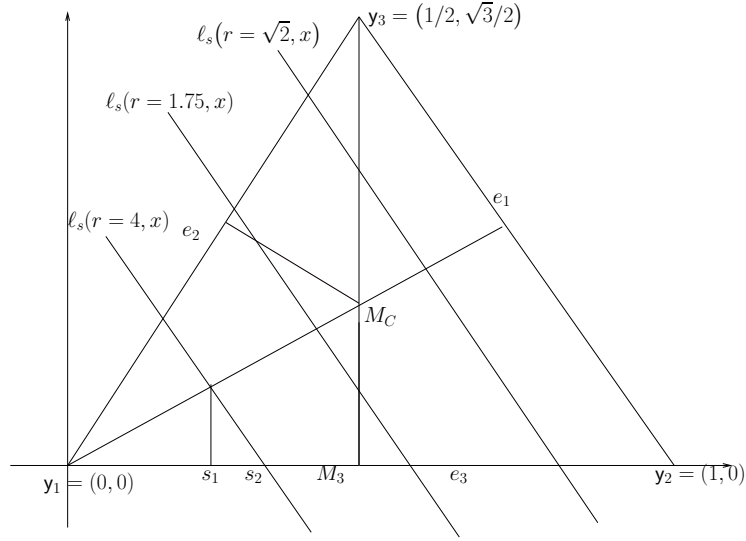}}
    \caption{The cases for relative position of $\ell_s(r,x)$ with various $r$ values.}
    \label{fig:ls-lam-cases}
\end{figure}

For $r \in [3/2,2)$, $\ell_s(r,x)$ crosses through $\overline{M_3M}_C$.
Let the $x$ coordinate of $\ell_s(r,x)\cap \overline{\y_1M}_C$ be $s_1$, then $s_1=3/(4\,r)$.
See Figure \ref{fig:ls-lam-cases} for the relative position of $\ell_s(r,x)$ and $T_s$.

 Then
\begin{eqnarray*}
\lefteqn{P(X_2 \in \NY^r(X_1), X_1 \in T_s)=
\int_0^{1/2}\int_0^{x/\sqrt{3}} \frac{A(\NY^r(x_1))}{A(T(\Y))^2}dydx}\\
& = &\int_0^{s_1}\int_0^{x/\sqrt{3}} \frac{A(\NY^r(x_1))}{A(T(\Y))^2}dydx+
\int_{s_1}^{1/2}\int_0^{\ell_s(r,x)} \frac{A(\NY^r(x_1))}{A(T(\Y))^2}dydx+
\int_{s_1}^{1/2}\int_{\ell_s(r,x)}^{x/\sqrt{3}} \frac{1}{A(T(\Y))}dydx\\
& = & -\frac{-36+r^4+64\,r-32\,r^2}{48\,r^2}.
\end{eqnarray*}
Hence for $r \in [3/2,2)$, $\mu(r)=-\frac{1}{8}\,r^2-8\,r^{-1}+\frac{9}{2}\,r^{-2}+4$.

For $r \in [2,\infty)$, $\ell_s(r,x)$ crosses through $\overline{\y_1M}_3$.
Let the $x$ coordinate of $\ell_s(r,x) \cap \overline{\y_1M}_3$ be $s_2$,
then $s_2=1/r$. See Figure \ref{fig:ls-lam-cases}.

Then
\begin{multline*}
P(X_2 \in \NY^r(X_1), X_1 \in T_s)=\int_0^{s_1}\int_0^{x/\sqrt{3}}
\frac{A(\NY^r(x_1))}{A(T(\Y))^2}dydx+\int_{s_1}^{s_2}\int_0^{\ell_s(r,x)}
\frac{A(\NY^r(x_1))}{A(T(\Y))^2}dydx\\
+\int_{s_1}^{s_2}\int_{\ell_s(r,x)}^{x/\sqrt{3}} \frac{1}{A(T(\Y))}dydx+
\int_{s_2}^{1/2}\int_{0}^{x/\sqrt{3}} \frac{1}{A(T(\Y))}dydx=\frac{1}{12}\,{\frac{-3+2\,r^2}{r^2}}.
\end{multline*}
Hence for $r \in [2,\infty)$, $\mu(r)=1-\frac{3}{2}\,r^{-2}$.

For $r=\infty $, $\mu(r)=1$ follows trivially.

To find $\Cov[h_{12},h_{13}] $, we introduce a related concept.

{\bf Definition:}
Let $(\Omega,\mathcal{M})$ be a measurable space and
consider the proximity map $N:\Omega \times \wp(\Omega) \rightarrow \wp(\Omega)$,
where $\wp(\cdot)$ represents the power set functional.
For $B \subset \Omega$, the {\em $\G_1$-region},
$\G_1(\cdot)=\G_1(\cdot,N):\Omega \rightarrow \wp(\Omega)$
associates the region $\G_1(B):=\{z \in \Omega: B \subseteq  N(z)\}$
with each set $B \subset \Omega$.
For $x \in \Omega$, we denote $\G_1(\{x\})$ as $\G_1(x)$.
Note that $\G_1$-region depends on proximity region $N(\cdot)$.

Furthermore, let $\G_1(\cdot,\NY^r)$ be the $\G_1$-region associated with
$\NY^r(\cdot)$, let $A_{ij}$ be the event that $\{X_iX_j \in \A\} =\{X_i \in \NY^r(X_j)\}$,
then $h_{ij}=I(A_{ij})+I(A_{ji})$. Let
$$
P^r_{2N}:=P(\{X_2,X_3\} \subset \NY^r(X_1)),\;\;
P^r_M:=P(X_2 \in \NY^r(X_1), X_3 \in \G_1(X_1,\NY^r),\;\;
P^r_{2G}:=P(\{X_2,X_3\} \subset \G_1(X_1,\NY^r)).
$$
 Then
$\Cov[h_{12},h_{13}]=\E[h_{12}\,h_{13}]-\E[h_{12}]\E[h_{13}]$ where
\begin{eqnarray*}
\E[h_{12}\,h_{13}] & = & \E[(\I(A_{12})+\I(A_{21}))\,(\I(A_{13})+\I(A_{31})] \\
                  & = & P(A_{12} \cap A_{13})+P(A_{12} \cap A_{31})+P(A_{21} \cap A_{13})+P(A_{21}\cap A_{31}). \\
                  &=&P(\{X_2,X_3\} \subset \NY^r(X_1))+2\,P(X_2 \in \NY^r(X_1), X_3 \in \G_1(X_1,\NY^r))+P(\{X_2,X_3\} \subset \G_1(X_1,\NY^r))\\
 &=&P^r_{2N}+2\,P^r_M+P^r_{2G}.
\end{eqnarray*}
So $\nu(r)=\Cov[h_{12},h_{13}]  = \left(P^r_{2N}+2\,P^r_M+P^r_{2G}\right)-[2\,\mu(r)]^2.$

Furthermore, for any $x_1=(u,v) \in T(\Y)$, $\G_1(x_1,\NY^r)$ is a convex or nonconvex polygon.
Let $\xi_j(r,x)$ be the line  between $x_1$ and the vertex $\y_j$ parallel to the edge $e_j$
such that $r\,d(\y_j,\xi_j(r,x))=d(\y_j,\ell_r(x_1)) \text{ for } j=1,2,3.$
Then   $\G_1(x_1,\NY^r)\cap R(\y_j)$ is bounded by $\xi_j(r,x)$ and the median lines.

For $x_1=(u,v)$, $\xi_1(r,x)=-\sqrt{3}\,x+(v+\sqrt{3}\,u)/r,\; \xi_2(r,x)=
(v+\sqrt{3}r\,(x-1)+\sqrt{3}(1-u))/r \text{ and } \xi_3(r,x)=(\sqrt{3}(r-1)+2\,v)/(2\,r).$

 To find the covariance, we need to find the possible types of $\G_1(x_1,\NY^r)$ and
$\NY^r(x_1)$ for $r \in [1,\infty)$.  First we find the possible intersection points
of $\ell_V(x)$ with $\partial(T(\Y))$ and $\partial(R(\y_j))$ for $j=1,2,3$.
Let
$$
G_1=\xi_1(r,x)\cap e_3, \;\; G_2=\xi_2(r,x)\cap e_3, \;\;G_3=\xi_2(r,x)\cap e_1, \;\;
G_4=\xi_3(r,x)\cap e_1, \;\; G_5=\xi_3(r,x)\cap e_2, \;\; G_6=\xi_1(r,x)\cap e_2.
$$
Then, for example,
$G_5=\left(\frac{(\sqrt{3}r-\sqrt{3}+2\,y)\sqrt{3}}{6\,r},\frac{\sqrt{3}r-\sqrt{3}+2\,y}{2\,r}\right)$.
Furthermore, let \\
$ L_1=\xi_1(r,x)\cap \overline{M_1M}_C, \;\;L_2=\xi_2(r,x)\cap \overline{M_1M}_C,\;\;
L_3=\xi_2(r,x)\cap \overline{M_2M}_C,\;\;
L_4=\xi_3(r,x)\cap \overline{M_2M}_C,\;\; L_5=\xi_3(r,x)\cap \overline{M_3M}_C, \;\;
L_6=\xi_1(r,x)\cap \overline{M_3M}_C.$

Then for example $L_5=\left(-\frac{(\sqrt{3}r-3\,\sqrt{3}+6\,y)\sqrt{3}}{6\,r},
\frac{\sqrt{3}r-\sqrt{3}+2\,y}{2\,r}\right)$.
Then $\G_1(x_1,\NY^r)$ is a polygon whose vertices are a subset of the
$\y_j,M_C,M_j,\;j=1,2,3$ and $G_j,\,L_j,\;j=1,\ldots,6$.

See Figure \ref{fig:G1-NYr-Cases-2} for the prototypes of $\G_1(x_1,\NY^r)$ with $r \in [4/3,3/2)$.

\begin{figure} [t]
   \centering
   \scalebox{.25}{\input{G1ofxCase1.pstex_t}}
   \scalebox{.25}{\input{G1ofxCase2.pstex_t}}
   \scalebox{.25}{\input{G1ofxCase3.pstex_t}}
   \scalebox{.25}{\input{G1ofxCase4.pstex_t}}
   \scalebox{.25}{\input{G1ofxCase5.pstex_t}}
   \scalebox{.25}{\input{G1ofxCase7.pstex_t}}
   \caption{The prototypes of the six cases for $\G_1(x_1,\NY^r)$ for $x \in T(\y_1,M_3,M_{CC})$ for $r \in [4/3,3/2)$.}
\label{fig:G1-NYr-Cases-2}
\end{figure}

\begin{figure} [ht]
    \centering
   \scalebox{.4}{\input{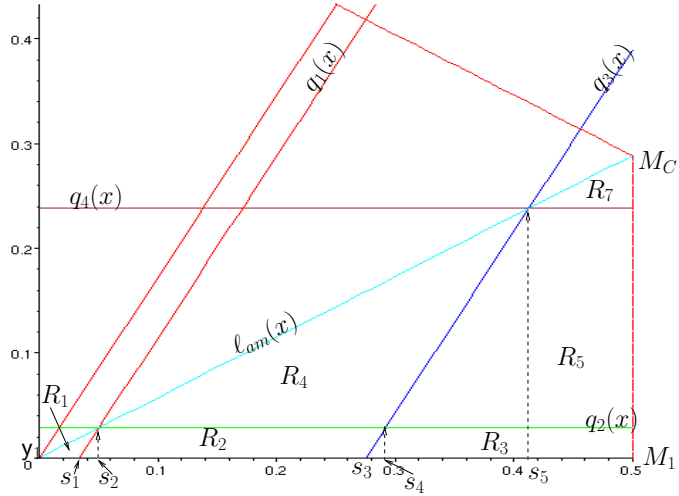}}
    \caption{The regions corresponding to the six cases for $r \in [4/3,3/2)$}
    \label{regions for N_nu6}
\end{figure}

We partition $[1,\infty)$ with respect to the types of $\NY^r(x_1)$ and
$\G_1(x_1,\NY^r)$ into $[1,4/3),\,[4/3,3/2),\,[3/2,2),\,[2,\infty)$.
For demonstrative purposes we pick the interval $[4/3,3/2)$.
For $r \in [\frac{4}{3},\frac{3}{2})$, there are six cases regarding $\G_1(x_1,\NY^r)$ and one case for $\NY^r(x_1)$.
Each case $j$ corresponds to the region $R_j$ in Figure \ref{regions for N_nu6}
where  $s_1=1-2\,r/3, \; s_2=3/2-r,\; s_3=1-r/2, \; s_4=3/2-5\,r/6, \; s_5=3/2-3\,r/4.$

Let $\mathscr P(a_1,a_2,\ldots, a_n)$ denote the polygon with vertices $a_1,a_2,\ldots, a_n $,
then, for $x_1=(x,y)\in R_j$, $j=1,\ldots,6$, $\G_1(x_1,\NY^r)$ are
$\mathscr P(G_1,M_1,M_C,M_3,G_6)$, $\mathscr P(G_1,M_1,L_2,L_3,M_C,M_3,G_6)$,
$\mathscr P(G_1,G_2,G_3,M_2,M_C,M_3,G_6)$, $\mathscr P(G_1,M_1,L_2,L_3,L_4,L_5,M_3,G_6)$,
$\mathscr P(G_1,G_2,G_3,M_2,L_4,L_5,M_3,G_6)$ and $\mathscr P(G_1,G_2,G_3,G_4,G_5,G_6)$,
respectively.

The explicit forms of $R_j$, $j=1,\ldots,6$ are as follows:\\
\\
$R_1=\{(x,y)\in  [0,s_1]\times [0,\ell_{am}(x)] \cup [s_1, s_2]\times [q_1(x), \ell_{am}(x)]\}$\\
$R_2=\{(x,y)\in [s_1,s_2] \times [0,q_1(x)]\cup [s_2,s_3] \times [0,q_2(x)] \cup [s_3,s_4] \times [q_3(x),q_2(x)]\}$\\
$R_3=\{(x,y)\in [s_3,s_4]\times [0,q_3(x)] \cup [s_4,1/2] \times [0,q_2(x)]\}$\\
$R_4=\{(x,y)\in [s_2,s_4] \times [q_2(x),\ell_{am}(x)] \cup  [s_4,s_6] \times [q_3(x),\ell_{am}(x)]\}$\\
$R_5=\{(x,y)\in [ s_4,s_6] \times [q_2(x),q_3(x)] \cup [s_6,1/2] \times [q_2(x),q_4(x)]\}$\\
$R_6=\{(x,y)\in [s_6,1/2] \times [q_4(x),\ell_{am}(x)]\}$,\\
\\
where $\ell_{am}(x)=x/\sqrt{3}$, $q_1(x)=(2\,r-3)/\sqrt{3}+\sqrt{3}\,x$, $q_2(x)=\sqrt{3}\,(1/2-r/3)$,
$q_3(x)=\sqrt{3}\,(x-1+r/2)$, and $q_4(x)=\sqrt{3}(1/2-r/4)$.

Then $P(\{X_2,X_3\} \subset \NY^r(X_1))=\frac{781}{19440}\,r^4$.
(We use the same limits of integration in $\mu(r)$ calculations
with the integrand being $A(\NY^r(x_1))^2/A(T(\Y))^3$.

Next, by symmetry, $P(\{X_2,X_3\} \subset \G_1(X_1,\NY^r))=6\,P(\{X_2,X_3\}
\subset \G_1(X_1,\NY^r),\; X_1 \in T(\y,M_3,M_C)).$ Then
 $$P(\{X_2,X_3\} \subset \G_1(X_1,\NY^r),\; X_1 \in T(\y,M_3,M_C))=\sum_{j=1}^6 P(\{X_2,X_3\}
\subset \G_1(X_1,\NY^r),\; X_1 \in R_j).$$
For example, for $x_1 \in R_4$,
\begin{eqnarray*}
\lefteqn{P(\{X_2,X_3\} \subset \G_1(X_1,\NY^r),\; X_1 \in R_4) =
\int_{s_2}^{s_4}\int_{q_2(x)}^{\ell_{am}(x)} \frac{A(\G_1(x_1,\NY^r))^2}{A(T(\Y))^3}dydx}\\
& &+\int_{s_4}^{s_6}\int_{q_3(x)}^{\ell_{am}(x)} \frac{A(\G_1(x_1,\NY^r))^2}{A(T(\Y))^3}dydx =
\frac{9637\,r^4-89640\,r^3+288360\,r^2-362880\,r+155520}{349920\,r^2}.
\end{eqnarray*}
where $A(\G_1(x_1,\NY^r))=\frac{\sqrt{3}(9\,r^2+18-24\,r+4\,\sqrt{3}r
\,y-18\,x+6\,{x}^2+14\,{y}^2+12\,r\,x-8\,x\,\sqrt{3}y-6\,\sqrt{3}y)}{12\,r^2}$.

Similarly, we calculate for $j=1,2,3,5,6$ and get
\begin{eqnarray*}
P(\{X_2,X_3\} \subset \G_1(X_1,\NY^r))&=&6\,\Biggl(\frac{-47880\,r^5-38880\,r^2+25687\,r^6-
1080\,r^4+60480\,r^3+3888}{349920\,r^4}\Biggr)\\
&=&\frac{-47880\,r^5-38880\,r^2+25687\,r^6-1080\,r^4+60480\,r^3+3888}{58320\,r^4}.
\end{eqnarray*}

Furthermore, $P(X_2 \in \NY^r(X_1),\;X_3\in \G_1(X_1,\NY^r),\; X_1 \in T(\y,M_3,M_C))=
\sum_{j=1}^6 P(X_2 \in \NY^r(X_1),\;X_3\in \G_1(X_1,\NY^r),\; X_1 \in R_j).$

For example, $x_1 \in R_4$, we get $-{\frac{1}{466560}}\,r^2(207360+404640\,r^2-
483840\,r-142920\,r^3+17687\,r^4)$
by using the same integration limits as above, with the integrand
being $A(\NY^r(x_1))\,A(\G_1(x_1,\NY^r))/A(T(\Y))^3$.

Similarly, we calculate for $j=1,2,3,5,6$ and get
{\small
\begin{eqnarray*}
P(X_2 \in \NY^r(X_1),\;X_3\in \G_1(X_1,\NY^r))&=&6\,\Biggl({\frac{5467}{2799360}}\,
r^6-{\frac{35}{2592}}\,r^5+{\frac{37}{1296}}\,r^4-{\frac{13}{648}}\,r^2+{\frac{83}{12960}}\Biggr)\\
&=&{\frac{5467}{466560}}\,r^6-{\frac{35}{432}}\,r^5+{\frac{37}{216}}\,r^4-{\frac{13}{108}}\,r^2+
{\frac{83}{2160}}.
\end{eqnarray*}
}

So,
{\small $\E[h_{12}\,h_{13}]=[5467\,r^{10}-37800\,r^9+89292\,r^8+46588\,r^6-191520\,r^5+
13608\,r^4+241920\,r^3-155520\,r^2+15552]/[233280\,r^4].$
}

Thus, for $r \in [4/3,3/2)$, {\small $\nu(r)=[5467\,r^{10}-37800\,r^9+61912\,r^8+46588\,r^6-
191520\,r^5+13608\,r^4+241920\,r^3-155520\,r^2+15552]/[233280\,r^4].$ }

\section*{Appendix 2: $\mu(r,\epsilon)$ for Segregation and Association Alternatives}

Derivation of $\mu(r,\epsilon)$ involves detailed geometric calculations and partitioning
of the space of $(r,\epsilon,x_1)$ for $r \in [1,\infty)$, $\epsilon \in [0,\sqrt{3}/3)$, and $x_1 \in T_s$.

\subsection*{$\mu_S(r,\epsilon)$ Under Segregation Alternatives}
Under segregation, we compute $\mu_S(r,\epsilon)$ explicitly.
For $\epsilon \in [0,\sqrt{3}/8)$, $\mu_S(r,\epsilon)=\sum_{j=1}^7 \mu_{1,j}(r,\epsilon)\,
\I(r \in \mathcal I_j)$
where
{\small
\begin{align*}
\mu_{1,1}(r,\epsilon)&=-\frac{576\,r^2\epsilon^4-1152\,\epsilon^4-37\,r^2+288\,\epsilon^2}
{216\,(2\,\epsilon+1)^2(2\,\epsilon-1)^2},\\
\mu_{1,2}(r,\epsilon)&=-[576\,r^4\epsilon^4-1152\,r^2\epsilon^4+91\,r^4+
512\,\sqrt{3}r^3\epsilon+2592\,r^2\epsilon^2+1536\,\sqrt{3}r\epsilon^3+1152\,\epsilon^4\\
&-768\,r^3-2304\,\sqrt{3}r^2\epsilon-6912\,r\epsilon^2-2304\,\sqrt{3}\epsilon^3+1728\,r^2+
3456\,\sqrt{3}r\epsilon+5184\,\epsilon^2\\
&-1728\,r-1728\,\sqrt{3}\epsilon+648]/[216\,r^2(2\,\epsilon+1)^2(2\,\epsilon-1)^2],\\
\mu_{1,3}(r,\epsilon)&=-[192\,r^4\epsilon^4-384\,r^2\epsilon^4+9\,r^4+864\,r^2\epsilon^2+
512\,\sqrt{3}r\epsilon^3+384\,\epsilon^4-2304\,r\epsilon^2-768\,\sqrt{3}\epsilon^3\\
&-288\,r^2+1728\,\epsilon^2+576\,r-324]/[72\,r^2(2\,\epsilon+1)^2(2\,\epsilon-1)^2],\\
\mu_{1,4}(r,\epsilon)&=-[192\,r^4\epsilon^4-384\,r^2\epsilon^4-9\,r^4-96\,\sqrt{3}r^3\epsilon+
288\,r^2\epsilon^2-128\,\epsilon^4+144\,r^3+576\,\sqrt{3}r^2\epsilon+256\\
&\sqrt{3}\epsilon^3-720\,r^2-1152\,\sqrt{3}r\epsilon-576\,\epsilon^2+1152\,r+
768\,\sqrt{3}\epsilon-612]/[72\,r^2(2\,\epsilon+1)^2(2\,\epsilon-1)^2],\\
\mu_{1,5}(r,\epsilon)&=-\frac{48\,r^4{\epsilon}^4-96\,r^2{\epsilon}^4+72\,r^2{\epsilon}^2-
32\,\epsilon^4+64\,\sqrt{3}{\epsilon}^3-18\,r^2-144\,\epsilon^2+27}{18\,r^2(2\,\epsilon+1)^2(2\,\epsilon-1)^2},\\
\mu_{1,6}(r,\epsilon)&=\frac{48\,r^4{\epsilon}^4+256\,r^3{\epsilon}^4-
128\,\sqrt{3}r^3{\epsilon}^3+288\,r^2{\epsilon}^4-192\,\sqrt{3}r^2{\epsilon}^3+
72\,r^2{\epsilon}^2+18\,r^2+48\,\sqrt{3}\epsilon-45}{18\,(2\,\epsilon+1)^2(2\,\epsilon-1)^2r^2},\\
\mu_{1,7}(r,\epsilon)&=1,
\end{align*}
}
with the corresponding intervals $\mathcal I_1=[1,3/2-\sqrt{3}\,\epsilon)$,
$\mathcal I_2=[3/2-\sqrt{3}\,\epsilon,3/2)$, $\mathcal I_3=[3/2,2-4\,\epsilon/\sqrt{3})$,
$\mathcal I_4=[2-4\,\epsilon/\sqrt{3},2)$, $\mathcal I_5=[2,\sqrt{3}/(2\,\epsilon)-1)$,
$\mathcal I_6=[\sqrt{3}/(2\,\epsilon)-1,\sqrt{3}/(2\,\epsilon))$, and
$\mathcal I_7=[\sqrt{3}/(2\,\epsilon),\infty)$.

For $\epsilon \in [\sqrt{3}/8,\sqrt{3}/6)$,
$\mu_S(r,\epsilon)=\sum_{j=1}^7 \mu_{2,j}(r,\epsilon)\,\I(r \in \mathcal I_j)$ where
$\mu_{2,j}(r,\epsilon)=\mu_{1,j}(r,\epsilon)$ for $j=1,2,4,5,6$, and for $j=3,7$,
{\small
\begin{align*}
\mu_{2,3}(r,\epsilon)&=-[576\,r^4{\epsilon}^4-1152\,r^2{\epsilon}^4+37\,r^4+
224\,\sqrt{3}r^3\epsilon+864\,r^2{\epsilon}^2-384\,\epsilon^4-336\,r^3-576\,\sqrt{3}r^2\epsilon\\
&+768\,\sqrt{3}{\epsilon}^3+432\,r^2-1728\,\epsilon^2+576\,\sqrt{3}\epsilon-216]/
[216\,r^2(2\,\epsilon+1)^2(2\,\epsilon-1)^2],\\
\mu_{2,7}(r,\epsilon)&=1,
\end{align*}
}
with the corresponding intervals $\mathcal I_1=[1,3/2-\sqrt{3}\,\epsilon)$,
$\mathcal I_2=[3/2-\sqrt{3}\,\epsilon,2-4\,\epsilon/\sqrt{3})$,
$\mathcal I_3=[2-4\,\epsilon/\sqrt{3},3/2)$,
$\mathcal I_4=[3/2,2)$,
$\mathcal I_5=[2,\sqrt{3}/(2\,\epsilon)-1)$,
$\mathcal I_6=[\sqrt{3}/(2\,\epsilon)-1,\sqrt{3}/(2\,\epsilon))$,
and $\mathcal I_5=[\sqrt{3}/(2\,\epsilon),\infty)$.

For $\epsilon \in [\sqrt{3}/6,\sqrt{3}/4)$,
$\mu_S(r,\epsilon)=\sum_{j=1}^6 \mu_{3,j}(r,\epsilon)\,\I(r \in \mathcal I_j)$
where $\mu_{3,1}(r,\epsilon)=\mu_{1,2}(r,\epsilon)$ and
{\small
\begin{align*}
\mu_{3,2}(r,\epsilon)&=-[576\,r^4\epsilon^4-1152\,r^2\epsilon^4+37\,r^4+224\,\sqrt{3}r^3\epsilon+
864\,r^2\epsilon^2-384\,\epsilon^4-336\,r^3-576\,\sqrt{3}r^2\epsilon\\
&+768\,\sqrt{3}\epsilon^3+432\,r^2-1728\,\epsilon^2+576\,\sqrt{3}\epsilon-216]/
[216\,r^2(2\,\epsilon+1)^2(2\,\epsilon-1)^2],\\
\mu_{3,3}(r,\epsilon)&=[576\,r^2\epsilon^4+3072\,r\epsilon^4-1536\,\sqrt{3}r\epsilon^3+
3456\,\epsilon^4-2304\,\sqrt{3}\epsilon^3-37\,r^2-224\,\sqrt{3}r\epsilon\\
&+864\,\epsilon^2+336\,r+576\,\sqrt{3}\epsilon-432]/[216\,(2\,\epsilon+1)^2(2\,\epsilon-1)^2],\\
\mu_{3,4}(r,\epsilon)&=[192\,r^4\epsilon^4+1024\,r^3\epsilon^4-512\,\sqrt{3}r^3\epsilon^3+
1152\,r^2\epsilon^4-768\,\sqrt{3}r^2\epsilon^3+9\,r^4+96\,\sqrt{3}r^3\epsilon+288\,r^2\epsilon^2\\
&-144\,r^3-576\,\sqrt{3}r^2\epsilon+720\,r^2+1152\,\sqrt{3}r\epsilon-1152\,r-576\,\sqrt{3}\epsilon+
540]/[72\,r^2(2\,\epsilon+1)^2(2\,\epsilon-1)^2],\\
\mu_{3,5}(r,\epsilon)&=\frac{48\,r^4\epsilon^4+256\,r^3\epsilon^4-128\,\sqrt{3}r^3\epsilon^3+
288\,r^2\epsilon^4-192\,\sqrt{3}r^2\epsilon^3+72\,r^2\epsilon^2+18\,r^2+48\,\sqrt{3}\epsilon-
45}{18\,r^2(2\,\epsilon+1)^2(2\,\epsilon-1)^2},\\
\mu_{3,6}(r,\epsilon)&=1,
\end{align*}
}
with the corresponding intervals
$\mathcal I_1=[1,2-4\,\epsilon/\sqrt{3})$,
$\mathcal I_2=[2-4\,\epsilon/\sqrt{3},\sqrt{3}/(2\,\epsilon)-1)$,
$\mathcal I_3=[\sqrt{3}/(2\,\epsilon)-1,3/2)$,
$\mathcal I_4=[3/2,2)$, $\mathcal I_5=[2,\sqrt{3}/(2\,\epsilon))$,
and $\mathcal I_5=[\sqrt{3}/(2\,\epsilon),\infty)$.

For $\epsilon \in [\sqrt{3}/4,\sqrt{3}/3)$,
$\mu_S(r,\epsilon)=\sum_{j=1}^3 \mu_{4,j}(r,\epsilon)\,\I(r \in \mathcal I_j)$ where
{\small
\begin{align*}
\mu_{4,1}(r,\epsilon)&=-\frac{9\,r^2\epsilon^2+2\,\sqrt{3}r^2\epsilon+48\,r\epsilon^2+
r^2-16\,\sqrt{3}r\epsilon-90\,\epsilon^2-12\,r+36\,\sqrt{3}\epsilon}{18\,(3\,\epsilon-\sqrt{3})^2},\\
\mu_{4,2}(r,\epsilon)&=-[9\,r^4\epsilon^4-4\,\sqrt{3}r^4\epsilon^3+48\,r^3\epsilon^4-
48\,\sqrt{3}r^3\epsilon^3-90\,r^2\epsilon^4+36\,r^3\epsilon^2+96\,\sqrt{3}r^2\epsilon^3-126\,r^2\epsilon^2\\
&-32\,\sqrt{3}r\epsilon^3-48\,\epsilon^4+36\,\sqrt{3}r^2\epsilon+144\,r\epsilon^2+
96\,\sqrt{3}\epsilon^3-18\,r^2-72\,\sqrt{3}r\epsilon-216\,\epsilon^2+36\,r\\
&+72\,\sqrt{3}\epsilon-27]/[2\,(3\,\epsilon-\sqrt{3})^4r^2],\\
\mu_{4,3}(r,\epsilon)&=1,
\end{align*}
}
with the corresponding intervals
$\mathcal I_1=[1,3-2\,\epsilon/\sqrt{3})$,
$\mathcal I_2=[3-2\,\epsilon/\sqrt{3},\sqrt{3}/\epsilon-2)$,
and $\mathcal I_3=[\sqrt{3}/\epsilon-2,\infty)$.

\subsection*{ $\mu_A(r,\epsilon)$ Under Association Alternatives}

Under association, we compute $\mu_A(r,\epsilon)$ explicitly.
For $\epsilon \in [0,(7\,\sqrt{3}-3\,\sqrt{15})/12 \approx .042)$,
$\mu_A(r,\epsilon)=\sum_{j=1}^6 \mu_{1,j}(r,\epsilon)\,\I(r \in \mathcal I_j)$ where
{\small
\begin{align*}
\mu_{1,1}(r,\epsilon)&=-[3456\,\epsilon^4r^4+9216\,\epsilon^4r^3-3072\,\sqrt{3}{\epsilon}^3r^4-
17280\,\epsilon^4r^2-3072\,\sqrt{3}{\epsilon}^3r^3+2304\,\epsilon^2r^4\\
&+4608\,\sqrt{3}{\epsilon}^3r^2-2304\,\epsilon^2r^3+6336\,\epsilon^4+6144\,\sqrt{3}{\epsilon}^3r+
6912\,\epsilon^2r^2+512\,\sqrt{3}\epsilon\,r^3\\
&-101\,r^4-6144\,\sqrt{3}{\epsilon}^3-11520\,\epsilon^2r-1536\,\sqrt{3}\epsilon\,r^2+256\,r^3+
5760\,\epsilon^2+1536\,\sqrt{3}\epsilon\,r\\ &-384\,r^2-512\,\sqrt{3}\epsilon+256\,r-
64]/[24\,(6\,\epsilon+\sqrt{3})^2(6\,\epsilon-\sqrt{3})^2r^2],\\
\mu_{1,2}(r,\epsilon)&=-[1728\,\epsilon^4r^4-1536\,\sqrt{3}{\epsilon}^3r^4-31104\,\epsilon^4r^2+
1152\,\epsilon^2r^4+15552\,\epsilon^4+10368\,\epsilon^2r^2-37\,r^4\\
&-20736\,\epsilon^2r+10368\,\epsilon^2]/[24\,(6\,\epsilon+\sqrt{3})^2(6\,\epsilon-\sqrt{3})^2r^2],\\
\mu_{1,3}(r,\epsilon)&=[-2592\,\epsilon^4r^4-2304\,\sqrt{3}{\epsilon}^3r^4-46656\,\epsilon^4r^2+
1728\,\epsilon^2r^4+10656\,\epsilon^4-9216\,\sqrt{3}{\epsilon}^3r\\
&+9072\,\epsilon^2r^2-432\,\sqrt{3}\epsilon\,r^3-15\,r^4+12288\,\sqrt{3}{\epsilon}^3-13824\,\epsilon^2r+
1728\,\sqrt{3}\epsilon\,r^2-216\,r^3\\
&+4032\,\epsilon^2-2304\,\sqrt{3}\epsilon\,r+432\,r^2+1024\,\sqrt{3}\epsilon-384\,r+128]/
[36\,(6\,\epsilon+\sqrt{3})^2(6\,\epsilon-\sqrt{3})^2r^2],\\
\mu_{1,4}(r,\epsilon)&=-\frac{1728\,\epsilon^4r^4-1536\,\sqrt{3}{\epsilon}^3r^4-
31104\,\epsilon^4r^2+1152\,\epsilon^2r^4-5184\,\epsilon^4+2592\,\epsilon^2r^2-37\,r^4-
3456\,\epsilon^2}{24\,(6\,\epsilon+\sqrt{3})^2(6\,\epsilon-\sqrt{3})^2r^2},\\
\mu_{1,5}(r,\epsilon)&=\frac{9\,(1152\,\epsilon^4r^2+192\,\epsilon^4-192\,\epsilon^2r^2-
r^4+128\,\epsilon^2+32\,r^2-64\,r+36)}{8\,(6\,\epsilon+\sqrt{3})^2(6\,\epsilon-\sqrt{3})^2r^2},\\
\mu_{1,6}(r,\epsilon)&=-\frac{9\,(r+6)(r-2)^3}{8\,(6\,\epsilon+\sqrt{3})^2(6\,\epsilon-\sqrt{3})^2r^2},
\end{align*}
}
with the corresponding intervals $\mathcal I_1=\Bigl[1,\frac{1+2\,\sqrt{3}\,\epsilon}{1-
\sqrt{3}\,\epsilon}\Bigr)$, $\mathcal I_2=\Bigl[\frac{1+2\,\sqrt{3}\,\epsilon}{1-
\sqrt{3}\,\epsilon},\frac{4\,(1-\sqrt{3}\,\epsilon}{3}\Bigr)$, $\mathcal I_3=
\Bigl[\frac{4\,(1-\sqrt{3}\,\epsilon}{3},\frac{4\,(1+2\,\sqrt{3}\,\epsilon}{3}\Bigr)$,
$\mathcal I_4=\Bigl[\frac{4\,(1+2\,\sqrt{3}\,\epsilon}{3},\frac{3}{2\,(1-\sqrt{3}\,\epsilon)}\Bigr)$,
$\mathcal I_5=\Bigl[\frac{3}{2\,(1-\sqrt{3}\,\epsilon)},2\Bigr)$
and $\mathcal I_6=[2,\infty)$.

For $\epsilon \in [(7\,\sqrt{3}-3\,\sqrt{15})/12,\sqrt{3}/12)$,
$\mu_A(r,\epsilon)=\sum_{j=1}^6 \mu_{2,j}(r,\epsilon)\,\I(r \in \mathcal I_j)$
where $\mu_{2,j}(r,\epsilon)=\mu_{1,j}(r,\epsilon)$ for $j=1,3,4,5,6$ and
{\small
\begin{align*}
\mu_{2,2}(r,\epsilon)&=[-3456\,\epsilon^2r^4+111\,r^4-5184\,\epsilon^4r^4+
4608\,\sqrt{3}\epsilon^3r^4-336\,\sqrt{3}\epsilon\,r^3-168\,r^3-13824\,\epsilon^4r^3\\
&+4608\,\sqrt{3}\epsilon^3r^3+3456\,\epsilon^2r^3+144\,r^2-6912\,\sqrt{3}\epsilon^3r^2-
3888\,\epsilon^2r^2+576\,\sqrt{3}\epsilon\,r^2\\
&+25920\,\epsilon^4r^2+3168\,\epsilon^4+2880\,\epsilon^2-256\,\sqrt{3}\epsilon-32-
3072\,\sqrt{3}\epsilon^3]/[36\,(\sqrt{3}+6\,\epsilon)^2(-6\,\epsilon+\sqrt{3})^2r^2]
\end{align*}
}
with the corresponding intervals
$\mathcal I_1=\Bigl[1,\frac{4\,(1-\sqrt{3}\,\epsilon)}{3} \Bigr)$,
$\mathcal I_2=\Bigl[\frac{4\,(1-\sqrt{3}\,\epsilon)}{3},\frac{1+2\,\sqrt{3}\,\epsilon}{1-\sqrt{3}\,\epsilon}\Bigr)$,
$\mathcal I_3=\Bigl[\frac{1+2\,\sqrt{3}\,\epsilon}{1-\sqrt{3}\,\epsilon},\frac{4\,(1+2\,\sqrt{3}\,\epsilon}{3} \Bigr)$,
$\mathcal I_4=\Bigl[\frac{4\,(1+2\,\sqrt{3}\,\epsilon}{3},\frac{3}{2\,(1-\sqrt{3}\,\epsilon)}\Bigr)$,
$\mathcal I_5=\Bigl[\frac{3}{2\,(1-\sqrt{3}\,\epsilon)},2\Bigr)$
and $\mathcal I_6=[2,\infty)$.

For $\epsilon \in [\sqrt{3}/12,\sqrt{3}/3)$,
$\mu_A(r,\epsilon)=\sum_{j=1}^3 \mu_{3,j}(r,\epsilon)\,\I(r \in \mathcal I_j)$ where
\begin{align*}
\mu_{3,1}(r,\epsilon)&=\frac{2\,r^2-1}{6\,r^2},\\
\mu_{3,2}(r,\epsilon)&=[432\,\epsilon^4r^4+1152\,\epsilon^4r^3-576\,\sqrt{3}{\epsilon}^3r^4+
1296\,\epsilon^4r^2-960\,\sqrt{3}{\epsilon}^3r^3+864\,\epsilon^2r^4-864\,\sqrt{3}{\epsilon}^3r^2\\
&+576\,\epsilon^2r^3-192\,\sqrt{3}\epsilon\,r^4-360\,\epsilon^4+648\,\epsilon^2r^2+
64\,\sqrt{3}\epsilon\,r^3+48\,r^4+192\,\sqrt{3}{\epsilon}^3-144\,\sqrt{3}\epsilon\,r^2\\
&-64\,r^3-504\,\epsilon^2+72\,r^2+88\,\sqrt{3}\epsilon-25]/[16\,(3\,\epsilon-\sqrt{3})^4r^2],\\
\mu_{3,3}(r,\epsilon)&=-\frac{-54\,\epsilon^2r^2+36\,\sqrt{3}\epsilon\,r^2+15\,\epsilon^2-
18\,r^2+2\,\sqrt{3}\epsilon+20}{6\,(-3\,\epsilon+\sqrt{3})^2r^2},
\end{align*}
with the corresponding intervals $\mathcal I_1=\Bigl[1,\frac{1+2\,\sqrt{3}\,
\epsilon}{2\,(1-\sqrt{3}\,\epsilon)}\Bigr)$, $\mathcal I_3=\Bigl[\frac{1+
2\,\sqrt{3}\,\epsilon}{2\,(1-\sqrt{3}\,\epsilon)},\frac{3}{2\,(1-\sqrt{3}\,
\epsilon)}\Bigr)$, $\mathcal I_5=\Bigl[\frac{3}{2\,(1-\sqrt{3}\,\epsilon)},\infty\Bigr)$.

\section*{Appendix 3: $\mu(r,\epsilon)$ and $\nu(r,\epsilon)$ for Segregation and
Association Alternatives with Sample $\epsilon$ values}
With $\epsilon=\sqrt{3}/4$, $r\in [1,2)$,
$ \mu_S(r,\sqrt{3}/4)=
\begin{cases}
          -{\frac{67}{54}}\,r^2+{\frac{40}{9}}\,r-3 &\text{for} \quad r \in [1,3/2)\\
           {\frac{7\,r^4-48\,r^3+122\,r^2-128\,r+48}{2\,r^2}} &\text{for} \quad r \in [3/2,2)
\end{cases}$ and\\
 $\nu_S(r,\sqrt{3}/4)=\sum_{j=1}^5\nu_j(r,\sqrt{3}/4)\,\I(\mathcal I_j)$
where
{\small
\begin{align*}
\nu_1(r,\sqrt{3}/4)&=-[14285\,r^7-28224\,r^6-233266\,r^5+1106688\,r^4-2021199\,r^3+1876608\,r^2\\
&-880794\,r+165888]/[3645\,r],\\
\nu_2(r,\sqrt{3}/4)&=-[14285\,r^{10}-28224\,r^9-233266\,r^8+1106688\,r^7-1234767\,r^6-3431808\,r^5\\
&+14049126\,r^4-22228992\,r^3+18895680\,r^2-8503056\,r+1594323]/[3645\,r^4],\\
\nu_3(r,\sqrt{3}/4)&=-[14285\,r^{10}-28224\,r^9-233266\,r^8+1106688\,r^7-2545713\,r^6+5903280\,r^5\\
&-13456044\,r^4+20636208\,r^3-18305190\,r^2+8503056\,r-1594323]/[3645\,r^4],\\
\nu_4(r,\sqrt{3}/4)&=[104920\,r^8-111072\,r^7+1992132\,r^6-15844032\,r^5+50174640\,r^4+6377292\\
&-34012224\,r+73220760\,r^2-81881280\,r^3+1909\,r^{10}-27072\,r^9]/[14580\,r^4],\\
\nu_5(r,\sqrt{3}/4)&=-[-1187904\,r^5+1331492\,r^6+433304\,r^2+611163\,r^{10}-850240\,r^9-198144\,r\\
&+955392\,r^4-705536\,r^3-387680\,r^{11}+1118472\,r^8-1308960\,r^7+175984\,r^{12}\\
&-46176\,r^{13}+5120\,r^{14}+56016]/[20\,r^4],
\end{align*}
}
and the corresponding intervals are $\mathcal I_1=[1,\frac{9}{8}),\; \mathcal I_2=[9/8,9/7),\;
\mathcal I_3=[9/7,4/3),\; \mathcal I_4=[4/3,3/2),\; \mathcal I_5=[3/2,2)$.

With $\epsilon=\sqrt{3}/12$,
$\mu_A(r,\sqrt{3}/12)=
\begin{cases}
          \frac{6\,r^4-16\,r^3+18\,r^2-5}{18\,r^2} &\text{for} \quad r \in [1,2)\\
           -{\frac{37}{18}}\,r^{-2}+1 &\text{for} \quad r \in [2,\infty)
\end{cases}$ and
 $\nu_A(r,\sqrt{3}/12)=\sum_{j=1}^3\nu_j(r,\sqrt{3}/12)\,\I(\mathcal I_j)$
where
{\small
\begin{align*}
\nu_1(r,\sqrt{3}/12)&=[10\,r^{12}-96\,r^{11}+240\,r^{10}+192\,r^9-1830\,r^8+3360\,r^7-
2650\,r^6+240\,r^5+1383\,r^4\\
&-1280\,r^3+540\,r^2-144\,r+35]/[405\,r^6],\\
\nu_2(r,\sqrt{3}/12)&=[10\,r^{12}-96\,r^{11}+240\,r^{10}+192\,r^9-1670\,r^8+2784\,r^7-
2650\,r^6+2400\,r^5-1047\,r^4\\
&-1280\,r^3+1269\,r^2-144\,r+35]/[405\,r^6],\\
\nu_3(r,\sqrt{3}/12)&=\frac{537\,r^4-683\,r^2-2448\,r+1315}{405\,r^6}.
\end{align*}
}
The corresponding intervals are $\mathcal I_1=[1,3/2),\; \mathcal I_2=[3/2,2),\; \mathcal I_3=[2,\infty)$.

\section*{Appendix 4: Proof of Corollary 1}
In the multiple triangle case,
\begin{multline*}
\mu(r,J)=\E[\rho_n(r)]=\frac{1}{n\,(n-1)}\sum\hspace*{-0.1 in}\sum_{i < j \hspace*{0.25 in}}
\hspace*{-0.1 in} \,\E[h_{ij}]=\frac{1}{2}\E[h_{12}]=\E[I(A_{12})] =P(A_{12})=P(X_2 \in \NY^r(X_1)).
\end{multline*}
But, by definition of $N_{\Y}^r(\cdot)$, $P(X_2 \in \NY^r(X_1))=0$ if $X_1$ and $X_2$ are in
 different triangles. So by the law of total probability
\begin{eqnarray*}
\mu(r,J)&:=&P(X_2 \in \NY^r(X_1))= \sum_{j=1}^{J}P(X_2 \in \NY^r(X_1)\,|\,\{X_1,X_2\}
\subset T_j)\,P(\{X_1,X_2\} \subset T_j)\\
&=& \sum_{j=1}^{J}\mu(r)\,P(\{X_1,X_2\} \subset T_j) \;\;\;\text{ (since $P(X_2 \in
\NY^r(X_1)\,|\,\{X_1,X_2\} \subset T_j)=\mu(r)$)}\\
&=& \mu(r) \, \sum_{j=1}^{J}(A(T_j) / A(C_H(\Y)))^2\;\;\; \text{ (since $P(\{X_1,X_2\}
\subset T_j)=(A(T_j) / A(C_H(\Y)))^2$)}
\end{eqnarray*}
Letting $w_j:=A(T_j) / A(C_H(\Y))$, we get $\mu(r,J)=\mu(r)\cdot(\sum_{j=1}^{J}w_j^2)$
where $\mu(r)$ is given by equation (\ref{eq:Asymean}).

Furthermore, the asymptotic variance is
\begin{eqnarray*}
\nu(r,J)&=&\E[h_{12}\,h_{13}]-\E[h_{12}]\E[h_{13}]\\
& = & P(\{X_2,X_3\} \subset \NY^r(X_1))+2\,P(X_2 \in \NY^r(X_1), X_3 \in \G_1(X_1,\NY^r))\\
& &+P(\{X_2,X_3\} \subset \G_1(X_1,\NY^r))-4\,(\mu(r,J))^2.
\end{eqnarray*}
Then for $J>1$, we have
\begin{eqnarray*}
P(\{X_2,X_3\} \subset \NY^r(X_1))&=&\sum_{j=1}^{J}P(\{X_2,X_3\} \subset \NY^r(X_1)\,|\,
\{X_1,X_2,X_3\} \subset T_j)\, P(\{X_1,X_2,X_3\} \subset T_j)\\
& = &\sum_{j=1}^{J}P^r_{2N}\, (A(T_j) / A(C_H(\Y)))^3 =P^r_{2N}\, \bigl(\sum_{j=1}^{J}w_j^3 \bigr).
\end{eqnarray*}
Similarly, $P(X_2 \in \NY^r(X_1), X_3 \in \G_1(X_1,\NY^r))=P^r_{M}\,\bigl(\sum_{j=1}^{J}w_j^3 \bigr)
\text{  and  }P(\{X_2,X_3\} \subset \G_1(X_1,\NY^r))=P^r_{2G}\,\bigl(\sum_{j=1}^{J}w_j^3 \bigr)$,
hence, $\nu(r,J)=(P^r_{2N}+2\,P^r_M+P^r_{2G})\,\bigl(\sum_{j=1}^{J}w_j^3 \bigr)-4\,(\mu(r,J))^2
=\nu(r)\,\bigl(\sum_{j=1}^{J}w_j^3 \bigr)+4\,\mu(r)^2\,\Bigl(\sum_{j=1}^{J}w_j^3-
\bigl(\sum_{j=1}^{J}w_j^2 \bigr)^2\Bigr),$ so conditional on $\mathcal W$, if $\nu(r,J)>0$
then $\sqrt{n}\,(\rho_n(r)-\mu(r,J)) \stackrel {\mathcal L}{\longrightarrow}
\mathcal N(0,\nu(r,J))$. $\blacksquare$

\end{document}